\documentclass[12pt,a4paper]{article}

\usepackage[latin1]{inputenc}
\usepackage{amsmath}
\usepackage{amsfonts}
\usepackage{amssymb}
\usepackage{latexsym}
\usepackage{wasysym}
\usepackage{appendix}
\usepackage{setspace}
\usepackage{graphicx}		
\usepackage{subfigure}
\usepackage{epsfig}
\usepackage{rotating}
\numberwithin{equation}{section}
\usepackage{lscape}
\newcommand{\bigslant}[2]{{\raisebox{.2em}{$#1$}\left/\raisebox{-.2em}{$#2$}\right.}} 
\usepackage[margin=2.50cm,nohead]{geometry}

\begin{document}

\begin{flushright}
CERN-PH-TH/2011-107, FTUV-11-1706, 
MAN/HEP/2011/06\\
June 2011
\end{flushright}

\medskip

\begin{center}
{\bf \Large{Vacuum Topology of the Two Higgs Doublet Model} }
\end{center}

\bigskip

\begin{center}
\large{Richard A. Battye$^a$, Gary D. Brawn$^a$ and Apostolos
  Pilaftsis$^{b,\, c,\, d}$}  
\end{center}

\begin{flushleft}
$^a$\textit{Jodrell Bank Centre for Astrophysics, School of Physics
    and Astronomy, University of Manchester, Manchester M13 9PL,
    United Kingdom}\\[2mm] 
$^b$\textit{Theory Division, CERN, CH-1211
    Geneva 23, Switzerland}\\[2mm] 
$^c$\textit{School of Physics and
    Astronomy, University of Manchester, Manchester M13 9PL, United
    Kingdom}\\[2mm] 
$^d$\textit{Department of Theoretical Physics and
    IFIC, University of Valencia, E-46100, Valencia, Spain}
\end{flushleft}

\bigskip\bigskip

\centerline{\bf ABSTRACT}
\vspace{2mm}

\noindent
We perform  a systematic study of generic  accidental Higgs-family and
CP  symmetries   that  could  occur   in  the  two-Higgs-doublet-model
potential, based on a Majorana scalar-field formalism which realizes a
subgroup  of  $\mathrm{GL(8,\mathbb{C})}$.    We  derive  the  general
conditions  of convexity  and stability  of the  scalar  potential and
present   analytical  solutions  for   two  non-zero   neutral  vacuum
expectation  values of the  Higgs doublets  for a  typical set  of six
symmetries, in terms of  the gauge-invariant parameters of the theory.
By~means  of a  homotopy-group analysis,  we identify  the topological
defects  associated with  the  spontaneous symmetry  breaking of  each
symmetry, as well  as the massless Goldstone bosons  emerging from the
breaking of the continuous symmetries. We find the existence of domain
walls  from  the breaking  of  $\mathrm{Z}_2$,  CP1  and CP2  discrete
symmetries,       vortices      in       models       with      broken
$\mathrm{U(1)}_{\mathrm{PQ}}$ and CP3 symmetries and a global monopole
in   the  $\mathrm{SO(3)}_{\mathrm{HF}}$-broken   model.   The~spatial
profile of the  topological defect solutions is studied  in detail, as
functions of the potential  parameters of the two-Higgs doublet model.
The~application  of our  Majorana scalar-field  formalism  in studying
more  general  scalar  potentials  that  are not  constrained  by  the
$\mathrm{U(1)_Y}$ hypercharge  symmetry is discussed.   In particular,
the same formalism  may be used to properly  identify seven additional
symmetries that may take place in a $\mathrm{U(1)_Y}$-invariant scalar
potential.

\thispagestyle{empty}

\newpage
\pagestyle{empty}
\tableofcontents

\newpage
\pagestyle{plain}
\setcounter{page}{1}

\newpage
\section{Introduction}

The standard  theory of  electroweak interactions, the  Standard Model
(SM)    \cite{Glashow:1961tr,Weinberg:1967tq,Salam:1968rm},    is    a
renormalizable theory  with a minimal particle  content which realizes
the                           Higgs                          mechanism
\cite{Higgs:1964ia,Englert:1964et,Higgs:1964pj,Guralnik:1964eu}      to
account  for  the origin  of  mass of  the  charged  fermions and  the
$\mathrm{W}^{\pm}$  and  $\mathrm{Z}$  bosons.  The SM  describes  the
experimental  data  collected over  the  years  at  the LEP  collider,
Tevatron  and in a  number of  low-energy experiments  with remarkable
success~\cite{Langacker:2009my}. In spite  of its conspicuous success,
however, several  key questions remain unanswered within  the SM, such
as the stability of the  gauge boson masses under quantum corrections,
the possible  unification of the  strong with the  electroweak forces,
the  Dark  Matter problem  and  the existence  of  new  sources of  CP
violation  to  account  for  the  observed  baryon  asymmetry  in  the
Universe.

Supersymmetric  theories softly  broken  at the  TeV  scale provide  a
natural framework to successfully  address all the above problems (for
a  recent  review,   see~\cite{Ibrahim:2007fb}).  In  particular,  the
Minimal Supersymmetric extension of the Standard Model (MSSM) requires
the existence of one more Higgs doublet~$\phi_2$ in addition to the SM
Higgs doublet~$\phi_1$,  so as to  maintain the holomorphicity  of the
superpotential and  ensure the  cancellation of the  chiral anomalies.
In the MSSM, CP-even~\cite{Okada:1990vk,Ellis:1990nz,Haber:1990aw} and
CP-odd~\cite{Pilaftsis:1998pe,Pilaftsis:1998dd,Demir:1999hj} radiative
corrections to  the scalar potential  can be very  significant, giving
rise          to           an          effective          CP-violating
potential~\cite{Pilaftsis:1999qt,Choi:2000wz,Carena:2000yi,Frank:2006yh}
which   acquires   the   form   of   the   Two-Higgs   Doublet   Model
(2HDM)~\cite{Lee:1973iz}~\footnote[1]{Historically,  the bilinear mass
  operator,  $\big(m^2_{12}\,\phi^\dagger_1\phi_2 +  {\rm H.c.}\big)$,
  was missing  in the original  article by T.D.~Lee~\cite{Lee:1973iz}.
  However,  it is  worth mentioning  that this  dimension-two operator
  plays an important  role in the renormalization of  the general 2HDM
  potential~\cite{Weinberg:1973ua},  including the  renormalization of
  possible CP-odd tadpole graphs~\cite{Pilaftsis:1998pe}.}.

Recently, a classification of  all the possible accidental symmetries that
could     occur      in     a     2HDM      potential     has     been
attempted~\cite{Ivanov:2007de,Ferreira:2009wh,Ma:2009ax,
  Ferreira:2010yh}.   Such a partial  classification was  motivated by
the use of a  gauge-invariant bilinear scalar-field formalism based on
the
$\mathrm{SL(2,\mathbb{C})}$~group~\cite{Nishi:2006tg,Ivanov:2006yq,
  Nishi:2007nh},         or         its         SU(2)         subgroup
\cite{Maniatis:2006fs,Maniatis:2007vn,Ferreira:2009wh}~\footnote[2]{Note
  that   the  largest  possible   symmetry  group   of  the   2HDM  is
  O(8)~\cite{Deshpande:1977rw},  giving  rise  to  a large  number  of
  symmetry breaking patterns, beyond  the restricted set considered so
  far  which realize  O(3) and  its maximal  subgroups.}.   The latter
subgroup   emerges  as   a  reparameterization   group  of   the  2HDM
potential~\cite{Ginzburg:2004vp}     in     the     restricted     two
Higgs-doublet-field  basis $\phi_{1,2}$~\footnote[3]{As  we  will see,
  however, the maximal reparameterization  group of the 2HDM potential
  is  $\mathrm{GL(8,\mathbb{R})}$, which  acts  on the  8 real  scalar
  fields contained in the two Higgs doublets $\phi_{1,2}$ and includes
  gauge transformations.}, upon  canonical renormalization of possible
loop-induced  Higgs-mixing kinetic  terms~\cite{Pilaftsis:1997dr}.  In
detail,  the 2HDM  potential  may exhibit  accidental symmetries,  for
given  choices  of  its  theoretical  parameters,  and  following  the
terminology in~\cite{Ferreira:2009wh,Ferreira:2010yh}, there exist two
classes  of symmetries.   The first  class of  symmetries  involve the
transformation of  the two Higgs doublets $\phi_{1,2}$,  but not their
complex  conjugates~$\phi^*_{1,2}$, and are  called Higgs  Family (HF)
symmetries.   The second class  linearly maps  the fields~$\phi_{1,2}$
into their CP-conjugates~$\phi^*_{1,2}$ and are therefore termed CP
symmetries.

Three physically interesting HF symmetries  of the 2HDM that have been
discussed  extensively  in  the  literature  are:  the  $\mathrm{Z}_2$
discrete  symmetry~\cite{Glashow:1976nt},  the Peccei--Quinn  symmetry
$\mathrm{U(1)}_{\rm  PQ}$~\cite{Peccei:1977hh}  and  the  HF  symmetry
$\mathrm{SO(3)}_{\rm
  HF}$~\cite{Deshpande:1977rw,Ivanov:2007de,Ma:2009ax,Ferreira:2010yh}
which involves  an $\mathrm{SU(2)}_{\rm HF}/\mathrm{Z}_2$  rotation of
the Higgs doublets $\phi_{1,2}$. Likewise, three typical CP symmetries
of the 2HDM  that have received much attention are:  the CP1 symmetry, which
realizes   the   canonical    CP   transformation   $\phi_{1(2)}   \to
\phi^*_{1(2)}$~\cite{Lee:1973iz,Deshpande:1977rw,Branco:1980sz},    the
CP2          symmetry,          where         $\phi_{1(2)}          \to
(-)\phi^*_{2(1)}$~\cite{Davidson:2005cw},  and the  CP3  symmetry, which
combines   CP1    with   an   $\mathrm{SO(2)}_{\rm   HF}/\mathrm{Z}_2$
transformation               of               the               fields
$\phi_{1,2}$~\cite{Ivanov:2007de,Ferreira:2009wh,Ma:2009ax,Ferreira:2010yh}.

In this paper,  we introduce a Majorana scalar-field  basis where both
the  HF and  CP  symmetries can  be  realized by  acting  on the  same
representation  of   Higgs  fields.    To~this  end,  we   extend  the
aforementioned  gauge-invariant  bilinear   formalism  to  the  larger
complex  linear  group   $\mathrm{GL(8,\mathbb{C})}$,  which  is  then
reduced by  a Majorana constraint and  gauge invariance. Specifically,
$\mathrm{GL(8,\mathbb{C})}$ is the  reparameterization group acting on
the 8-dimensional complex field multiplet $\Phi$ that contains the two
Higgs   doublets  $\phi_{1,2}$   and   their  hypercharge   conjugates
$i\sigma^2\phi^*_{1,2}$ as components,  where $\sigma^2$ is the second
Pauli  matrix. The  multiplet  $\Phi$ satisfies  the Majorana  constraint
$\Phi = {\rm C}\, \Phi^*$ which, together with the constraint of ${\rm
  SU(2)_L     \otimes     U(1)_Y}$     gauge    invariance,     reduces
$\mathrm{GL(8,\mathbb{C})}$   into   two   subgroups   isomorphic   to
$\mathrm{GL(4,\mathbb{R})}$, where  ${\rm C}$ is  a charge-conjugation
matrix  defined in Section  2.  The  first subgroup  is related  to the HF
transformations and  the second one to the generalized CP transformations
on the  Majorana field  multiplet $\Phi$. Therefore,  we refer  to the
above description as the Majorana scalar-field formalism, or in short,
the Majorana formalism.  

As we  will explicitly demonstrate  in Section \ref{Beyond  the 2HDM},
the $\mathrm{GL(8,\mathbb{C})}$ Majorana  formalism has the analytical
advantage that  scalar potentials being only constrained  by the ${\rm
  SU(2)_L}$  gauge group,  but not  by  ${\rm U(1)}_{\rm  Y}$, can  be
described in a  similar quadratic form as in  the usual ${\rm SU(2)_L}
\otimes {\rm U(1)}_{\rm  Y}$ gauge-invariant~2HDM.  In particular, the
same formalism can be used  to identify symmetries of ${\rm U(1)}_{\rm
  Y}$-invariant 2HDM  potentials that are larger than  ${\rm O(3)}$ in
the bilinear field space, such as O(8) and ${\rm O(4)\otimes O(4)}$ in
the  real  field  space~\cite{Deshpande:1977rw}.   As~we will  see  in
Section~\ref{Beyond  the 2HDM},  these  latter symmetries  fail to  be
captured      by     the      restricted     framework      of     the
$\mathrm{SL(2,\mathbb{C})}$  bilinear approach  adopted in  the recent
literature.

In  this  article,  we  also  derive the  complete  set  of  algebraic
conditions  for  the  convexity   of  the  general  CP-violating  2HDM
potential  and  its  boundedness  from below,  by  applying  Sylvester's
criterion  (see, e.g.~\cite{Sylvester}).   These  algebraic conditions
extend  previous  partial  results  obtained  in  the  literature  for
particular            forms           of            the           2HDM
potential~\cite{Deshpande:1977rw,ElKaffas:2006nt,Grzadkowski:2009bt}
and may have  a geometric interpretation in terms  of conical sections
as                  presented                 in~\cite{Ivanov:2006yq}.
Following~\cite{Ivanov:2006yq,Maniatis:2006fs}, we employ the Lagrange
multiplier  method  to  analytically  calculate all  non-zero  neutral
vacuum   expectation    value   (VEV)   solutions    for   the   Higgs
doublets~$\phi_{1,2}$,  associated  with the  six  generic  HF and  CP
symmetries mentioned  above. The non-zero VEV  solutions are expressed
entirely  in terms of  the gauge-invariant  parameters of  the theory,
thereby  obtaining the analytical  dependence of  possible non-trivial
topological  features in  the vacuum  manifold. As  a  cross-check, we
verify that our solutions  satisfy the minimization conditions derived
by  more  traditional  methods   as  explicitly  given,  for  example,
in~\cite{Pilaftsis:1999qt}.

In order to get a topologically  stable solution in the 2HDM, both the
VEVs  of the  two Higgs  doublets should  be non-zero,  such  that the
topological  configuration cannot  be  removed away  by ${\rm  SU(2)_L
  \otimes  U(1)_Y}$ gauge  transformations. We  use  an homotopy-group
analysis to determine the nature of the topological defects associated
with  the  spontaneous  symmetry  breaking  of  each  symmetry.   More
explicitly,  topological defects,  such  as domain  walls, strings  or
vortices and monopoles,  are created, when a symmetry  group ${\rm G}$
of  the Lagrangian,  which may  be either  local, global  or discrete,
breaks down into  a subgroup ${\rm H}$, in a way  such that the vacuum
manifold  ${\cal  M}  =  {\rm  G/H}$  is  not  trivial.   Knowing  the
topological  properties of the  vacuum manifold  ${\cal M}$  under its
homotopy  groups, $\Pi_n  ({\cal M})$,  determines the  nature  of the
topological  defects~\cite{Vilenkin,Vachaspati}.   Thus, domain  walls
arise for  $\Pi_0 ({\cal  M}) \neq {\bf  I}$, strings or  vortices for
$\Pi_1 ({\cal M})  \neq {\bf I}$, monopoles if  $\Pi_2 ({\cal M}) \neq
{\bf  I}$  and  textures  if   $\Pi_{n  >  2}  ({\cal  M})  \neq  {\bf
  I}$~\cite{Vilenkin,Hindmarsh:1994re},   where  ${\bf   I}$   is  the
identity element.   After having identified the precise  nature of the
topological  solution,  we  then  study quantitatively  their  spatial
profile, as a function of the potential parameters. The results of our
analysis  may  be used  in  future  studies  to derive  cosmo\-logical
constraints on the 2HDM, or  on inflationary models with related SU(2)
group
structure~\cite{Garbrecht:2006az,Battye:2008wu,Jeannerot:2003qv,Kibble:1982dd,Kibble:1982ae}.

The layout of the paper is as follows. Section 2 briefly reviews basic
aspects of a general tree-level 2HDM potential, on which we derive the
sufficient and necessary conditions  for its convexity and boundedness
from  below.    In  the  same  section,  we   introduce  the  Majorana
scalar-field formalism  for describing the 2HDM potential,  as well as
possible extended  scalar potentials that  are not constrained  by the
${\rm U(1)_Y}$  hypercharge group. In  addition, we present  the group
structure of the  vacuum manifolds for a set of six  generic HF and CP
symmetries. In  Section 3, we  calculate the neutral  vacuum solutions
for the three HF symmetries,  ${\rm Z_2}$, ${\rm U(1)_{PQ}}$ and ${\rm
  SO(3)_{HF}}$,    and   identify   their    topological   properties.
Correspondingly, Section~4 discusses  the neutral vacuum solutions and
their topology, for the CP symmetries: CP1, CP2 and CP3. In Section 5,
we perform  a quantitative analysis  of all the  topological solutions
found above, in terms of the fundamental parameters of the theory.  We
present  several key  features of  the topological  defects, including
their spatial profile and  energy density.  In Section~\ref{Beyond the
  2HDM},  we  show how  the  Majorana  scalar-field  formalism can  be
extended to  study ${\rm  U(1)_Y}$-violating 2HDM potentials.  We also
show how  the same extended  version of the  formalism can be  used to
identify  further accidental  symmetries that  could take  place  in a
${\rm U(1)_Y}$-invariant 2HDM  potential.  Finally, Section 7 contains
our   conclusions.    Some  technical   details   of  our   analytical
calculations are presented in Appendices A--D.

\bigskip

\section{Two Higgs Doublet Model Potential}

In this  section we  first review the  2HDM potential in  the bilinear
field  formalism~\cite{Nishi:2006tg,Ivanov:2006yq,Maniatis:2006fs}. We
then derive the conditions for  convexity and stability of the general
2HDM potential, and briefly explain the Lagrange multiplier method for
finding the neutral VEV solutions  for the two Higgs doublets. We then
proceed by introducing our Majorana scalar-field formalism and present
the group  structure of the six  generic symmetries that  may occur in
the 2HDM potential. Finally,  we discuss the general group-theoretical
properties of  the vacuum  manifold, which enable  us to  identify the
exact nature of the topological defects in the 2HDM.

Let us start  our discussion by writing down  the tree-level structure
of the general 2HDM potential ${\rm V}$:
\begin{eqnarray}
\label{V2HDM}
\mathrm{V} &=& -\mu_1^2 (\phi_1^{\dagger} \phi_1) - \mu_2^2
(\phi_2^{\dagger} \phi_2) - m_{12}^2 (\phi_1^{\dagger} \phi_2) -
m_{12}^{*2}(\phi_2^{\dagger} \phi_1) \nonumber \\ 
&&+ \lambda_1 (\phi_1^{\dagger} \phi_1)^2 + \lambda_2
(\phi_2^{\dagger} \phi_2)^2 + \lambda_3 (\phi_1^{\dagger}
\phi_1)(\phi_2^{\dagger} \phi_2) + \lambda_4 (\phi_1^{\dagger}
\phi_2)(\phi_2^{\dagger} \phi_1) \nonumber \\ 
&&+ \frac{\lambda_5}{2} (\phi_1^{\dagger} \phi_2)^2 +
\frac{\lambda_5^{*}}{2} (\phi_2^{\dagger} \phi_1)^2 + \lambda_6
(\phi_1^{\dagger} \phi_1) (\phi_1^{\dagger} \phi_2) + \lambda_6^{*}
(\phi_1^{\dagger} \phi_1)(\phi_2^{\dagger} \phi_1) \nonumber \\ 
&& + \lambda_7 (\phi_2^{\dagger} \phi_2) (\phi_1^{\dagger} \phi_2) +
\lambda_7^{*} (\phi_2^{\dagger} \phi_2) (\phi_2^{\dagger} \phi_1)\; . 
\end{eqnarray}
It is easy to see that the 2HDM potential $\mathrm{V}$ contains 3 mass
parameters $\mu_1^2$, $\mu_2^2$ and  $m_{12}^2$ and 7 quartic  couplings 
$\lambda_{1,2,\dots,7}$. For the potential $\rm V$ to be Hermitian, the 
parameters $\mu_{1,2}^2$ and $\lambda_{1,2,\dots,4}$ are constrained to 
be real, whereas $m_{12}^2$ and $\lambda_{5,6,7}$ are in general complex. 
In order to evaluate the  VEVs of the Higgs doublets  $\phi_1$ and $\phi_2$, 
we have to  calculate first the  extremization conditions by  solving the
two coupled cubic equations
\begin{subequations}
\begin{align}
 \label{eq:Phi 1 Minimization} 
\frac{\partial \mathrm{V}}{\partial \phi_1^{\dagger}}\ &=\
\left[-\mu_1^2 + 2\lambda_1 (\phi_1^{\dagger} \phi_1) + \lambda_3
 (\phi_2^{\dagger} \phi_2) + \lambda_6 (\phi_1^{\dagger} \phi_2) +
 \lambda_6^{*} (\phi_2^{\dagger} \phi_1) \right] \phi_1 \nonumber\\ 
&\ + \left[-m_{12}^2 + \lambda_4 (\phi_2^{\dagger} \phi_1) + \lambda_5
 (\phi_1^{\dagger} \phi_2) + \lambda_6 (\phi_1^{\dagger} \phi_1) +
 \lambda_7 (\phi_2^{\dagger} \phi_2) \right] \phi_2\ =\ 0\; ,\\[3mm]
 \label{eq:Phi 2 Minimization} 
\frac{\partial \mathrm{V}}{\partial \phi_2^{\dagger}}\ &=\
\left[-\mu_2^2 + 2\lambda_2 (\phi_2^{\dagger} \phi_2) + \lambda_3
 (\phi_1^{\dagger} \phi_1) + \lambda_7 (\phi_1^{\dagger} \phi_2) +
 \lambda_7^{*} (\phi_2^{\dagger} \phi_1) \right] \phi_2 \nonumber\\ 
&\ + \left[-m_{12}^{*2} + \lambda_4 (\phi_1^{\dagger} \phi_2) +
 \lambda_5^{*} (\phi_2^{\dagger} \phi_1) + \lambda_6^{*}
 (\phi_1^{\dagger} \phi_1) + \lambda_7^{*} (\phi_2^{\dagger} \phi_2)
 \right] \phi_1\ =\ 0\; . 
\end{align}
\end{subequations}
Finding analytical solutions to the above coupled cubic equations for
the VEVs  of $\phi_{1,2}$, in  terms of the  gauge-invariant potential
parameters,  is a  formidable task  within the  2HDM. This  problem is
usually  avoided in  the  literature,  by assuming  that  the VEVs  of
$\phi_{1,2}$  are the  input parameters,  for a  given set  of quartic
couplings,  whereas the  potential  mass parameters  are derived  from
these (see,  e.g.~\cite{Pilaftsis:1999qt}). Nevertheless, it  would be
highly preferable, particularly in the  study of topological  defects, to  devise a
method,  in  which  the  VEVs  of  $\phi_{1,2}$  can  be  analytically
expressed,  in terms  of the  gauge-invariant mass  terms  and quartic
couplings of the 2HDM potential.

An analytical method which can address this problem is the
bilinear           scalar-field          formalism          introduced
in~\cite{Nishi:2006tg,Ivanov:2006yq,Maniatis:2006fs}.   According   to
this formalism,  the 2HDM  potential ${\rm V}$  given in~\eqref{V2HDM}
can now be expressed in full by the 4-dimensional vector
\begin{equation}
{\rm R}^\mu\ \equiv\ \phi^\dagger \sigma^\mu \phi\ =\ \left( \begin {array}{c}
 \phi_1^{\dagger} \phi_1+\phi_2^{\dagger}
 \phi_2\\
\noalign{\medskip}\phi_1^{\dagger} \phi_2+\phi_2^{\dagger}
 \phi_1\\\noalign{\medskip}-i\left[\phi_1^{\dagger}
 \phi_2-\phi_2^{\dagger}
 \phi_1\right]\\
\noalign{\medskip}\phi_1^{\dagger}
 \phi_1-\phi_2^{\dagger} \phi_2 
\end {array} \right)\; , 
\label{eq:r^mu definition}
\end{equation}
where $\phi = (\phi_1\,,\,\phi_2)^{\rm T}$ and $\sigma^\mu$ (with $\mu
= 0,1,2,3$) denote the two-by-two identity and the three Pauli
matrices:
\begin{equation}
\sigma^0 = \left( \begin {array}{cc} 1&0\\\noalign{\medskip}0&1
\end {array} \right)\;,\quad \sigma^1 = \left( \begin {array}{cc}
 0&1\\\noalign{\medskip}1&0 
\end {array} \right) \;,\quad \sigma^2 = \left( \begin {array}{cc}
 0&-i\\\noalign{\medskip}i&0 
\end {array} \right) \;,\quad 
\sigma^3 = \left( \begin {array}{cc} 1&0\\\noalign{\medskip}0&-1
\end {array} \right)\; .
\end{equation}
It is  obvious that the  scalar-field multiplet $\phi$ spans  an ${\rm
  SL}(2,\mathbb{C})$ group space similar  to the spinorial Weyl space.
Hence, the vector  ${\rm R}^\mu$ becomes a proper 4-vector  in the Minkowski
space,   described   by   the    flat   metric   $\eta_{\mu   \nu}   =
\mathrm{diag}(1,-1,-1,-1)$. In terms of the 4-vector ${\rm R}^\mu$, the 2HDM
potential reads:
\begin{equation}
 \label{eq:Pot before LM}
\mathrm{V}\ =\ -\;\frac{1}{2} \mathrm{M}_\mu {\rm R}^\mu\: +\: \frac{1}{4}
\mathrm{L}_{\mu \nu} {\rm R}^\mu {\rm R}^\nu\: +\: \mathrm{V}_0\; ,
\end{equation}
where $\mathrm{M}_{\mu}$ and $\mathrm{L}_{\mu \nu}$ are given by
\begin{subequations}
\begin{align}
\mathrm{M}_{\mu} \! &= \! \left( \begin {array}{cccc} \mu_1^2 +
 \mu_2^2\,,&2\mathrm{Re}(m_{12}^2)\,,&-2\mathrm{Im}(m_{12}^2)\,,&\mu_1^2 -
 \mu_2^2\end {array} \right)\; , \\[3mm] 
\mathrm{L}_{\mu \nu} &= \left( \begin {array}{cccc} \lambda_1 +
 \lambda_2 + \lambda_3&\mathrm{Re}(\lambda_6 +
 \lambda_7)&-\mathrm{Im}(\lambda_6 + \lambda_7)&\lambda_1 -
 \lambda_2\\\noalign{\medskip}\mathrm{Re}(\lambda_6 +
 \lambda_7)&\lambda_4+\mathrm{Re}(\lambda_5)& 
 -\mathrm{Im}(\lambda_5)&\mathrm{Re}(\lambda_6 - \lambda_7) 
\\\noalign{\medskip}-\mathrm{Im}(\lambda_6 +
\lambda_7)&-\mathrm{Im}(\lambda_5)&\lambda_4 -
\mathrm{Re}(\lambda_5)&-\mathrm{Im}(\lambda_6 -
\lambda_7)\\\noalign{\medskip}\lambda_1 -
\lambda_2&\mathrm{Re}(\lambda_6 - \lambda_7)&-\mathrm{Im}(\lambda_6 -
\lambda_7)&\lambda_1+\lambda_2-\lambda_3 
\end {array} \right)\; . 
\end{align}
\end{subequations}
Notice that we have added a constant term ${\rm V}_0$ to the scalar
potential ${\rm V}$ in \eqref{eq:Pot before LM}, which is adjusted
such that the minimum of the potential ${\rm V}_{\rm min}$ is set to
zero, thereby accounting for the vanishing small cosmological
constant.

\subsection{Convexity and Stability Conditions}

An obvious  advantage of the  bilinear scalar-field formalism  is that
the 2HDM scalar potential  ${\rm V}$ in~\eqref{V2HDM} has been reduced
from  a fourth  order polynomial  in $\phi_{1,2}$  to a  polynomial of
second  degree  in  ${\rm  R}^\mu$, as  given  in~\eqref{eq:Pot  before
  LM}.  We can  now  calculate  the neutral  vacuum  solutions of  the
potential ${\rm  V}({\rm R}^\mu)$, which amounts to  finding the local
extrema of ${\rm  V}({\rm R}^\mu)$, for which ${\rm  R}^\mu$ is a null
vector,  i.e.~$R^2  =  {\rm   R}^\mu  {\rm  R}_\mu  =  4(\phi_1^\dagger
\phi_1)(\phi_2^\dagger         \phi_2)        -        4(\phi_1^\dagger
\phi_2)(\phi_2^\dagger  \phi_1)  =  0$.   To  enforce  the  null  norm
restriction  on ${\rm  R}^\mu$, we  introduce the  Lagrange multiplier
$\zeta$ and modify the potential ${\rm V}$ of \eqref{eq:Pot before LM}
to
\begin{equation}
\mathrm{V}_\zeta\ =\ -\; \frac{1}{2} \mathrm{M}_\mu {\rm R}^\mu\: +\: 
\frac{1}{4} \mathrm{N}_{\mu \nu} {\rm R}^\mu {\rm R}^\nu\: +\: \mathrm{V}_0\; ,
\end{equation}
with $\mathrm{N}_{\mu \nu} = \mathrm{L}_{\mu \nu} - \zeta \eta_{\mu
 \nu}$. More explicitly, the modified quartic-coupling matrix ${\rm
 N}_{\mu \nu}$ is given by
\begin{equation}
\mathrm{N}_{\mu \nu} = \left( \begin {array}{cccc} \lambda_1 +
 \lambda_2 + \lambda_3-\zeta&\mathrm{Re}(\lambda_6 +
 \lambda_7)&-\mathrm{Im}(\lambda_6 + \lambda_7)&\lambda_1 -
 \lambda_2\\\noalign{\medskip}\mathrm{Re}(\lambda_6 +
 \lambda_7)&\lambda_4+\mathrm{Re}(\lambda_5)+\zeta&
 -\mathrm{Im}(\lambda_5)&\mathrm{Re}(\lambda_6 - \lambda_7) 
\\\noalign{\medskip}-\mathrm{Im}(\lambda_6 +
\lambda_7)&-\mathrm{Im}(\lambda_5)&\lambda_4 -
\mathrm{Re}(\lambda_5)+\zeta&-\mathrm{Im}(\lambda_6 -
\lambda_7)\\\noalign{\medskip}\lambda_1 -
\lambda_2&\mathrm{Re}(\lambda_6 - \lambda_7)&-\mathrm{Im}(\lambda_6 -
\lambda_7)&\lambda_1+\lambda_2-\lambda_3+\zeta 
\end {array} \right)\; . 
\end{equation}
Consequently, within the bilinear scalar-field formalism, the
extremization conditions for the neutral vacuum solutions of the 2HDM
potential are given by $\partial {\rm V}_\zeta/\partial {\rm R}^\mu = 0$ and
$\partial {\rm V}_\zeta /\partial \zeta = 0$, or equivalently by
\begin{subequations}
\begin{align}
\mathrm{M}_{\mu}\ &=\ \mathrm{N}_{\mu \nu}
{\rm R}^\nu\; , \label{eq:Extremization Condition 1} \\ 
{\rm R}_\mu {\rm R}^\mu\ &=\ 0\; .\label{eq:Neutral Vacuum Condition 2}
\end{align}
\end{subequations}
For an extremal point to be a local minimum, we require that the
Hessian $\mathrm{H}$ derived from the scalar potential ${\rm
 V}(\phi_{1,2})$ be positive definite. The Hessian ${\rm H}$ is, in
general, an $8\times 8$-dimensional matrix obtained by double
differentiation with respect to all 8 scalar fields contained in the
two Higgs doublets $\phi_{1,2}$, evaluated at the neutral VEVs
$v^0_{1,2}$ of $\phi_{1,2}$ and their possible relative phase $\xi$
(for exact notation, see Section~\ref{Points on the Vacuum Manifold}).
However, for the given HF and CP symmetries, it is sufficient to
examine the positivity of ${\rm H}$ derived in the restricted
3-dimensional space of $v^0_{1,2}$ and $\xi$. Having identified
all local minima, we then compare the values of the 2HDM potential
${\rm V}$ at these minima. The lowest value obtained for ${\rm V}$
singles out the global minimum, provided ${\rm V}$ itself is bounded
from below. It is therefore important to derive the constraints on the theoretical
parameters for having a scalar potential which is convex and therefore
bounded from below. To ensure this, we require that the matrix
$\mathrm{L}_{\mu \nu}$ be positive definite~\cite{Ivanov:2006yq}. The
latter can be enforced by applying Sylvester's criterion which
yields the following general restrictions:
\begin{subequations}
\begin{align}
\lambda_{1} + \lambda_{2} + \lambda_{3}\ &>\ 0\; ,\label{eq:2HDM Convexity 1}\\
(\lambda_{1} + \lambda_{2} + \lambda_{3})(\lambda_4 + R_5) - (R_6 + 
R_7)^2\ &>\ 0\; , \label{eq:2HDM Convexity 2}\\ 
(\lambda_{1} + \lambda_{2} + \lambda_{3})(\lambda_4^2 - |\lambda_5|^2)
- \lambda_4 \left[(R_6 + R_7)^2 + (I_6 + I_7)^2 \right] & \nonumber
\\ 
- 2 I_5 \left(R_6 + R_7 \right) \left(I_6 + I_7 \right) + R_5
\left[(R_6 + R_7)^2 - (I_6 + I_7)^2 \right]\ &>\ 0\; . \label{eq:2HDM
 Convexity 3} 
\end{align}
\end{subequations} 
In the above, we used the shorthand notation: $R_k =
\mathrm{Re}(\lambda_k)$ and $I_k = \mathrm{Im}(\lambda_k)$. In
addition to~\eqref{eq:2HDM Convexity 1}--\eqref{eq:2HDM Convexity 3},
we require that the determinant of $\mathrm{L}_{\mu \nu}$, which is given
analytically in~\eqref{detL}, be positive as well, i.e.~${\rm
 det}\,[\mathrm{L}_{\mu \nu}] > 0$.

We may now  observe that if ${\rm R}^\mu {\rm R}_\mu  > 0$, this would
imply that $\zeta = 0$. This is since  the 2HDM potential should not modify by
the      addition     of     the      Lagrange     multiplier~$\zeta$,
i.e.~$\mathrm{V}_\zeta~=~\mathrm{V}$.  Hence,  possible solutions with
$\zeta  = 0$  usually signify  a charged-breaking  vacuum for  the six
HF/CP symmetries  considered here and  they are therefore  rejected in
our analysis.  As a consequence,  there are two distinct sets of ${\rm
  U(1)}_{\rm  em}$-preserving minima  that  could occur  in the  2HDM,
depending on whether $\mathrm{det} [\mathrm{N}_{\mu \nu}]$ vanishes or
not.   If  ${\rm  N}_{\mu\nu}$  is  not  singular,  i.e.~$\mathrm{det}
[\mathrm{N}_{\mu \nu}]  \neq 0$, the vector $\rm R^\mu$  can be obtained
by simply inverting~\eqref{eq:Extremization Condition 1}, i.e.
\begin{equation}
{\rm R}^\mu\ =\ \left(\mathrm{N}^{-1} \right)^{\mu \nu} \mathrm{M}_{\nu}\; , 
\end{equation}
and  the Lagrange  multiplier must  guarantee that  ${\rm  R}^\mu {\rm
  R}_\mu = 0$, i.e.
\begin{equation}
 \label{eq:Lagrange Multiplier Condition}
\left(\mathrm{N}^{-1} \right)_{\mu \alpha } \mathrm{M}^\alpha
\mathrm{M}_\beta \left(\mathrm{N}^{-1} \right)^{\beta \mu}\ =\ 0\; .
\end{equation}
As we will  see in Sections~3 and~4, the  neutral vacuum solutions for
the generic  HF and CP symmetries  under study (with  exception of the
CP1 symmetry) imply  that at least one of the  VEVs of $\phi_{1,2}$ is
zero, when $\mathrm{det} [\mathrm{N}_{\mu  \nu}] \neq 0$.  Such vacuum
solutions  are  uninteresting,  since  they  do  not  lead  to  stable
topological defects.

The   second  set  of   neutral  vacua   occurs,  when   the  modified
quartic-coupling  matrix  ${\rm  N}_{\mu\nu}$ is  singular,  i.e.~when
$\mathrm{det}[\mathrm{N}_{\mu \nu}]  = 0$. In this  case, the Lagrange
multiplier $\zeta$ takes on a specific value which leads to a singular
matrix ${\rm N}_{\mu\nu}$. If this happens, the undetermined component
of ${\rm  R}^\mu$ is calculated  by requiring that the  neutral vacuum
condition ${\rm R}_\mu {\rm R}^\mu =  0$ is met.  In this second class
of solutions,  both the  VEVs of the  Higgs doublets can  be non-zero,
leading to the interesting topological solutions which we study.

For  each   of  the  neutral   vacuum  solutions  we  obtain   by  the
Lagrange multiplier and Hessian methods outlined above, we cross-check
that   they  also   satisfy   the  convexity   and  the   conventional
extremization       conditions~\eqref{eq:Phi      1      Minimization}
and~\eqref{eq:Phi  2 Minimization}.  In  this way,  we  ensure that  a
stable and  global neutral  vacuum is found  for the  2HDM potential.
Since the matrix $\mathrm{N}_{\mu  \nu}$ plays an instrumental role in
our  analysis,  Appendix~\ref{The  Determinant  of  N} contains
analytical expressions  for its determinant, as well  as solutions for
the  Lagrange  multiplier  $\zeta$  that  give  rise  to  a  vanishing
determinant, i.e.~$\mathrm{det}[\mathrm{N}_{\mu \nu}] = 0$.

\subsection{The Majorana Formalism}\label{subsec:Majorana}

It would be interesting to introduce a formalism where both the HF and
CP symmetries can be realized by acting on the same representation of
scalar fields. For this purpose, we extend the gauge-invariant
bilinear formalism based on the $\mathrm{SL(2,\mathbb{C})}$ group to
the larger complex linear group $\mathrm{GL(8,\mathbb{C})}$ (see
also~\cite{Nishi:2011gc} for a related discussion). Specifically, this
latter group is acting on the 8-dimensional complex field multiplet
\begin{equation}
 \label{eq:Phi Definition}
\Phi\ =\ \left( \begin {array}{c} \phi_1\\\noalign{\medskip}
 \phi_2\\\noalign{\medskip} i \sigma^2 \phi_1^*\\\noalign{\medskip} i
 \sigma^2 \phi_2^* \end {array} \right)\; .
\end{equation}
Notice that under a ${\rm SU(2)_L}$ gauge transformation ${\rm U_L}$,
all doublet components of the multiplet $\Phi$ transform in the same way,
i.e.~$\Phi \to {\rm U_L}\, \Phi$, with
\begin{equation}
 \label{eq:SUL}
\mathrm{U_L}\ =\ \exp \Big[ i\, \theta^i\,
 \Big(\sigma^0\otimes \sigma^0 \otimes \sigma^i/2\Big)\Big]\ =\
\sigma^0\otimes \sigma^0 \otimes \exp \Big[ i\, \theta^i \sigma^i/2\Big]\; ,
\end{equation}
where  the summation  convention over  the repeated  group  indices $i
=1,2,3$ is  assumed, with  $\sigma^{1,2,3}/2$ being the  generators of
the ${\rm SU(2)_L}$ gauge  group and $\theta^{1,2,3} \in [0,4\pi)$ are
  the associated group parameters.

In order to describe the 2HDM potential, we introduce the 4-vector
$\widetilde{\rm R}^\mu$:
\begin{equation}
 \label{Rmu}
\widetilde{\mathrm{R}}^\mu\ =\ \Phi^\dagger \Sigma^\mu \Phi\; ,
\end{equation}
where $\Sigma^\mu$ in the full 8-dimensional field space must have the
form:   $\Sigma^\mu  =  \Sigma^{\mu}_{\;\alpha   \beta}  \sigma^\alpha
\otimes    \sigma^\beta   \otimes    \sigma^0$,    as   required    by
$\mathrm{SU}(2)_\mathrm{L}$  gauge   invariance.  Moreover,  as  shown
explicitly  in  Appendix~\ref{The  Form   of  S},  the  imposition  of
$\mathrm{U}(1)_\mathrm{Y}$ invariance and  a Majorana constraint to be
discussed  below further reduces the  form of  the  4-vector matrices
$\Sigma^{\mu}$ to
\begin{equation}
\Sigma^{\mu}\ =\ \frac{1}{2}\left( \begin
 {array}{cc}\sigma^{\mu}& {\bf 0}_2\\
\noalign{\medskip}{\bf 0}_2 &(\sigma^\mu)^{\mathrm{T}} \end
 {array} \right) \otimes \sigma^0\; ,
\end{equation}
where ${\bf 0}_2$ is the two-by-two null matrix.  Consequently, in the
Majorana     scalar-field     formalism,     we     obtain     for
$\mathrm{U}(1)_\mathrm{Y}$-invariant 2HDM potentials that
\begin{equation}
\widetilde{\mathrm{R}}^\mu\ =\ {\rm R}^\mu\ .
\end{equation}
However, we should stress  here that if the $\mathrm{U}(1)_\mathrm{Y}$
symmetry is lifted from the 2HDM potential, the 4-vector ${\rm R}^\mu$
needs  to be  promoted to  a  6-vector ${\rm  R^A}$ (with  ${\rm A}  =
0,1,\dots,5$)  and  the corresponding  structure  of $\Sigma^{\rm  A}$
becomes  non-trivial.   In  this  respect, the  Majorana  scalar-field
formalism  has  the  analytical  advantage in  expressing  the  scalar
potential of an U(1)-violating 2HDM  via a similar quadratic form with
respect  to  ${\rm  R^A}$  as in~\eqref{eq:Pot  before  LM}  for~${\rm
  R}^\mu$.   We discuss and demonstrate this application further in 
  Section~\ref{Beyond the 2HDM}.

Under charge conjugation, the multiplet $\Phi$ exhibits the following
property:
\begin{equation}
 \label{MajoranaPhi}
\Phi\ =\ \mathrm{C}\, \Phi^*\;,
\end{equation}
where $\mathrm{C} = \sigma^2  \otimes \sigma^0 \otimes \sigma^2$, with
${\rm  C}  =  {\rm  C}^{-1}$.   Hence,  $\Phi$  satisfies  a  Majorana
constraint, very analogous to the one obeyed by Majorana fermions. For
this  reason,  we  call   this  formalism  the  Majorana  scalar-field
formalism. In addition, the Majorana multiplet $\Phi$ transforms under
the reparameterization group $\mathrm{GL(8,\mathbb{C})}$ as
\begin{equation}
\Phi^\prime\ =\ \mathrm{M}\, \Phi \; ,
\end{equation}
with $\mathrm{M} \in \mathrm{GL(8,\mathbb{C})}$. However, as we will
see below, the form of $\mathrm{M}$ cannot be general, but it is
constrained by three basic conditions: (i)~the conservation of ${\rm
 SU(2)_L}$ symmetry by the transformation matrices ${\rm M}$;
(ii)~the Majorana condition~\eqref{MajoranaPhi} for any
$\mathrm{GL(8,\mathbb{C})}$-transformed multiplet $\Phi^\prime$;
(iii)~the conservation of ${\rm U(1)_Y}$ symmetry by the
transformation matrices ${\rm M}$. Applying these three constraints
on ${\rm M}$, the 4-vector matrix $\Sigma^\mu$ is found to transform
as
\begin{equation}
 \label{eq:Lambda M equation}
e^{\sigma/8}\, \Lambda^{\mu}_{\;\nu} \Sigma^{\nu}\ =\ \mathrm{M}^{\dagger}
\Sigma^{\mu} \mathrm{M}\; ,
\end{equation}
implying that ${\rm R}^\mu$ transforms into 
\begin{equation}
 \label{eq:Rprime}
\mathrm{R}^{\prime\,\mu}\ =\ e^{\sigma/8}\, \Lambda^{\mu}_{\;\nu}\,
\mathrm{R}^{\nu}\; ,
\end{equation}
where $e^\sigma = {\rm det}\,[\mathrm{M}^{\dagger} \mathrm{M}] > 0$
and $\Lambda^{\mu}_{\;\nu} \in {\rm SO}(1,3)$.

Since ${\rm M} \in \mathrm{GL(8,\mathbb{C})}$, the matrix $\mathrm{M}$
can then be represented in the full 8-dimensional scalar-field basis
$\Phi$ by the triple tensor product:
\begin{equation}
 \label{Mtransform}
\mathrm{M}\ =\ M_{\mu \nu \lambda}\, 
\sigma^{\mu} \otimes \sigma^{\nu} \otimes \sigma^\lambda\; .
\end{equation}
As  was  mentioned  in  the  introduction,  there  are  two  types  of
$\mathrm{GL(8,\mathbb{C})}$     transformations~$\mathrm{M}$    acting
on~$\Phi$. The first one is a HF transformation, where the transformed
multiplet~$\Phi'$  transforms in  the same  way under  ${\rm SU(2)_L}$
as~$\Phi$, whereas the second one is a CP transformation where~$\Phi'$
transforms    in    the   same    way    as   the    charge-conjugated
multiplet~$\Phi^*$.  Thus,  for a  HF  transformation compatible  with
${\rm SU(2)_L}$ gauge invariance, we must have that $\mathrm{M} = {\rm
  U_L^\dagger}\,  {\rm M}\,  {\rm U_L}$,  where ${\rm  U_L}$  is given
in~\eqref{eq:SUL}.   Instead,    for   a   general    CP   and   ${\rm
  SU(2)_L}$-invariant transformation, we  must demand that $\mathrm{M}
=  {\rm  U_L^T}\,  {\rm  M}\,  {\rm  U_L}$.  Consequently,  the  ${\rm
  SU(2)_L}$-invariant  tensorial forms for the two types of transformation, 
  which we denote as $\mathrm{M}_\pm$, are
\begin{subequations}
\begin{align}
 \label{eq:HFSU2}
\mathrm{HF} &:\quad 
\mathrm{M}_+\ =\ M_{\mu \nu}\, 
\sigma^{\mu} \otimes \sigma^{\nu} \otimes \sigma^0\; , \\ 
 \label{eq:CPSU2}
\mathrm{CP} &:\quad 
\mathrm{M}_-\ =\ M_{\mu \nu}\, 
\sigma^{\mu} \otimes \sigma^{\nu} \otimes (-i\sigma^2)\; , 
\end{align}
\end{subequations}
where we have used that  ${\rm V^T}\, i\sigma^2\,{\rm V} = i\sigma^2$,
for any ${\rm V} \in {\rm SU(2)}$.

It is now interesting to discuss the remaining two constraints imposed
on  the above  ${\rm SU(2)_L}$-invariant  structure of  ${\rm M}_\pm$,
resulting  from  the  Majorana condition~\eqref{MajoranaPhi}  and  the
conservation of  the ${\rm U(1)_Y}$ hypercharge symmetry.  The requirement that
the  Majorana condition~\eqref{MajoranaPhi}  should  consistently hold
for   the  multiplet  $\Phi$   and  the   HF/CP-transformed  multiplet
$\Phi^\prime   =  {\rm   M}_\pm\,  \Phi$   produces   the  non-trivial
constraint:
\begin{equation}
  \label{eq:MajM}
\mathrm{M}^*_\pm\ =\ \mathrm{C\,M_\pm\,C}\; .
\end{equation}
This last constraint reduces the form of the tensor $M_{\mu \nu}$
defined in~\eqref{eq:HFSU2} and~\eqref{eq:CPSU2} to
\begin{equation}
 \label{Mmunu}
M_{\mu \nu}\ =\ \left( \begin {array}{cccc}
 M_{00}&M_{01}&iM_{02}&M_{03}\\
\noalign{\medskip}iM_{10}&iM_{11}&M_{12}&iM_{13} \\
\noalign{\medskip}iM_{20}&iM_{21}&M_{22}&iM_{23} \\
\noalign{\medskip}iM_{30}&iM_{31}&M_{32}&iM_{33} 
\end {array} \right)\; ,
\end{equation}
where all the components $M_{00}, M_{01}, M_{02}, \dots, M_{33}$ are real numbers. More details of this
calculation are given in~Appendix~\ref{The Majorana Constraint on M}.
Thus, we observe that the Majorana condition applied to ${\rm M}$
reduces the reparameterization group from $\mathrm{GL}(8,\mathbb{C})$
to two subgroups isomorphic to $\mathrm{GL}(4,\mathbb{R})$, acting on
a complex vector space.

The  HF  and CP  transformation  matrices  ${\rm  M}_\pm$ should  also
respect the  $\mathrm{U(1)}_{\mathrm{Y}}$ hyper\-charge symmetry  of the
theory.   Following  a similar  line  of steps  as  for  the ${\rm  SU
  (2)_L}$-gauge invariance case, we  require that $\mathrm{M}_+ = {\rm
  U_Y^*}\,  {\rm  M}_+\, {\rm  U_Y}$,  for  a  HF transformation,  and
$\mathrm{M}_- = {\rm  U_Y}\, {\rm M}_-\, {\rm U_Y}$,  for a general CP
transformation, where
\begin{equation}
 \label{eq:UY}
\mathrm{U_Y}\ =\ \exp \Big[ i\, \theta_{\rm Y}/2\,
 \Big(\sigma^3\otimes \sigma^0 \otimes \sigma^0\Big)\Big]\ =\ 
\exp \Big(i\, \theta_{\rm Y}\, \sigma^3/2 \Big)
\otimes \sigma^0 \otimes \sigma^0\; ,
\end{equation}
in  the $\mathrm{GL}(8,\mathbb{C})$ representation,  with $\theta_{\rm
  Y}  \in  [0,4\pi)$.  Evidently,   the  above  two  constraints  from
  requiring  $\mathrm{U(1)}_{\mathrm{Y}}$  invariance  result  in  the
  commutator and anti-commutator conditions
\begin{subequations}
\begin{align}
 \label{eq:U star con}
\left[\mathrm{M}_+,\, \sigma^3 \otimes \sigma^0 \otimes \sigma^0
 \right]\ &=\ 0\; ,\\ 
 \label{eq:U no star con} 
\left\{\mathrm{M}_-,\, \sigma^3 \otimes \sigma^0 \otimes \sigma^0 
\right\}\ &=\ 0\; ,
\end{align}
\end{subequations}
for the HF and CP transformations, respectively. Since ${\rm M}_+ =
M_{\mu \nu} \sigma^\mu \otimes \sigma^\nu \otimes \sigma^0$, the
commutator relation~\eqref{eq:U star con} becomes $M_{\mu \nu}\left[
 \sigma^\mu,\, \sigma^3 \right] \otimes \sigma^\nu \otimes \sigma^0 =
0$. It is not difficult to see that only $\mu = 0,\,3$ satisfy the
last commutator relation, whereas $M_{1 \alpha} = M_{2 \alpha} = 0$,
for $\alpha = 0,1,2,3$. Then, $M_{\mu \nu}$ takes on the form:
\begin{equation}
M_{\mu \nu}\ =\ \left( \begin {array}{cccc}
 M_{00}&M_{01}&iM_{02}&M_{03}\\\noalign{\medskip}0&0&0&0 
\\\noalign{\medskip}0&0&0&0\\\noalign{\medskip}iM_{30}&iM_{31}&M_{32}&iM_{33}
\end {array} \right)\; , 
\end{equation}
leading to the following structure for the HF transformation matrix
$\mathrm{M}_{+}$: 
\begin{equation}
 \label{eq:M+}
\mathrm{M}_{+}\ =\ 
\left( \begin {array}{cc} \mathrm{T}_+& {\bf 0}_2\\
\noalign{\medskip}{\bf 0}_2 &\mathrm{T}_+^* \end {array} \right) 
\otimes \sigma^0\; ,
\end{equation}
where 
\begin{equation}
 \label{eq:M+1}
\mathrm{T}_+\ =\ 
\left( \begin {array}{cc} M_{00} + M_{03} + iM_{30} + iM_{33}&M_{01} +
 M_{02} + iM_{31} - iM_{32}\\\noalign{\medskip}M_{01} - M_{02} +
 iM_{31} + iM_{32}&M_{00} - M_{03} + iM_{30} - iM_{33} \end {array}
\right) \; .
\end{equation}
is a general complex $2\times 2$ matrix. The matrix
form~\eqref{eq:M+} for ${\rm M}_+$ also provides closure in the
4-vector space of ${\rm R}^\mu$, through the relation:
\begin{equation}
 \label{eq:M + Lambda Main Relation}
\mathrm{M}_+^\dagger \Sigma^\mu \mathrm{M}_+\ =\
e^{\sigma_+/8} \left(\Lambda_+\right)^\mu_{\;\nu} \Sigma^\nu\; , 
\end{equation}
where $e^{\sigma_+} = {\rm det}\,[\mathrm{T}_+^* \mathrm{T}_+] > 0$
and $\left(\Lambda_+\right)^\mu_{\;\nu} \in {\rm SO}(1,3)$.

Correspondingly, the anti-commutator relation given in~\eqref{eq:U no
 star con} leads to the constraint: $M_{\mu \nu} \left\{
\sigma^\mu,\, \sigma^3 \right\} \otimes \sigma^\nu \otimes
(-i\sigma^2) = 0$. One can readily observe that only $\mu = 1,\,2$
satisfy the last anti-commutation relation, whilst $M_{0 \alpha} =
M_{3 \alpha} = 0$, for $\alpha = 0,1,2,3$. Thus, $M_{\mu \nu}$
acquires the form:
\begin{equation}
M_{\mu \nu}\ =\ \left( \begin {array}{cccc}
 0&0&0&0\\\noalign{\medskip}iM_{10}&iM_{11}&M_{12}&iM_{13} 
\\\noalign{\medskip}iM_{20}&iM_{21}&M_{22}&iM_{23}\\
\noalign{\medskip}0&0&0&0 \end {array} \right)\; . 
\end{equation}
The resulting matrix $\mathrm{M}_{-}$ for general CP transformations
is given by
\begin{equation}
 \label{eq:M-}
\mathrm{M}_-\ =\ \left( \begin {array}{cc} {\bf 0}_2 & \mathrm{T}_-
 \\\noalign{\medskip} -\mathrm{T}^*_-& {\bf 0}_2 
\end {array} \right) \otimes (-i\sigma^2)\; ,
\end{equation}
where $\mathrm{T}_-$ is a complex two-by-two matrix given by
\begin{equation}
 \label{eq:M-1}
\mathrm{T}_-\ =\ \left( \begin {array}{cc} M_{10} + M_{13} - iM_{20} -
 iM_{23}&M_{11} - M_{12} - iM_{21} +
 iM_{22}\\\noalign{\medskip}M_{11} + M_{12} - iM_{21} -
 iM_{22}&M_{10} - M_{13} - iM_{20} + iM_{23} 
\end {array} \right)\; . 
\end{equation}
As before, the block-off diagonal form of ${\rm M}_-$ provides closure
in the 4-vector space of ${\rm R}^\mu$, since
\begin{equation}
 \label{eq:M - Lambda Main Relation} 
\mathrm{M}_-^\dagger \Sigma^\mu
\mathrm{M}_-\ =\ e^{\sigma_-/8} \left(\Lambda_-\right)^\mu_{\;\nu} \Sigma^\nu\; ,
\end{equation}
with $e^{\sigma_-} = {\rm det}\,[\mathrm{T}_-^* \mathrm{T}_-] > 0$ and
$\left(\Lambda_-\right)^\mu_{\;\nu} \in {\rm SO}(1,3)$. 

In addition we note that mixed transformations involving both $\mathrm{M}_+$ and $\mathrm{M}_-$ do not provide closure within the 4-vector space of ${\rm R}^\mu$, i.e.
\begin{equation}
\mathrm{M}^\dagger_+ \Sigma^\mu \mathrm{M}_- \not\propto \Lambda^\mu_{\;\nu}
\Sigma^\nu\; .
\end{equation}
Hence, two distinct SO(1,3) spaces exist which are compatible with
$\mathrm{U(1)}_{\mathrm{Y}}$   invariance.    We   denote   these   by
$\left(\Lambda_+\right)^\mu_{\;\nu}$                                and
$\left(\Lambda_-\right)^\mu_{\;\nu}$,   and  their   respective  field
transformation  matrices  by  $\mathrm{M}_+$  and  $\mathrm{M}_-$.  Of
course, combined transformation of  different types are also possible,
resulting in a composite transformation described by $\mathrm{M}_+$ or
$\mathrm{M}_-$, as shown in Table \ref{table:M+ M- Table}.

\begin{table}[t!]
\begin{center}
\begin{tabular}{c|c|c}
\hline
First Transformation Type & Second Transformation Type & Composite Type \\
\hline
\hline

$\mathrm{M}_+$ & $\mathrm{M}_+$ & $\mathrm{M}_+$ \\

\hline

$\mathrm{M}_+$ & $\mathrm{M}_-$ & $\mathrm{M}_-$ \\

\hline

$\mathrm{M}_-$ & $\mathrm{M}_+$ & $\mathrm{M}_-$ \\

\hline

$\mathrm{M}_-$ & $\mathrm{M}_-$ & $\mathrm{M}_+$ \\

\hline
\end{tabular}

\caption{\it Transformation properties after two successive operations of 
${\rm M}_\pm$.\label{table:M+ M- Table}} 
\end{center}
\end{table}

In summary, the HF and CP transformation matrices~$\mathrm{M}_\pm$ may
be written down in the following tensorial forms:
\begin{subequations}
\begin{align}
 \label{eq:Mtensor+}
\mathrm{HF} &:\ 
\mathrm{M}_+\ =\ \bigg[\, \frac{(\sigma^0 + \sigma^3)}{2} \otimes
 \mathrm{T}_+\: +\: 
\frac{(\sigma^0 - \sigma^3)}{2} \otimes \mathrm{T}^{*}_+\,\bigg]
\otimes \sigma^0\; , \\ 
 \label{eq:Mtensor-}
\mathrm{CP} &:\ 
\mathrm{M}_-\ =\ \bigg[\, 
\frac{(\sigma^1 + i \sigma^2)}{2} \otimes \mathrm{T}_-\: -\:
\frac{(\sigma^1 - i \sigma^2)}{2} \otimes
\mathrm{T}^{*}_-\,\bigg]\otimes (-i\sigma^2)\; . 
\end{align}
\end{subequations}
Given the above representation of the HF and CP transformations, we
observe that
\begin{equation}
\mathrm{M}_\mp\ =\ \mathrm{C}\, \mathrm{M}_\pm\; ,
\end{equation}
provided we set  ${\rm T}_- = {\rm T}^*_+$. This  means that a general
CP transformation  can be thought  of as a  combination of a HF  and a
standard CP transformation. This is also consistent with the geometric
interpretation  presented   in~\cite{Ferreira:2010yh}.  Likewise,  the
action of two successive CP  transformations is equivalent to a single
HF  transformation,  as  can be  seen  from  the  last line  of  Table
\ref{table:M+ M- Table}.

\begin{table}[t!]
\begin{center}
\begin{tabular}{c|c|c}
\hline
 & & \\
HF/CP Symmetry & Transformation Matrix ${\rm T}_+$ & Transformation
 Matrix ${\rm T}_-$ \\ 
 & in the Basis $(\phi_1\,,\,\phi_2)$  & in the Basis
 $(\phi_1\,,\,\phi_2)$ \\[2mm] 
\hline
\hline

$\mathrm{Z}_2$ & $\left( \begin {array}{cc}
 1&0\\\noalign{\medskip}0&1\end {array} 
 \right)\; , \; \left( \begin {array}{cc}
 1&0\\\noalign{\medskip}0&-1\end {array} \right)$ 
& \\ 

\hline

$\mathrm{U}(1)_{\mathrm{PQ}}$ & $\left( \begin {array}{cc}
 e^{-i\alpha}&0\\\noalign{\medskip}0&e^{i\alpha} \end {array}
\right)$ & \\ 
& $\alpha \in [0, \pi)$ & \\ 

\hline

$\bigslant{\mathrm{SU(2)}_{\mathrm{HF}}}{\mathrm{Z}_{2}} \cong
\mathrm{SO(3)}_{\mathrm{HF}}$ 
 & $\left( \begin {array}{cc} e^{-i \alpha}\cos \theta & e^{-i
 \beta}\sin\theta \\\noalign{\medskip}-e^{i \beta} \sin\theta&
 e^{i \alpha} \cos \theta \end {array} \right)$ & \\
 & $\theta, \alpha, \beta \in [0,\pi)$  & \\

\hline

CP1 & $\left( \begin {array}{cc} 1&0\\\noalign{\medskip}0&1\end {array}
 \right)$ & $\left( \begin {array}{cc} 1&0\\\noalign{\medskip}0&1\end {array}
 \right)$ \\ 

\hline

CP2 & $\left( \begin {array}{cc} 1&0\\\noalign{\medskip}0&1\end {array}
 \right)$ &
$\left( \begin {array}{cc} 0&1\\\noalign{\medskip}-1&0\end {array}
 \right) $ \\

\hline

CP3 & 
$\left( \begin {array}{cc} \cos \theta & \sin\theta
 \\\noalign{\medskip} -\sin\theta& \cos \theta \end {array} \right)$ &
$\left( \begin {array}{cc} \cos \theta & \sin\theta
 \\ \noalign{\medskip} -\sin\theta& \cos \theta \end {array} \right)$ \\ 
& $\theta \in \left[0, \pi \right)$ & $\theta \in \left[0, \pi \right)$ \\ 

\hline

\end{tabular}

\caption{\it Matrix representations of ${\rm T}_\pm$ for 6 generic HF
 and CP symmetries of the 2HDM.\label{table:Transformation Matrix Table}}
\end{center}
\end{table}

In  Table~\ref{table:Transformation  Matrix  Table},  we  display  the
matrix representations  of ${\rm  T_+}~({\rm T_\pm})$ for  the HF~(CP)
symmetries that we will be analyzing. In detail, the HF transformation
matrices  ${\rm   T}_+$  are  displayed   in  the  second   column  of
Table~\ref{table:Transformation   Matrix   Table}.   These   are   the
$\mathrm{Z}_2$     discrete     symmetry~\cite{Glashow:1976nt},    the
Peccei--Quinn  symmetry  $\mathrm{U(1)}_{\rm PQ}$~\cite{Peccei:1977hh}
and         the        HF         symmetry        $\mathrm{SO(3)}_{\rm
  HF}$~\cite{Deshpande:1977rw,Ivanov:2007de,Ma:2009ax,Ferreira:2010yh}
which  is  isomorphic   to  a  $\mathrm{SU(2)}_{\rm  HF}/\mathrm{Z}_2$
transformation of $\phi_{1,2}$. Table~\ref{table:Transformation Matrix
  Table} also  exhibits the transformation  matrices~${\rm T}_\pm$ for
three typical  CP symmetries of  the 2HDM potential: the  CP1 symmetry
which is equivalent to the standard CP transformation $\phi_{1(2)} \to
\phi^*_{1(2)}$~\cite{Lee:1973iz,Deshpande:1977rw,Branco:1980sz},    the
CP2          symmetry          where         $\phi_{1(2)}          \to
(-)\phi^*_{2(1)}$~\cite{Davidson:2005cw} and the CP3 symmetry which is
a  combination of CP1  with an  $\mathrm{SO(2)}_{\rm HF}/\mathrm{Z}_2$
transformation          of         the          Higgs         doublets
$\phi_{1,2}$~\cite{Ivanov:2007de,Ferreira:2009wh,Ma:2009ax,Ferreira:2010yh}.

Let  us  comment on  the  domains of  the  group  parameters shown  in
Table~\ref{table:Transformation  Matrix Table}. Specifically,  we have
considered     $\alpha     \in      [0,     \pi     )$     for     the
  $\mathrm{U}(1)_{\mathrm{PQ}}$ symmetry, $\theta \in  [ 0, \pi )$ for
    the CP3  symmetry, and $\alpha, \beta,  \theta \in [0,  \pi )$ for
      the  $\mathrm{SO(3)}_{\mathrm{HF}}$   symmetry.   The  parameter
      intervals for the potential symmetry groups are chosen so as to
      avoid double  covers of  the total symmetry  group $\mathrm{G}$,
      because   of  the  presence   of  the   SM  gauge   group  ${\rm
        SU(2)_L}\otimes   {\rm  U(1)_Y}$,   and   especially  of~${\rm
        U(1)_Y}$ hypercharge~\cite{Ferreira:2009wh}.

Another  important comment  is in  order here;  for each  CP symmetry,
there should be  a HF symmetry associated to it.  This arises when the
CP symmetry  is raised  to even powers  and guarantees closure  of the
symmetry group (cf.~Table~\ref{table:M+ M- Table}). For the CP1 and CP2 symmetries, an even number of applications of the symmetry results in the identity mapping, i.e. $\rm(CP1)^{2n} = \mathbf{I}$ and $\rm(CP2)^{2n} = \mathbf{I}$. However,
for  CP3, we  obtain a  non-trivial HF  symmetry,  i.e.~(CP3)$^{2\mathrm{n}} \cong
\mathrm{SO(2)}_\mathrm{HF}/\mathrm{Z}_{2}$.  Unlike the CP symmetries,
HF symmetries close within themselves, as shown in Table~\ref{table:M+
  M- Table}.  In Section~\ref{Points on the Vacuum  Manifold}, we will
discuss  further theoretical  issues related  to the  breaking  of the
symmetry group $\mathrm{G}$ into  a subgroup ${\rm H}$, as these issues
are  important  in  order  to  generate  the  entire  vacuum  manifold
associated to a given 2HDM potential.

\begin{table}[t!]
\begin{center}
\begin{tabular}{c|c}
\hline
 & \\[-1mm]
HF/CP Symmetry & $\mathcal{O}_{\mathrm{HF/CP}}$ Matrices in the Basis
$({\rm R}^1\,,\, {\rm R}^2\,,\, {\rm R}^3)$ \\[2mm]
\hline
\hline
$\mathrm{Z}_2$	&	$\left( \begin {array}{ccc}
 1&0&0\\ \noalign{\medskip}0&1&0 \\ \noalign{\medskip}0&0&1\end
 {array} \right)\,,\ \left( \begin {array}{ccc}
 -1&0&0\\ \noalign{\medskip}0&-1&0 \\ \noalign{\medskip}0&0&1\end
 {array} \right)$	\\ 

\hline

$\mathrm{U}(1)_{\mathrm{PQ}}$	&	$\left( \begin {array}{ccc}
 c_{2\alpha}&-s_{2\alpha}&0\\ \noalign{\medskip}s_{2\alpha}
 &c_{2\alpha} &0 \\ \noalign{\medskip}0&0&1\end {array} \right)$
 \\ 
&	$\alpha \in [0, \pi)$	\\ 

\hline

$\mathrm{SO(3)}_{\mathrm{HF}}$	&	$\left( \begin {array}{ccc}
 c_{2\alpha} c^2_\theta - c_{2 \beta} s^2_\theta&-s_{2\alpha}
 c^2_\theta - s_{2 \beta} s^2_\theta&-s_{2\theta}
 c_{\alpha + \beta}\\ \noalign{\medskip}s_{2\alpha} c^2_\theta - s_{2
 \beta} s^2_\theta&c_{2 \alpha} c^2_\theta + c_{2\beta}
 s^2_\theta&-s_{2 \theta} s_{\alpha+\beta} \\ \noalign{\medskip}s_{2
 \theta} c_{\alpha - \beta}&-s_{2\theta} s_{\alpha -\beta}&c_{2
 \theta}\end {array} \right)$		\\ 
&	$\theta, \alpha, \beta \in \left[0, \pi \right)$	\\

\hline
CP1	&	$\left( \begin {array}{ccc}
 1&0&0\\ \noalign{\medskip}0&1&0 \\ \noalign{\medskip}0&0&1\end
 {array} \right) \;,\ \left( \begin {array}{ccc}
 1&0&0\\ \noalign{\medskip}0&-1&0 \\ \noalign{\medskip}0&0&1\end
 {array} \right)$	\\ 

\hline

CP2	&	$\left( \begin {array}{ccc}
 1&0&0\\ \noalign{\medskip}0&1&0 \\ \noalign{\medskip}0&0&1\end
 {array} \right) \;,\ \left( \begin {array}{ccc}
 -1&0&0\\ \noalign{\medskip}0&-1&0 \\ \noalign{\medskip}0&0&-1\end
 {array} \right)$		\\ 

\hline

CP3	& $\left( \begin {array}{ccc} c_{2\theta} &0&s_{2
 \theta}\\ \noalign{\medskip}0&1&0 \\ \noalign{\medskip}-s_{2
 \theta}&0&c_{2\theta} \end {array} \right)\;, \ \left( \begin
 {array}{ccc} c_{2\theta} &0&s_{2 
 \theta}\\ \noalign{\medskip}0&-1&0 \\ \noalign{\medskip}-s_{2
 \theta}&0&c_{2\theta} \end {array} \right)$	\\ 
&	$\theta \in [0, \pi)$	\\ 
\hline
\end{tabular}
\caption{\it Matrix representation of $\mathcal{O}_{\mathrm{HF/CP}}$
 defined in~\eqref{eq:OG} for the 6 generic HF/CP symmetries
 of the 2HDM potential. Here we use the shorthand notation $c_\theta
 = \cos \theta$ and $s_\theta = \sin \theta$. \label{table:Lambda
 Matrix Table}}
\end{center}
\end{table}

If the 2HDM potential ${\rm V}$  is invariant under a particular HF or
CP    symmetry~${\rm   G_{HF/CP}}$,    realized   by    the   matrices
$(\Lambda_\pm)^\mu_{\;\nu}$,    then   the    theoretical   parameters
$\mathrm{M}_\mu$ and $\mathrm{L}_{\mu \nu}$ satisfy the relations:
\begin{subequations}
\begin{align}
\mathrm{M}_\nu\ &=\ \mathrm{M}_\mu \Lambda^{\mu}_{\;\nu}\; , \label{eq:Mass Vector Transformation Constraint} \\
\mathrm{L}_{\alpha \beta}\ &=\ \mathrm{L}_{\mu \nu}
\Lambda^{\mu}_{\;\alpha} \Lambda^{\nu}_{\;\beta}\; , \label{eq:Quartic Matrix Transformation Constraint}
\end{align}
\end{subequations}
Here,   for   convenience,   we   drop  the   subscript   $\pm$   from
$(\Lambda_\pm)^\mu_{\;\nu}$ and have implicitly assumed that $\sigma = 0$
or $e^{\sigma/4} =1$.  Hence, for each HF or  CP transformation acting
on  the  Majorana  field  multiplet~$\Phi$,  there  is  an  equivalent
transformation on  ${\rm R}^\mu$,  as given in  \eqref{eq:Rprime}. The
tensor $\Lambda^\mu_{\;\nu}$ in the ${\rm SO}(1,3)$ space has then the
following matrix form:
\begin{equation}
 \label{eq:OG}
\Lambda_{\mathrm{HF/CP}}\ =\ 
 \mathrm{diag}\,\Big(1\,,\ \mathcal{O}_{\mathrm{HF/CP}}\Big)\; ,
\end{equation}
where $\mathcal{O}_{\mathrm{HF/CP}}$ is a subgroup of ${\rm O}(3)$ for
the HF and CP symmetries under consideration.
In~Table~\ref{table:Lambda Matrix Table}, we give the matrix
representation of $\mathcal{O}_{\mathrm{HF/CP}}$, for the three HF and
the three CP symmetries, respectively.

\subsection{The Vacuum Manifold}\label{Points on the Vacuum Manifold}

After minimization of the 2HDM potential, the field multiplet $\Phi$
acquires, in general, a non-zero VEV, i.e.
\begin{equation}
\Phi\ =\ \left( \begin {array}{c} \phi_1\\\noalign{\medskip}\phi_2
 \\\noalign{\medskip} i \sigma^2 \phi_1^* \\\noalign{\medskip} i
 \sigma^2 \phi_2^*\end {array} \right)\ =\ \left( \begin {array}{c}
 \mathbf{V}_1\\\noalign{\medskip} \mathbf{V}_2 \\\noalign{\medskip}
 i \sigma^2 \mathbf{V}_1^* \\\noalign{\medskip} i \sigma^2
 \mathbf{V}_2^* \end {array} \right)\; , 
\end{equation}
where  $\mathbf{V}_{1,2}$  denote  the  VEVs  of  the  Higgs  doublets
$\phi_{1,2}$.   Employing the  freedom  of the  ${\rm SU(2)_L  \otimes
  U(1)_Y}$ gauge  transformations, the VEVs  $\mathbf{V}_{1,2}$ can be
parameterized as:
\begin{subequations}
\begin{align}
 \label{eq:V1 Ansatz}
\mathbf{V}_1 &= \frac{1}{ \sqrt{2} } \left( \begin {array}{c}
 0\\\noalign{\medskip}v^0_1\end {array} \right)\; ,\\
 \label{eq:V2 Ansatz} 
\mathbf{V}_2 &= \frac{1}{ \sqrt{2} } \left( \begin {array}{c}
 v^{+}_2\\
\noalign{\medskip}v^0_2 e^{i \xi} \end {array} \right)\; . 
\end{align}
\end{subequations}
where  the vacuum  manifold parameters  $v_1^0$, $v_2^0$,  $v^+_2$ and
$\xi$    are   all   real.    This    parameterization   of
$\mathbf{V}_{1,2}$ represents a single point of the vacuum manifold in
the $\Phi$-space,  which we denote as $\Phi_0$.  Under this particular
parameterization  of the VEVs  of the  two doublets  $\phi_{1,2}$, the
equivalent extremal  point in the  $\mathrm{R}^\mu$ basis in  terms of
the vacuum manifold parameters is:
\begin{equation}
\mathrm{R}^\mu_0\ =\ \left( \begin {array}{c} \frac{1}{2} (v_1^0)^2 +
  \frac{1}{2} (v_2^0)^2 + \frac{1}{2} (v_2^+)^2\\\noalign{\medskip}
  v_1^0 v_2^0 \cos \xi\\\noalign{\medskip} v_1^0 v_2^0 \sin
  \xi\\\noalign{\medskip} \frac{1}{2} (v_1^0)^2 - \frac{1}{2}(v_2^0)^2
  - \frac{1}{2} (v_2^+)^2 
\end {array} \right)\; . 
\label{eq:r^mu in terms of vacuum manifold parameters}
\end{equation}

Our aim is to determine  the entire vacuum manifold ${\cal M}_\Phi$ of
the  2HDM  potential,  which  amounts  to  finding  all  topologically
distinct points  of $\Phi$,  by appropriately acting  on~$\Phi_0$ with
the set ${\cal M}$ that leaves the minimum of the 2HDM potential ${\rm
  V}_{\rm min}$  invariant. Thus, our task  is to find  ${\cal M}$ and
its  topological  properties.  We  are  interested  in neutral  vacuum
solutions where  both VEVs  $v^0_{1,2}$ of $\phi_{1,2}$  are non-zero, 
i.e. situations where the vacuum component $v^+_2$ in 
\eqref{eq:V2 Ansatz} vanishes, $v^+_2=  0$, and both $v_{1,2}^0 \neq 0$.
As  a  consequence   of  the  latter,   the  VEVs
$\mathbf{V}_{1,2}$  are  invariant under  rotations  generated by  the
electromagnetic  operator ${\rm  Q}_{\rm em}  =  \frac{1}{2}\sigma^3 +
y_\phi\,\sigma^0$,   since  ${\rm   Q}_{\rm  em}   \mathbf{V}_{1,2}  =
(0,0)^{\rm  T}$,   where  $y_\phi  =   1/2$  is  the   hypercharge  of
$\phi_{1,2}$.  Hence,  if no  HF or CP  symmetries are present  in the
2HDM   potential,   a   non-trivial   transformation   of   the   VEVs
$\mathbf{V}_{1,2}$ can  only be  obtained by the  action of  the coset
set: ${\rm SU(2)_L \otimes U(1)_Y}/{\rm U(1)}_{\rm em}$.

If there is a HF (CP) symmetry group ${\rm G}_{\rm HF}$~$({\rm G}_{\rm
  CP})$ acting  on the scalar  potential~${\rm V}$, then one  needs to
know  whether there  is a  residual  HF (CP)  symmetry, ${\rm  H}_{\rm
  HF}$~$({\rm  H}_{\rm  CP})$ say,  which  survives after  spontaneous
symmetry breaking.   In such a breaking pattern:  ${\rm G}_{\rm HF/CP}
\to  {\rm  H}_{\rm HF/CP}$,  the  vacuum  manifold  point $\Phi_0$  is
invariant under the action of  the little group ${\rm H}_{\rm HF/CP}$,
such that
\begin{equation}
  \label{StabPhi0}
{\rm  H}_{\rm HF/CP}:\ \Phi_0\  \to\ \Phi'_0\  =\ {\rm M}_{\rm
  H}\,\Phi_0\ =\ \Phi_0\; ,
\end{equation}  
or  equivalently  ${\rm R}^\mu_0$  is  invariant  under ${\rm  H}_{\rm
  HF/CP}$, i.e.
\begin{equation}
  \label{StabR0}
{\rm  H}_{\rm HF/CP}:\ {\rm R}^\mu_0\ \to\ {\rm R}'^\mu_0 \ =\ 
(\Lambda^{\rm  H})^\mu_{\,\nu}\, {\rm  R}^\nu_0\ =\  {\rm R}^\mu_0\,,
\end{equation}
where  ${\rm  M}_{\rm   H}$  [$(\Lambda^{\rm  H})^\mu_{\,\nu}$]  is  a
representation  of the  unbroken group  ${\rm H}_{\rm  HF/CP}$  in the
${\rm GL}(8,\mathbb{C})$ [SO(1,3)] space.  As  we will see in the next
section,  this is  the case  for  the ${\rm  SO(3)_{HF}}$ model  which
breaks into  the subgroup  ${\rm SO(2)_{HF}} \cong  {\rm U(1)'_{PQ}}$.

Consequently, a non-trivial HF/CP  transformation of $\Phi_0$ or ${\rm
  R}^\mu_0$ can only  be performed in the coset  spaces: ${\rm G}_{\rm
  HF}/{\rm H}_{\rm HF}$  or ${\rm G}_{\rm CP}/{\rm H}_{\rm  CP}$. In a
group-theoretic language, the vacuum manifold points $\Phi_0$ or ${\rm
  R}^\mu_0$ satisfying~\eqref{StabPhi0}  and~\eqref{StabR0} are called
orbit stablizers  and the entire  vacuum manifold can be  generated by
the transitive action  of the total symmetry group ${\rm  G}$ on them, where
${\rm       G}       =       \mathrm{SU}(2)_{\mathrm{L}}       \otimes
\mathrm{U}(1)_{\mathrm{Y}}         \otimes         {\rm        G}_{\rm
  HF/CP}$~\footnote[4]{Throughout  our  study,  we  ignore  the  ${\rm
    SU(3)}_c$ colour gauge group  which remains unbroken by the colour
  singlet  VEVs of the  Higgs doublets  $\phi_{1,2}$.}.  Thus,  in the
${\rm  GL}(8,\mathbb{C})$  space, the  entire  vacuum  manifold for  a
potential with HF/CP symmetry may be described by the set
\begin{equation}
\mathcal{M}^{\rm HF/CP}_\Phi\ =\ \left\{\Phi \,:\; \Phi =
\mathcal{M}\,\Phi_0, \
\mathcal{M} \in ({\rm G}_{\rm HF/CP}/{\rm H}_{\rm HF/CP}) \otimes 
(\mathrm{SU}(2)_{\mathrm{L}} \otimes
\mathrm{U}(1)_{\mathrm{Y}}/\mathrm{U}(1)_{\rm em}) \right\}\; ,
\end{equation}
where $\Phi_0$  is the orbit  stabilizer which is invariant  under the
little  group $\mathrm{U}(1)_{\rm  em} \otimes  {\rm  H}_{\rm HF/CP}$.
The  topological properties of  $\mathcal{M}^{\rm HF/CP}_\Phi$  or its
generating set $\mathcal{M}$ under  its homotopy groups, $\Pi_n ({\cal
  M})$,     determines    the     nature     of    the     topological
defects~\cite{Vilenkin,Vachaspati}.    In  particular,  we   have  the
existence of domain walls for  $\Pi_0 ({\cal M}) \neq {\bf I}$, string
solutions for  $\Pi_1 ({\cal  M}) \neq {\bf  I}$, monopoles  if $\Pi_2
({\cal M}) \neq {\bf I}$ and  textures if $\Pi_{n > 2} ({\cal M}) \neq
{\bf  I}$~\cite{Vilenkin,Hindmarsh:1994re},  where  ${\bf I}$  is  the
identity element.

It is therefore vital to determine the representation of ${\cal M}$ in
the full 8-dimensional $\Phi$-space, for  a HF and a CP symmetry. With
this aim, we first note that  a general element ${\rm U}$ of the ${\rm
  SU(2)_L \otimes U(1)_Y}$ gauge group can always be written down as
\begin{equation}
 \label{eq:SMgroup}
{\rm U} \ =\ {\rm U_L\, U_Y}\ =\ \exp \bigg( i\, \theta_{\rm Y}\,
\frac{\sigma^3}{2}\bigg) \otimes \sigma^0 \otimes \exp\bigg( i\,
 \tilde{\theta}^1 \frac{\sigma^1}{2} + i\, \tilde{\theta}^2
 \frac{\sigma^2}{2}\bigg)\, \exp \bigg( i\, \tilde{\theta}^3
\frac{\sigma^3}{2}\,\bigg)\; .
\end{equation}
where  ${\rm  U_L}$  and   ${\rm  U_Y}$  are  given  in~\eqref{eq:SUL}
and~\eqref{eq:UY},  respectively.  Here,  we also  used  the so-called
Baker--Campbell--Haussdorf  formula to factor  out the  third rotation
due  to   the  generator  $\sigma^3/2$  of  ${\rm   U_L}$,  where  the
transformed group  parameters $\tilde{\theta}^{1,2,3}$ take  values in
the domain $[0,4\pi)$. Using~\eqref{eq:SMgroup}, one one can show that
  an element ${\rm U}^\perp$ of  the coset space ${\rm SU(2)_L \otimes
    U(1)_Y/U(1)_{\rm em}}$ may be represented in the $\Phi$-space as
\begin{equation}
 \label{eq:Uperp}
{\rm U}^\perp\ =\ \left(\frac{\sigma^0 + \sigma^3}{2} \right)
\otimes \sigma^0 \otimes \mathrm{U}_+\: +\: \left(\frac{\sigma^0 -
 \sigma^3}{2} \right) \otimes \sigma^0 \otimes \mathrm{U}_-\; ,
\end{equation}
with
\begin{equation}
 \label{eq:U+-}
{\rm U}_\pm\ =\ \exp \bigg( i\, \tilde{\theta}^1\,\frac{\sigma^1}{2} + 
i\, \tilde{\theta}^2\,
\frac{\sigma^2}{2}\bigg)\, \exp \bigg[ \pm i\, 
\bigg(\frac{\theta_{\rm Y} - \tilde{\theta}^3}{2}\bigg)\, 
\bigg(\frac{\sigma^0 \mp \sigma^3}{2}\bigg)\,\bigg]\; .
\end{equation}
Note that the elements ${\rm U}^\perp$ represent gauge transformations
of the  VEVs ${\bf V}_{1,2}$  orthogonal to the~${\rm  U(1)}_{\rm em}$
electromagnetic group.   In the $\Phi$-space, the latter  group can be
represented by an expression very analogous to~\eqref{eq:Uperp}, where
the $2\times 2$ matrices ${\rm U}_\pm$ are replaced with
\begin{equation}
  \label{eq:Uem}
{\rm U}^{\rm em}_\pm\ =\ \exp \bigg[ \pm i\, 
\bigg(\frac{\theta_{\rm Y} + \tilde{\theta}^3}{2}\bigg)\, 
\bigg(\frac{\sigma^0 \pm \sigma^3}{2}\bigg)\,\bigg]\; .
\end{equation}
Obviously, ${\rm  U}^\perp$ does  not account for  redundant rotations
within~${\rm U(1)}_{\rm  em}$, since $\frac{\sigma^0  + \sigma^3}{2}\,
{\bf   V}_{1,2}  =   (0\,,\,   0)^{\rm  T}$   and  $\frac{\sigma^0   -
  \sigma^3}{2}\, i\sigma ^2 {\bf  V}^*_{1,2} = (0\,,\, 0)^{\rm T}$. In
this  decomposition  of  the  electroweak gauge  group  ${\rm  SU(2)_L
  \otimes  U(1)_Y}$ into  the electromagnetic  group  ${\rm U(1)}_{\rm
  em}$ and  the coset space  ${\rm U}^\perp$, the  linear combinations
$\theta_\pm = \frac{1}{2}(\theta_{\rm Y} \pm \tilde{\theta}^3)$ should
be  regarded as  independent  parameters which  assume  values in  the
domain $\theta_\pm \in [0,2\pi)$.
 
Given  the  representation~\eqref{eq:Uperp}  for  ${\rm  U}^\perp$,  a
non-trivial  HF and  CP transformation  of the  vacuum  manifold point
$\Phi_0$ is given by the ${\rm GL}(8,\mathbb{C})$ matrices
\begin{subequations}
\begin{align}
 \label{eq:calM+}
\mathcal{M}_+\ = \ {\rm M}^\perp_+\, {\rm U}^\perp\ &=\
\left(\frac{\sigma^0 + \sigma^3}{2}
\right) \otimes \mathcal{T}_+ \otimes \mathrm{U}_+ \: 
+\:
\left(\frac{\sigma^0 - \sigma^3}{2} \right) \otimes \mathcal{T}^{*}_+
\otimes \mathrm{U}_-\; ,\\[3mm]
 \label{eq:calM-} 
 \mathcal{M}_-\ = \ {\rm M}^\perp_-\, {\rm U}^\perp\ &=\ 
\left(\frac{\sigma^1 + i\sigma^2}{2} \right) \otimes 
\mathcal{T}_- \otimes \Big[(-i\sigma^2)\, \mathrm{U}_-\Big] \nonumber\\
& \hspace{3cm} -\:
 \left(\frac{\sigma^1 - i\sigma^2}{2} \right) \otimes
 \mathcal{T}_-^{*} \otimes \Big[(-i\sigma_2)\, \mathrm{U}_+\Big]\; ,
\end{align} 
\end{subequations}
where $\mathcal{T}_{+} \in {\rm G}_{\rm HF}/{\rm H}_{\rm HF}$ and
$\mathcal{T}_\pm \in {\rm G}_{\rm CP}/{\rm H}_{\rm CP}$, with ${\cal
 T}_\pm$ being $2\times 2$ complex matrices. Similarly, ${\rm
 M}^\perp_+ \in {\rm G}_{\rm HF}/{\rm H}_{\rm HF}$ and ${\rm
 M}^\perp_\pm \in {\rm G}_{\rm CP}/{\rm H}_{\rm CP}$ are ${\rm
 GL}(8,\mathbb{C})$ matrices acting on the HF/CP coset spaces, whose
tensorial form is very analogous to those given in~\eqref{eq:Mtensor+}
and~\eqref{eq:Mtensor-}.

At this point,  it is important to reiterate that  a HF symmetry ${\rm
  G}_{\rm   HF}$  of   the   2HDM  potential   is   closed  under   HF
transformations ${\rm M}_+$ only,  whereas a CP symmetry requires both
types of  HF and CP transformations  ${\rm M}_\pm$ in  order to obtain
group closure,  according to Table~\ref{table:M+  M- Table}. Likewise,
the entire vacuum manifold for a 2HDM potential with a HF symmetry can
be  generated by  acting  only with  transformation  matrices of  type
${\cal M}_+$ given in~\eqref{eq:calM+}  on the initial vacuum manifold
point~$\Phi_0$.  Instead, for a  general CP-symmetric  2HDM potential,
the  complete  vacuum manifold  requires  the  use  of both  types  of
transformation      matrices     ${\cal     M}_\pm$      acting     on
$\Phi_0$~[cf.~\eqref{eq:calM+} and~\eqref{eq:calM-}].

As  was  already  mentioned   above,  we  may  obtain  an  alternative
description of the vacuum  manifold in the $\mathrm{R}^\mu$ space.  In
this bilinear field basis, the entire vacuum manifold can be generated
by  the transitive  action of  the full  group ${\rm  G}$ on  a single
vacuum manifold point $\mathrm{R}^\mu_0$, which is invariant under the
orbit  stabilizer  group  ${\rm H}_{\rm  HF/CP}$~[cf.~\eqref{StabR0}].
For this  purpose, we would need  to use the  $\Lambda_{\rm HF/CP}$ or
${\cal O}_{\rm  HF/CP}$ matrices presented  in~Table \ref{table:Lambda
  Matrix Table}  associated with  a given HF/CP  symmetry of  the 2HDM
potential.  The vacuum manifold is then given by the set
\begin{equation}
\mathcal{M}^{\rm HF/CP}_{\mathrm{R}^\mu}\ =\ \left\{\mathrm{R}^\mu \,:\; 
\mathrm{R}^\mu = \Lambda^\mu_{\;\nu}\, \mathrm{R}^\nu_0, \
\Lambda^\mu_{\;\nu} \in \Lambda_{\mathrm{HF/CP}}/\Lambda^{\rm H}_{\mathrm{HF/CP}}
\right\}\; ,
\end{equation}
where  $\Lambda^{\rm  H}_{\rm HF/CP}$  is  a  possible residual  HF/CP
symmetry that  remains intact after spontaneous  symmetry breaking. In
the      gauge-invariant      bilinear      field      basis,      the
$\mathrm{SU}(2)_{\mathrm{L}}    \otimes    \mathrm{U}(1)_{\mathrm{Y}}$
gauge-group  rotations   are  not  present,  so  the   nature  of  the
topological  defect  solution  depends  only on  the  homotopic  group
properties   of  the  coset   bilinear  field   spaces:  $\Lambda_{\rm
  HF}/\Lambda^{\rm  H}_{\rm  HF}$  or  $\Lambda_{\rm  CP}/\Lambda^{\rm
  H}_{\rm  CP}$. We  have checked  that the  analysis of  the homotopy
groups  of  the  vacuum   manifolds  in  the  Majorana-field  and  the
bilinear-field    bases,    $\mathcal{M}^{\rm    HF/CP}_{\Phi}$    and
$\mathcal{M}^{\rm   HF/CP}_{\mathrm{R}^\mu}$,   lead   to   identical
results. 

Finally,  we should  note  that the  breaking  of the  SM gauge  group
$\mathrm{SU}(2)_{\mathrm{L}}  \otimes  \mathrm{U}(1)_{\mathrm{Y}}$  to
${\rm  U(1)}_{\rm em}$  gives  rise  to a  vacuum  manifold, which  is
homeomorphic  to~$S^3$.    This  would  imply  that   $\Pi_3  [S^3]  =
\mathrm{Z}$,  which   would  be   indicative  for  the   formation  of
non-trivial  topological  configurations  called  textures.   However,
such local textures turn out to be gauge artifacts since they can
be removed by a gauge transformation~\cite{Vilenkin}. Global textures
and monopoles, whilst unstable due to Derrick's theorem, can be 
cosmologically interesting, for instance global monopoles can provide a mechanism for structure formation \cite{PhysRevLett.65.1709}. For  this
reason, our  focus will  be on non-trivial  topological configurations
that arise  from the  breaking of HF  or CP symmetries:  ${\rm G}_{\rm
  HF/CP} \to {\rm H}_{\rm HF/CP}$.

\section{Neutral Vacuum Solutions of the HF Symmetries}
\label{HF Symmetries}

We start our analysis by  considering the three generic HF symmetries:
$\mathrm{Z}_2$,            $\mathrm{U}(1)_{\mathrm{PQ}}$           and
$\mathrm{SO(3)}_{\mathrm{HF}}$.   These HF symmetries  impose specific
relations~\cite{Ferreira:2009wh}  among  the  parameters of  the  2HDM
potential,   which  are   presented  in   Table~\ref{table:HF  Parameter
  Conditions}. For  the $\mathrm{Z}_2$ symmetry,  the quartic coupling
$\lambda_5$ can always be made  real by a simple phase redefinition of
one of the two Higgs doublets $\phi_{1,2}$.

\begin{table}[t!]
\begin{center}
\begin{tabular}{c||cccccccccc}
\hline
 & & & & & & & & & & \\[-2mm]
Symmetry & 	$\mu_1^2$	&	$\mu_2^2$	&
$m_{12}^2$	&	$\lambda_1$	&	$\lambda_2$	&
$\lambda_3$	&	$\lambda_4$	&	$\lambda_5$	&
$\lambda_6$	&	$\lambda_7$ \\[2mm] 
\hline
\hline

$\mathrm{Z}_2$	&	--	&	--	&	0	&
--	&	--	&	--	&	--	&	Real
&	0	&	0	 \\ 

\hline

$\mathrm{U}(1)_{\mathrm{PQ}}$	&	--	&	--	&
0	&	--	&	--	&	--	&	--
&	0	&	0	&	0 \\ 

\hline

$\mathrm{SO(3)}_{\mathrm{HF}}$	&	--	&	$\mu_1^2$
&	0	&	--	&	$\lambda_1$	&	--
&	$2\lambda_1 - \lambda_3$	&	0	&	0
&	0\\ 

\hline
\end{tabular}
\caption{\it Parameter relations  in the  2HDM potential that  result from
  the imposition of the three  generic HF symmetries. A dash indicates
  the absence of a constraint.\label{table:HF Parameter Conditions} }
\end{center}
\end{table}

Given  the constraints  on  the  potential parameters  due  to the  HF
symmetries,  the  four  general  convexity  conditions  \eqref{eq:2HDM
  Convexity  1}--\eqref{eq:2HDM  Convexity  3}  and  \eqref{detL}  become
greatly   simplified.   These   four  conditions   are   exhibited  in
Table~\ref{table:HF Convexity Conditions}. In the ${\rm SO(3)}_{\rm HF}$
case, the convexity conditions are  not independent of each other and 
only one distinct condition survives.

\begin{table}[t!]
\begin{center}
\begin{tabular}{c||c|c|c}
\hline
Convexity Condition	& 	$\mathrm{Z}_2$	&	$\rm U(1)_{PQ}$	&
$\mathrm{SO(3)}_{\mathrm{HF}}$	\\ 
\hline
\hline

1	&	$\lambda_1 > 0$	&	$\lambda_1 > 0$	&
$2\lambda_1 > |\lambda_3|$	\\ 

\hline

2	&	$\lambda_2 > 0$	&	$\lambda_2 > 0$	&	--	\\

\hline

3	&	$2 \sqrt{\lambda_1 \lambda_2} > |\lambda_3|$	&
$2 \sqrt{\lambda_1 \lambda_2} > |\lambda_3|$	&	--	\\ 

\hline

4	&	$\lambda_4 > |\lambda_5|$	&	$\lambda_4 >
0$	&	--	\\ 

\hline
\end{tabular}
\caption{\it  The four convexity  conditions for  a bounded-from-below
  2HDM potential for each of the three HF symmetries. A dash signifies
  the  absence  of  any additional  constraints on the parameters.\label{table:HF  Convexity
    Conditions} }
\end{center}
\end{table}

We  will now  derive analytical  expressions for  the neutral  VEVs of
$\phi_{1,2}$ for  each of  the three HF  symmetries, by  utilizing the
Lagrange multiplier method. These results will enable us to study in
more detail possible topological defects that can emerge from a non-trivial vacuum topology
of the theory, as shown in Section~\ref{Topological  Defects of the 2HDM}.

\subsection{$\mathrm{Z}_2$ Symmetry}\label{Z2 Symmetry}

The discrete  $\mathrm{Z}_2$ symmetry  of the 2HDM  is defined  by the
following transformations of the two Higgs doublets $\phi_{1,2}$:
\begin{eqnarray}
\phi_1 &\to& \phi'_1\ =\ \phi_1 \; , \nonumber \\
\phi_2 &\to& \phi'_2\ =\ -\phi_2 \; . \nonumber
\end{eqnarray}
To solve the extremization condition \eqref{eq:Extremization Condition
  1},  we consider two  cases: (i)  $\mathrm{det} [\mathrm{N}_{\mu \nu}]
\neq 0$ and (ii) $\mathrm{det} [\mathrm{N}_{\mu \nu}] = 0$. In the first
case,  the  matrix $\mathrm{N}_{\mu  \nu}$  can  be  inverted and  the
4-vector ${\rm  R}^\mu$ can  be straightforwardly derived,  whereas in
the  second  case  $\mathrm{N}_{\mu  \nu}$  is not  invertible  and  a
slightly different  strategy needs  to be deployed  to determine~${\rm
  R}^\mu$.

Taking into  account the  parameter restrictions of  Table \ref{table:HF
  Parameter  Conditions} for  the  ${\rm Z}_2$  symmetry,  we may  now
calculate the determinant of $\mathrm{N}_{\mu \nu}$ (see also Appendix
\ref{The Determinant of  N}). This can be expressed  in the factorized
form:
\begin{equation}
   \label{eq:Z2 Det}
\mathrm{det} [\mathrm{N}_{\mu \nu}]\ =\ \left[\lambda_5^2 -
  \left(\lambda_4 + \zeta \right)^2 \right] \left[\left(\lambda_3 -
  \zeta \right)^2 - 4 \lambda_1 \lambda_2 \right]\; . 
\end{equation}
For   the    $\mathrm{Z}_2$   symmetry,   the    extremization   condition
$\mathrm{N}_{\mu  \nu} \mathrm{R}^\nu  = \mathrm{M}_\mu$  decomposes into
two separate matrix equations:
\begin{subequations}
\begin{align}
  \label{eq:r0 r3 Z2 equation}
\left( \begin {array}{cc} \lambda_1 + \lambda_2 + \lambda_3 -
  \zeta&\lambda_1 - \lambda_2\\\noalign{\medskip}\lambda_1 -
  \lambda_2&\lambda_1 + \lambda_2 - \lambda_3 + \zeta\end {array} 
 \right) \left( \begin {array}{c}
   \mathrm{R}^0\\\noalign{\medskip}\mathrm{R}^3\end {array} \right)\ &=\
   \left( \begin {array}{c} \mu_1^2 + \mu_2^2\\\noalign{\medskip}
     \mu_1^2 - \mu_2^2\end {array} \right)\; ,  \\[3mm] 
  \label{eq:r1 r2 Z2 equation}
\left( \begin {array}{cc} \lambda_4 + \lambda_5 + \zeta&
  0\\\noalign{\medskip}0&\lambda_4 - \lambda_5 + \zeta\end {array} 
 \right) \left( \begin {array}{c}
   \mathrm{R}^1\\\noalign{\medskip}\mathrm{R}^2\end {array} \right)\ &=\
   \left( \begin {array}{c} 0\\\noalign{\medskip}0\end {array}
     \right) \; .
\end{align}
\end{subequations}
Assuming that $\mathrm{N}_{\mu \nu}$ is non-singular, the above matrix
relations   can  be   inverted  and   the  individual   components  of
$\mathrm{R}^\mu$  for an arbitrary  point on  the vacuum  manifold are
found to be
\begin{subequations}
\begin{align}
\mathrm{R}^0\ &=\ \frac{2\lambda_2 \mu_1^2 + 2 \lambda_1 \mu_2^2 -
  (\lambda_3 - \zeta)(\mu_1^2 + \mu_2^2)}{4\lambda_1 \lambda_2 -
  (\lambda_3 - \zeta)^2}\ , \label{eq:r0 Z2}\\ 
\mathrm{R}^1\ &=\ 0\; , \\[2mm]
\mathrm{R}^2\ &=\ 0\; , \\
\mathrm{R}^3\ &=\ \frac{2\lambda_2 \mu_1^2 - 2 \lambda_1 \mu_2^2 +
  (\lambda_3 - \zeta)(\mu_1^2 - \mu_2^2)}{4\lambda_1 \lambda_2 -
  (\lambda_3 - \zeta)^2}\  . \label{eq:r3 Z2} 
\end{align}
\end{subequations}
From the defining  equation~\eqref{eq:r^mu definition} for the 4-vector
$\mathrm{R}^\mu$, the following analytical expressions for the VEVs of
the Higgs field bilinears are easily obtained:
\begin{subequations}
\begin{align}
\langle\phi_{1}^{\dagger}\phi_{1}\rangle\ &=\ \frac{2\lambda_2 \mu_1^2 -
  (\lambda_3 - \zeta) \mu_2^2}{4\lambda_1 \lambda_2 - (\lambda_3 -
  \zeta)^2} \ \geq\ 0\; , \label{eq:Z2 Det not zero VEV1}\\ 
\langle\phi_{2}^{\dagger}\phi_{2}\rangle\ &=\ \frac{2\lambda_1 \mu_2^2 -
  (\lambda_3 - \zeta) \mu_1^2}{4\lambda_1 \lambda_2 - (\lambda_3 -
  \zeta)^2} \ \geq\ 0\; , \\[2mm] 
\langle\phi_{1}^{\dagger}\phi_{2}\rangle\ &=\
\langle\phi_{2}^{\dagger}\phi_{1}\rangle\ =\ 0\; . 
  \label{eq:Z2 Det not zero  VEV2} 
\end{align}
\end{subequations}
In  order to  have  a neutral  vacuum  solution, we  must satisfy  the
condition~\eqref{eq:Lagrange Multiplier  Condition}, namely that ${\rm
  R}^\mu$  is a null  4-vector, with  ${\rm R}^\mu  {\rm R}_\mu  = 0$.
This restriction leads to 
\begin{equation}
\frac{\left[2\lambda_2 \mu_1^2 - (\lambda_3 - \zeta) \mu_2^2\right]
  \left[2\lambda_1 \mu_2^2 - (\lambda_3 - \zeta) \mu_1^2
    \right]}{4\lambda_1 \lambda_2 - (\lambda_3 - \zeta)^2}\  =\ 0\; , 
\label{eq:LM Z2 Conditions, det N not 0}
\end{equation}
which completely specifies the Lagrange  multiplier. More explicitly,
requiring that the numerator of  \eqref{eq:LM Z2 Conditions, det N not
  0} vanishes, we find two solutions for the Lagrange multiplier:
\begin{subequations}
\begin{align}
\zeta_1\ &=\ \lambda_3 - \frac{2 \lambda_1
  \mu_2^2}{\mu_1^2}\ , \label{eq:Z2 Zeta 1} \\ 
\zeta_2\ &=\ \lambda_3 - \frac{2 \lambda_2 \mu_1^2}{\mu_2^2} \ .
  \label{eq:Z2 Zeta 2}
\end{align} 
\end{subequations}
Using   the  specific   parameterization   \eqref{eq:V1  Ansatz}   and
\eqref{eq:V2 Ansatz}  for the VEVs  of $\phi_{1,2}$, we  can determine
the vacuum manifold parameters  ($v_1^0, v_2^0, v_2^{+}, \xi$) for the
two  values   $\zeta_{1,2}$  of  the  Lagrange   multiplier  given  in
\eqref{eq:Z2 Zeta 1} and \eqref{eq:Z2  Zeta 2}.  The results are given
in Table \ref{table:Z2 vacuum  manifold parameters det not 0}. Moreover,
we have verified that the  two solutions  $\zeta_{1,2}$ do  not lead  to a
singular matrix~$\mathrm{N}_{\mu \nu}$.

\begin{table}[t!]
\begin{center}
\begin{tabular}{c||c|c}
\hline
VEV parameter &  ~~$\zeta_1$~~ & ~~$\zeta_2$~~ \\
\hline
\hline

$v_1^0$ & $\displaystyle \sqrt{\frac{\mu_1^2}{\lambda_1}}$ & 0 \\

\hline

$v_2^0$ & 0 & $\displaystyle \sqrt{\frac{\mu_2^2}{\lambda_2}}$ \\

\hline

$v_2^+$ & 0 & 0 \\

\hline

$\xi$ & 0 & 0 \\

\hline
\end{tabular}
\caption{\it The  two neutral  vacuum solutions to  the $\mathrm{Z}_2$
  symmetric  2HDM  potential  for $\mathrm{det}[\mathrm{N}_{\mu  \nu}]
  \neq  0$.    The  Lagrange  multipliers   $\zeta_{1,2}$  are  given
  in~\eqref{eq:Z2  Zeta  1}  and  \eqref{eq:Z2  Zeta  2}.\label{table:Z2
    vacuum manifold parameters det not 0}}
\end{center}
\end{table}

In order  for a  set of  neutral vacuum solutions  to correspond  to a
local minimum  of the  potential, we require  that the Hessian  of the
$\mathrm{Z}_2$  invariant 2HDM  potential is  positive  definite.  The
general Hessian  of the  $\mathrm{Z}_2$ invariant 2HDM  potential with
respect to~$v^0_1$ and~$v^0_2$ is given by
\begin{equation}
\mathrm{H}\ =\ \left( \begin {array}{cc} -\mu_1^2 + 3\lambda_1 (v_1^0)^2
  + \frac{1}{2} \lambda_{345} (v_2^0)^2& \lambda_{345} v_1^0
  v_2^0\\\noalign{\medskip}\lambda_{345} v_1^0 v_2^0&-\mu_2^2 +
  3\lambda_2 (v_2^0)^2 + \frac{1}{2} \lambda_{345} (v_1^0)^2\end
  {array} \right)\; . 
\label{eq:Z2 Hessian}
\end{equation} 
Here we introduce the common summation conventions between the quartic
couplings  of the model:  $\lambda_{ab} =  \lambda_a +  \lambda_b$ and
$\lambda_{abc}  =  \lambda_a  +  \lambda_b +  \lambda_c$.   Thus,  the
positivity  of~${\rm H}$  leads to  additional constraints,  which are
listed   in   Table~\ref{table:Z2  Minimum   Conditions   Det  Not   0}.
Specifically,  the   first  condition  in   Table~\ref{table:Z2  Minimum
  Conditions Det Not 0} corresponds  to having a local minimum, whilst
the second  one is to ensure that  this minimum is the  lowest one. If
$\mu^2_1 = \mu^2_2$ and $\lambda_1 = \lambda_2$, the global minimum is
given by
\begin{equation}
  \label{eq:Zero VEV Z2 Pot Min}
\mathrm{V}_0\ =\ -\; \frac{\mu_{1,2}^4}{4 \lambda_{1,2}}\ .
\end{equation}

\begin{table}[t!]
\begin{center}
\begin{tabular}{c||c|c}
\hline
Condition	&	$\zeta_1$	&	$\zeta_2$ \\
\hline
\hline

1	&	$\displaystyle \mu_1^2 > 0$	& $\displaystyle
\mu_2^2 > 0$ 	 \\ 

\hline

2	&	$\displaystyle \frac{\mu_1^2}{\mu_2^2} > \frac{2
  \lambda_1}{\lambda_{345}}$ &	$\displaystyle \frac{\mu_1^2}{\mu_2^2}
< \frac{\lambda_{345}}{2 \lambda_2}$ \\ 

\hline
\end{tabular}
\caption{\it Minimization conditions  for two neutral vacuum solutions
  in    a    $\mathrm{Z}_2$    symmetric    2HDM    potential,    with
  $\mathrm{det}[\mathrm{N}_{\mu  \nu}] \neq  0$.  The  first condition
  corresponds to having a local minimum and the second one is for this
  minimum to  be the lowest.~\label{table:Z2 Minimum  Conditions Det Not
    0}}
\end{center}
\end{table}

As can  be seen from Table~\ref{table:Z2 vacuum  manifold parameters det
  not 0}, when the  determinant of $\mathrm{N}_{\mu \nu}$ is non-zero,
at least  one of the VEVs  of the Higgs doublets  $\phi_{1,2}$ must be
zero, in order to have a  neutral vacuum solution.  As we will discuss
in~Section  \ref{Z2   Topology},  such   solutions  do  not   lead  to
topological defects, such  as domain walls in this  case, and they are
not of  interest for the present  study. We now turn  our attention to
the  neutral   vacuum  solutions  that  can  occur   when  the  matrix
$\mathrm{N}_{\mu \nu}$  becomes singular for a specific  choice of the
Lagrange multiplier.

\subsubsection{Neutral Vacuum Solutions from a Singular Matrix N}
\label{sec:Z2LMsingular}

We now consider the possibility that the matrix $\mathrm{N}_{\mu \nu}$
has  no   inverse,  by  requiring   that  its  determinant   given  in
\eqref{eq:Z2  Det} vanishes.  Equating separately  the two  factors in
\eqref{eq:Z2 Det} to zero, we obtain four solutions:
\begin{subequations}
\begin{align}
\zeta_{1,\,\pm}\ &=\ -\lambda_4 \pm \lambda_5\; , 
\label{eq:zeta 1 Z2, N singular} \\
\zeta_{2,\,\pm}\ &=\ \pm 2 \sqrt{\lambda_1 \lambda_2} +
\lambda_3 \label{eq:zeta 2 Z2, N singular}\, . 
\end{align}
\end{subequations}
Since  the extremization  condition for  the  $\mathrm{Z}_2$ invariant
potential splits  into two separate matrix  equations, \eqref{eq:r0 r3
  Z2 equation}  and \eqref{eq:r1 r2  Z2 equation}, the  application of
either of the above four Lagrange multipliers only results in one of
the matrices in the equations becoming singular. For the solution
$\zeta_{2,\,\pm}$, it is the  $2\times 2$ matrix in~\eqref{eq:r0 r3 Z2
  equation}   which  becomes   singular.   However,   since   the  RHS
of~\eqref{eq:r0 r3  Z2 equation}  is in general  a non-zero  vector in
this case,  unless $\mu^2_1  = \mu^2_2 =  0$, this matrix  equation is
overdetermined.  Unless  the parameters $\mu^2_{1,2}$  and the quartic
couplings $\lambda_{1,2,3}$ satisfy an unnatural fine-tuning relation,
the matrix equation~\eqref{eq:r0  r3 Z2 equation} becomes incompatible
for the Lagrange multiplier $\zeta_{2,\,\pm}$. We therefore reject the
second   solution~$\zeta_{2,\,\pm}$    and   focus   on    the   first
solution~$\zeta_{1,\,\pm}$.

For the Lagrange  multiplier solution~$\zeta_{1,\,\pm}$, the matrix in
\eqref{eq:r1  r2  Z2 equation}  becomes  singular,  whilst the  matrix
equation  \eqref{eq:r0 r3  Z2 equation}  can be  inverted  in general,
using standard linear algebra methods.  Evaluating the singular matrix
in~\eqref{eq:r1  r2  Z2  equation},   we  observe  that  the  solution
$\zeta_{1,\,+}$ yields  $\mathrm{R}^1 = 0$,  but leaves $\mathrm{R}^2$
undetermined.    Likewise,   the   solution  $\zeta_{1,\,-}$   renders
$\mathrm{R}^2 =  0$, but  $\mathrm{R}^1 \neq 0$  in general.   The two
solutions are  related by a reparameterization of  the doublets, since
$\Phi_2  \to  i  \Phi_2$  implies $\zeta_{1,\,+}  \to  \zeta_{1,\,-}$.
Therefore, only one  solution of the Lagrange multipliers  needs to be
considered.

Having the  above in mind,  we consider the  solution $\zeta_{1,\,-}$,
where  $\lambda_5$  enters  additively  in  all  resulting  equations.
Substituting $\zeta_{1,\,-}$ into \eqref{eq:r0 r3 Z2 equation} gives
\begin{subequations}
\begin{align}
\mathrm{R}^0\ &=\ \frac{2\lambda_1 \mu_2^2 + 2 \lambda_2 \mu_1^2 -
  \lambda_{345}(\mu_1^2 + \mu_2^2)}{4\lambda_1 \lambda_2 -
  \lambda_{345}^2}\ , \label{eq:LM1 r0 Z2}\\ 
\mathrm{R}^3\ &=\ \frac{2\lambda_2 \mu_1^2 - 2 \lambda_1 \mu_2^2 +
  \lambda_{345}(\mu_1^2 - \mu_2^2)}{4\lambda_1 \lambda_2 -
  \lambda_{345}^2}\ . \label{eq:LM1 r3 Z2} 
\end{align}
\end{subequations}
In  terms of  field bilinear  VEVs, $\mathrm{R}^0$  and $\mathrm{R}^3$
imply that
\begin{subequations}
\begin{align}
\langle\phi_{1}^{\dagger}\phi_{1} \rangle\ &=\ \frac{2\lambda_2 \mu_1^2 -
  \lambda_{345} \mu_2^2}{4\lambda_1 \lambda_2 - \lambda_{345}^2}\ >\ 0\; , 
\label{eq:LM1 VEV1 Z2} \\[3mm] 
\langle\phi_{2}^{\dagger}\phi_{2} \rangle\ &=\ \frac{2\lambda_1 \mu_2^2 -
  \lambda_{345} \mu_1^2}{4\lambda_1 \lambda_2 -
  \lambda_{345}^2}\ >\ 0\; . 
\label{eq:LM1 VEV2 Z2} 
\end{align}
\end{subequations}
In  addition,   the  constraint  ${\rm  R}^2  =   0$  translates  into
$\langle\phi_{1}^{\dagger}\phi_{2}              \rangle              =
\langle\phi_{2}^{\dagger}\phi_{1}   \rangle$,   which   can  only   be
satisfied  if the  phase $\xi$  is a  multiple of  $\pi$,  i.e.~$\xi =
n\pi$, with $n$ being an integer.

In order  to uniquely fix  the undetermined component ${\rm  R}^1$, we
require now that ${\rm  R}^\mu$ is a null vector, i.e.~$\mathrm{R}_\mu
\mathrm{R}^\mu = 0$.  Employing this last condition, we find that
\begin{equation}
  \label{eq:LM1 r1 Z2}
(\mathrm{R}^1)^2\ =\ \frac{4\left[2\lambda_2 \mu_1^2 - \lambda_{345}
      \mu_2^2 \right] \left[2\lambda_1 \mu_2^2 - \lambda_{345} \mu_1^2
      \right] }{\left[4\lambda_1 \lambda_2 -
      \lambda_{345}^2\right]^2}\ .
\end{equation}
Comparing \eqref{eq:LM1 r0 Z2}, \eqref{eq:LM1 r3 Z2} and \eqref{eq:LM1
  r1 Z2} with  the $\mathrm{R}^\mu$ parameterization in~\eqref{eq:r^mu
  in terms of vacuum manifold parameters}  with $v_2^+ = 0$ and $\xi =
0$, we obtain
\begin{subequations}
\begin{align}
v_1^0\ &=\ \sqrt{\frac{4\lambda_2 \mu_1^2 - 2\lambda_{345}
    \mu_2^2}{4\lambda_1 \lambda_2 - \lambda_{345}^2}} \label{eq:Z2
  Zeta 1 V1}\ , \\  
v_2^0\ &=\ \sqrt{\frac{4\lambda_1 \mu_2^2 - 2\lambda_{345}
    \mu_1^2}{4\lambda_1 \lambda_2 - \lambda_{345}^2}}\ . \label{eq:Z2 Zeta 1 V2}
\end{align}
\end{subequations}
By analogy, we may calculate the vacuum manifold parameters related to
the    Lagrange    multiplier    $\zeta_{1,\,+}   =    -\lambda_4    +
\lambda_5$. These are found simply by replacing $\lambda_{345}$ in all
equations with $\bar{\lambda}_{345}$,  where we extended the summation
convention  as:   $\bar{\lambda}_{abc}  =  \lambda_a   +  \lambda_b  -
\lambda_c$.   As  we  discuss  in Section~\ref{Points  on  the  Vacuum
  Manifold}, the space of the  entire vacuum manifold is generated via
the transitive action  of the total symmetry group  on this particular
set of the vacuum manifold  parameters.  We have also checked that the
VEVs  of the  Higgs  doublets $\phi_{1,2}$  obtained  by the  Lagrange
multiplier method  satisfy the  extremization conditions given  by the
usual     tadpole      equations~\eqref{eq:Phi     1     Minimization}
and~\eqref{eq:Phi 2 Minimization}.

\begin{figure}[t!]
\begin{center}
\includegraphics[scale=0.5]{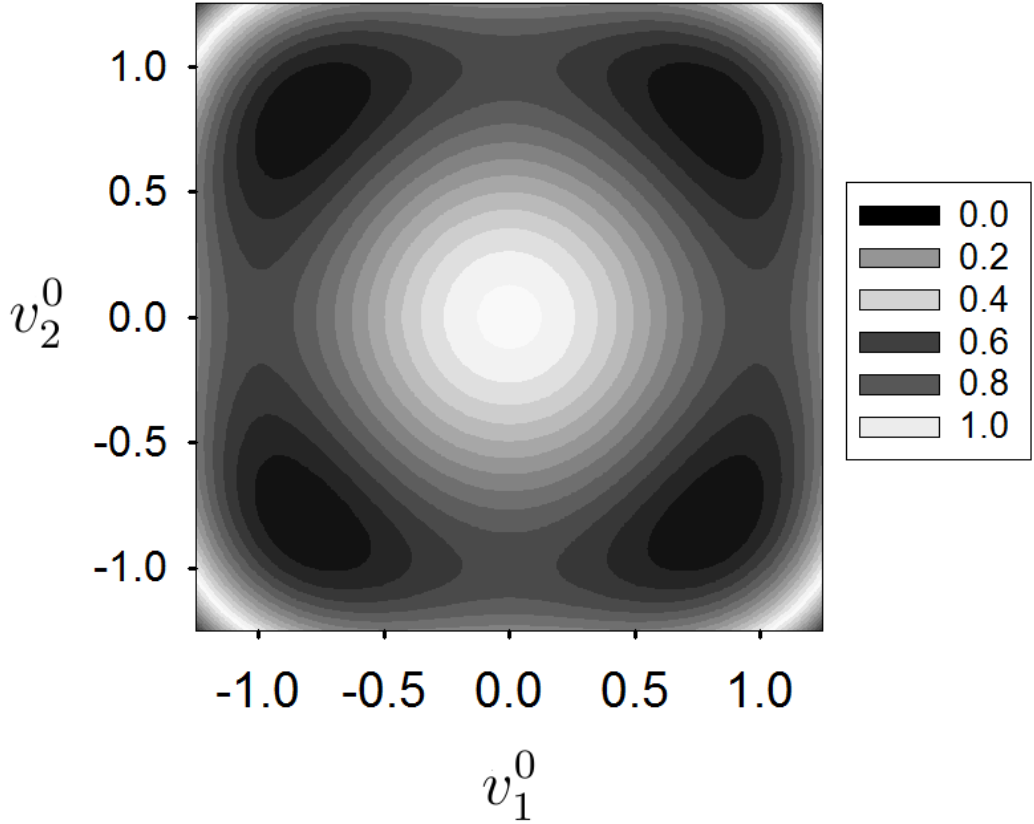}
\label{fig:Z2PotentialContour1}
\caption{\it   Contour  plot   depicting   the  shape   of  the $\rm Z_2$ invariant  2HDM
  potential~${\rm  V}$  for  the  parameter set  $\{\mu_1^2,  \mu_2^2,
  \lambda_1,  \lambda_2, \lambda_{345}  \} =  \{1, 1,  1, 1,  1\}$, in
  arbitrary mass units  and normalized such that ${\rm  V}_{\rm min} =
  0$.  The four degenerate and disconnected global minima are shown in
  black around  the central  local maximum. The  four minima  form two
  pairs;   the  members   within  each   pair  are   related   by  the
  $\mathrm{Z}_2$ symmetry and the two pairs are related to one another
  by ${\rm U}(1)_{\rm Y}$.}\label{fig:Z2potential}
\end{center}
\end{figure}

To  determine whether  the  above extremal  solutions represent  local
minima as well, we require  that the Hessian ${\rm H}$ in \eqref{eq:Z2
  Hessian},   evaluated   at   the   extremal  points,   is   positive
definite. This requirement generates two conditions:
\begin{subequations}
\begin{align}
  \label{eq:Z2 Zeta 2 Minima Condition 1} 
\lambda_1 \, \left(\frac{4\lambda_2 \mu_1^2 - 2\lambda_{345}
    \mu_2^2}{4\lambda_1 \lambda_2 - \lambda_{345}^2}\right) \ &>\ 0\; ,\\[2mm]
  \label{eq:Z2 Zeta 2 Minima Condition 2}
\frac{\Big( 4\lambda_2 \mu_1^2 - 2\lambda_{345} \mu_2^2 \Big)\,
  \Big( 4\lambda_1 \mu_2^2 - 2\lambda_{345} \mu_1^2
    \Big)}{4\lambda_1 \lambda_2 - \lambda_{345}^2}\ &>\ 0\; .  
\end{align}
\end{subequations}
These two inequalities are equivalent to the positivity conditions for
the squared VEVs in \eqref{eq:LM1 VEV1 Z2} and \eqref{eq:LM1 VEV2 Z2},
provided $4\lambda_1 \lambda_2 > \lambda_{345}^2$ and $\lambda_1 > 0$.
The  constraint  $\lambda_1  >  0$  represents one  of  the  convexity
conditions    for   the    ${\rm   Z}_2$-symmetric    2HDM   potential
(see~Table~\ref{table:HF    Convexity   Conditions}).     However,   the
restriction  $4\lambda_1  \lambda_2 >  \lambda_{345}^2$  has not  been
accounted  before and  creates  two additional  inequalities from  the
numerators   of  the   fractions  given   in~\eqref{eq:LM1   VEV1  Z2}
and~\eqref{eq:LM1  VEV2 Z2}.  These can  be summarized  in  the double
inequality
\begin{equation}
\frac{\lambda_{345}}{2\lambda_2}\ <\ \frac{\mu_1^2}{\mu_2^2}\ <\
\frac{2 \lambda_1}{\lambda_{345}}\ .
\label{eq:Z2 Zeta1 Global Minimum Condition}  
\end{equation}
Comparing   this   double  inequality   with   the   second  line   in
Table~\ref{table:Z2  Minimum Conditions Det  Not 0},  we see  that local
minima with  $v^0_{1,2} =  0$ and $v^0_{1,2}  \neq 0$  cannot coexist.
The value  of the potential at  the local minimum  associated with the
Lagrange multiplier $\zeta_{1,\,-}$ is given by
\begin{equation}
\mathrm{V}_0\ =\ \frac{\lambda_{345} \mu_1^2 \mu_2^2 - \lambda_1
  \mu_2^4 - \lambda_2 \mu_1^4}{4 \lambda_1 \lambda_2 -
  \lambda_{345}^2} \ .
\label{eq:LM1 Z2 Pot Min}
\end{equation}
The corresponding value  ${\rm V}_0$ for the local  minimum related to
$\zeta_{1,+}  = -  \lambda_4 +  \lambda_5$ is  obtained by  making the
substitution $\lambda_{345}  \to \bar{\lambda}_{345}$ in \eqref{eq:LM1
  Z2 Pot  Min}.  Between  these two solutions,  the lowest  minimum is
given by $\zeta_{1,+} = -  \lambda_4 + \lambda_5$, if $\lambda_5 > 0$,
and by  $\zeta_{1,-} = - \lambda_4  - \lambda_5$, if  $\lambda_5 < 0$.
Hence, the potential at the lowest minimum is given by
\begin{equation}
\mathrm{V}_0\ =\ \frac{\left(\lambda_3 + \lambda_4 - |\lambda_5| \right)
  \mu_1^2 \mu_2^2 - \lambda_1 \mu_2^4 - \lambda_2 \mu_1^4}{4 \lambda_1
  \lambda_2 - \left(\lambda_3 + \lambda_4 - |\lambda_5| \right)^2 }\ . 
\label{eq:LM1 Z2 Absolute Pot Min}
\end{equation}
Note  that this  lowest  minimum becomes  a  global one  of the  ${\rm
  Z}_2$-symmetric 2HDM potential, if~\eqref{eq:Z2 Zeta1 Global Minimum
  Condition}  is fulfilled.  Otherwise, the  global minimum  is given
by~\eqref{eq:Zero VEV  Z2 Pot  Min}.  A numerical  example of  a ${\rm
  Z}_2$-symmetric 2HDM potential, where both $v^0_{1,2}$ are non-zero,
is shown in Figure~\ref{fig:Z2potential}.

\subsubsection{$\mathrm{Z}_2$ Topology}
\label{Z2 Topology}

It is now  important to determine the topology  of the vacuum manifold
for the $\mathrm{Z}_2$ invariant  2HDM potential, applying some of the
general  results  presented   in  Section~\ref{Points  on  the  Vacuum
  Manifold}. In the symmetric phase, the $\mathrm{Z}_2$ invariant 2HDM
potential    is    governed    by    the    total    symmetry    group
$\mathrm{G}_{\mathrm{Z}_2}       \equiv      \mathrm{Z}_2      \otimes
\mathrm{SU}(2)_{\mathrm{L}}    \otimes    \mathrm{U}(1)_{\mathrm{Y}}$,
including  the electroweak  gauge group.   After  spontaneous symmetry
breaking of the electroweak gauge group, we have
\begin{equation}
  \label{eq:Electroweak Symmetry Breaking Scheme}
\mathrm{SU}(2)_{\mathrm{L}}  \otimes  \mathrm{U}(1)_{\mathrm{Y}} \simeq
S'^3 \times S'^1\ \to\ {\rm U(1)}_{\rm em} \simeq S^1\; .
\end{equation}
In the  above, we used  the well-known homeomorphisms  between compact
groups  and  $n$-spheres denoted  as  $S^n$  (or $S'^n$):~${\rm  U(1)}
\simeq S^1$ and ${\rm SU(2)} \simeq S^3$.  According to our discussion
in Section~\ref{Points on the Vacuum  Manifold}, in the absence of any
HF/CP symmetry  in the  theory, the vacuum  manifold of the  2HDM will
then be homeomorphic to the coset space $(S'^3\times S'^1)/S^1$, which
in    turn    is     homeomorphic    to    $S^3$,    i.e.~$(S'^3\times
S'^1)/S^1~\simeq~S^3$.

In   the   present  case,   there   exists   an  additional   discrete
$\mathrm{Z}_2$ symmetry  acting on  the 2HDM, which  can break  to the
identity, i.e.~$\mathrm{Z}_2 \to  {\bf I}$, after electroweak symmetry
breaking.  If this happens, the breaking pattern of the total symmetry
group proceeds as follows:
\begin{equation}
  \label{vac:Z2}
\mathrm{G}_{\mathrm{Z}_2}\ \equiv\ \mathrm{Z}_2 \otimes
\mathrm{SU}(2)_{\mathrm{L}} \otimes \mathrm{U}(1)_{\mathrm{Y}}\ \to
\ \mathrm{H}_{\mathrm{Z}_2}\ \equiv\ {\bf I} \otimes
\mathrm{U}(1)_{\mathrm{em}}\; .
\end{equation}
As a  consequence, the  topology of the  vacuum manifold will  then be
described   by   the    coset   space   ${\cal   M}^{\mathrm{Z}_2}_\Phi   =
\mathrm{G}_{\mathrm{Z}_2}/\mathrm{H}_{\mathrm{Z}_2}$.

In order  to generate the complete  set ${\cal M}^{\mathrm{Z}_2}_\Phi$
of the  vacuum manifold points in  the $\Phi$-space, we first need to
find an initial point $\Phi_0$ of the Majorana scalar-field multiplet,
which     remains     invariant     under     the     little     group
$\mathrm{H}_{\mathrm{Z}_2}$.    Then,  ${\cal  M}^{\mathrm{Z}_2}_\Phi$
will     be    generated     by    the     transitive     action    of
$\mathrm{G}_{\mathrm{Z}_2}$ on  $\Phi_0$.  In the  parameterization of
the  Higgs-doublet VEVs  ${\rm V}_{1,2}$  of \eqref{eq:V1  Ansatz} and
\eqref{eq:V2 Ansatz}, the Majorana scalar-field vacuum point $\Phi_0$,
which  is   invariant  under  $\mathrm{H}_{\mathrm{Z}_2}   \cong  {\rm
  U(1)}_{\rm em}$, is given by $v^+_2 = 0$ and $\xi = 0$.

Let us first consider the non-trivial case where $v^0_{1,2} \neq 0$ as
discussed  in   Section~\ref{sec:Z2LMsingular}.   The  general  vacuum
manifold point~$\Phi$  is given by 
\begin{equation}
\Phi\ =\ {\cal M}^{\rm Z_2}_+\, \Phi_0\; ,
\end{equation}
where the HF transformation matrix  ${\cal M}^{\rm Z_2}_+$ is stated in
\eqref{eq:calM+} and ${\cal T}_+ = \mathrm{T}_+ = \{ \sigma^0\, , \,\sigma^3\}$
are   the   $2\times   2$   HF  transformation   matrices   given   in
Table~\ref{table:Transformation  Matrix Table}  under the  ${\rm Z}_2$
symmetry.   It  is interesting  to  see  the  different roles  of  the
$\mathrm{Z}_2$ symmetry  and the ${\rm  U(1)_Y}$ hypercharge symmetry,
according to the more intuitive chart:
\begin{equation}
 \begin {array}{ccc} \left( \begin {array}{c}
     \mathbf{V}_1\\\noalign{\medskip}\mathbf{V}_2\end {array}
     \right)&\stackrel{\mathrm{U}(1)_\mathrm{Y}}{\longleftrightarrow}&
     \left( \begin 
       {array}{c} -\mathbf{V}_1\\\noalign{\medskip}-\mathbf{V}_2\end
       {array} \right)\\\noalign{\medskip}\mathrm{Z}_2
       \updownarrow&&\updownarrow\mathrm{Z}_2 
\\\noalign{\medskip}\left( \begin {array}{c}
  \mathbf{V}_1\\\noalign{\medskip}-\mathbf{V}_2\end {array}
  \right)&\stackrel{\mathrm{U}(1)_\mathrm{Y}}{\longleftrightarrow}
  &\; \left( \begin {array}{c}
    -\mathbf{V}_1\\\noalign{\medskip}\mathbf{V}_2\end {array}
    \right)\; .\end {array}  
\label{eq:U1 Cannot Undo Z2}
\end{equation}
Observe  that  for  ${\rm  Z}_2$-symmetric  2HDM  scenarios  with  two
non-zero  VEVs  $v^0_{1,2}  \neq  0$,  we  cannot  move  via  a  ${\rm
  U}(1)_{\rm  Y}$   transformation  from  one   vacuum  configuration,
e.g.~$(v^0_1\,,\,v^0_2)$,   to    its   ${\rm   Z}_2$-symmetric   one,
i.e.~$(v^0_1\,,\,  -v^0_2)$  or  $(-v^0_1\,,\, v^0_2)$.   However,  if
$v^0_1$ or  $v^0_2$ were zero, then  such a transformation  would be possible,
and  the discrete  vacua  will  be connected  via  a continuous  ${\rm
  U}(1)_{\rm Y}$  gauge transformation. In the latter  case, there are
no  topological  defects,  such  as domain  walls  or  superconducting
condensates     similar      to     the     ones      discussed     by
Hodges~\cite{Hodges:1988qg},  even  though  such  scenarios  might  be
interesting as they  predict stable scalars which may  act as DM (see,
e.g.~\cite{Grzadkowski:2009bt}).

On the  other hand, the  $\mathrm{Z}_2$ invariant 2HDM, where  the two
VEVs are non-zero, can lead to non-trivial topological solutions, such
as domain  walls~\footnote[5]{Here we assume that there are no other
  sources  that violate  the  $\mathrm{Z}_2$ symmetry  of the  theory,
  e.g.,~either       by       Yukawa       couplings,      or       by
  anomalies~\cite{Dvali:1994wv}.}.    The  vacuum   manifold   in  the
$\Phi$-space may be given by
\begin{equation}
\mathcal{M}^{\mathrm{Z}_2}_{\Phi}\ \simeq\ \mathrm{Z}_2 \times S^3\; ,
\end{equation}
where the second  factor $S^3$ comes from the  breaking pattern of the
electroweak  gauge group  as  given in~\eqref{eq:Electroweak  Symmetry
  Breaking Scheme}.  Thus, the action  of the zeroth homotopy group on
this  vacuum manifold  is non-trivial, since~$\Pi_0 \left[\mathrm{Z}_2
  \times S^3 \right] =  \Pi_0 \left[\mathrm{Z}_2 \right] \times \Pi_0
\left[ S^3  \right] \neq {\bf I}$, with  $\Pi_0 \left[ S^3  \right] =
     {\bf  I}$~\cite{TopologyGeometryGaugeFields}.   This  leaves  the
     possibility   for  the   formation   of  domain   walls  in   the
     $\mathrm{Z}_2$  symmetric 2HDM, whose  spatial profile is
     studied in Section~\ref{Topological Defects of the 2HDM}.

\subsection{U$(1)_{\mathrm{PQ}}$ Symmetry}\label{U(1) Symmetry}

We  now analyze  the  Peccei--Quinn  symmetry of  the  2HDM, which  is
defined  by the following  transformations of  the two  Higgs doublets
$\phi_{1,2}$:
\begin{eqnarray}
\phi_1 &\to& \phi'_1\ =\ e^{-i \alpha}\, \phi_1\; , \nonumber \\
\phi_2 &\to& \phi'_2\ =\ e^{i \alpha}\, \phi_2\; , \nonumber
\end{eqnarray}
where $\alpha \in [0,\pi)$. The  study of the neutral vacuum solutions
  of  the  U$(1)_{\mathrm{PQ}}$  invariant  2HDM proceeds  in  a  very
  analogous fashion to the  $\mathrm{Z}_2$ invariant 2HDM discussed in
  the   previous  section,   since  the   only   additional  parameter
  restriction in the U$(1)_{\mathrm{PQ}}$ invariant theory is that one
  now has $\lambda_5 = 0$.  Therefore, we only quote a few key results
here.

For  neutral vacuum  solutions  resulting from  a non-singular  matrix
$\mathrm{N}_{\mu  \nu}$, the VEVs  are given  by \eqref{eq:Z2  Det not
  zero VEV1}  and \eqref{eq:Z2 Det  not zero VEV2}, with  $\lambda_5 =
0$, i.e.
\begin{subequations}
\begin{align}
\langle \phi_{1}^{\dagger}\phi_{1} \rangle\ &=\ \frac{2\lambda_2 \mu_1^2
  - (\lambda_3 - \zeta) \mu_2^2}{4\lambda_1 \lambda_2 - (\lambda_3-
  \zeta)^2} \ >\ 0\; , \\ 
\langle \phi_{2}^{\dagger}\phi_{2} \rangle\ &=\ \frac{2\lambda_1 \mu_2^2
  - (\lambda_3 - \zeta) \mu_1^2}{4\lambda_1 \lambda_2 - (\lambda_3-
  \zeta)^2} \ >\ 0\; , \\ 
\langle \phi_{1}^{\dagger}\phi_{2} \rangle\ &=\ \langle
\phi_{2}^{\dagger}\phi_{1} \rangle\ =\ 0\; .  
\end{align}
\end{subequations}
There are  two Lagrange multiplier solutions for  this situation which
are given by  \eqref{eq:Z2 Zeta 1} and \eqref{eq:Z2  Zeta 2}.  Because
of this  close similarity, the vacuum manifold  parameters are exactly
the  same  as those  detailed  in  Table  \ref{table:Z2 vacuum  manifold
  parameters det not 0} of Section \ref{Z2 Symmetry}. Correspondingly,
the conditions  for each solution  to correspond to a  minima
are given  in Table \ref{table:Z2 Minimum  Conditions Det Not  0}. As in
the $\mathrm{Z}_2$ case,  the U$(1)_{\mathrm{PQ}}$-invariant 2HDM must
also have  at least  one doublet with  a zero VEV,  when $\mathrm{det}
[\mathrm{N}_{\mu  \nu}]  \neq 0$,  which  only  leads to  topologically
trivial configurations.  We are, therefore, only interested in neutral vacuum
solutions for which the matrix $\mathrm{N}_{\mu \nu}$ is singular.

\subsubsection{Neutral Vacuum Solutions from a Singular Matrix N}
\label{sec:U1LMsingular}

In  order for  the matrix  $\mathrm{N}_{\mu \nu}$  to have  no inverse
in the case of the Peccei-Quinn symmetry,  we require that the expression given
in  \eqref{eq:Z2 Det} with  $\lambda_5 =  0$ be  equal to  zero.  This
requirement leads to two candidate solutions:
\begin{subequations}
\begin{align}
\zeta_1\ &=\ -\lambda_4\; , \label{eq:U1 zeta 1} \\
\zeta_{2,\pm}\ &=\ \pm 2 \sqrt{\lambda_1 \lambda_2} + \lambda_3 \; .
\label{eq:U1 zeta 2}
\end{align}
\end{subequations}
However, for the  same reasons as in the ${\rm Z}_2$  case, we have to
reject  the  second  solution  $\zeta_{2,\pm}$,  as  it  leads  to  an
incompatible matrix equation, unless  there is a particular fine-tuned
relation between the parameters of the 2HDM. Therefore, we only focus
on the first solution $\zeta_1$.

Under this choice for  the Lagrange multiplier both $\mathrm{R}^1$ and
$\mathrm{R}^2$ remain  undetermined, since  the $2\times 2$  matrix in
\eqref{eq:r1 r2  Z2 equation} becomes the null matrix.  The remaining
components of the vector $\mathrm{R}^\mu$ are found using \eqref{eq:r0
  r3 Z2 equation} and have the form:
\begin{subequations}
\begin{align}
\mathrm{R}^0\ &=\ \frac{2\lambda_1 \mu_2^2 + 2 \lambda_1 \mu_1^2 -
  \lambda_{34}(\mu_1^2 + \mu_2^2)}{4\lambda_1 \lambda_2 -
  \lambda_{34}^2}\ , \label{eq:LM1 r0 U1} \\ 
\mathrm{R}^3\ &=\ \frac{2\lambda_2 \mu_1^2 - 2 \lambda_2 \mu_2^2 +
  \lambda_{34}(\mu_1^2 - \mu_2^2)}{4\lambda_1 \lambda_2 -
  \lambda_{34}^2}\ . \label{eq:LM1 r3 U1} 
\end{align}
\end{subequations}
From   these  expressions,  we   obtain  by   means  of~\eqref{eq:r^mu
  definition} the VEVs of the scalar-field bilinears
\begin{subequations}
\begin{align}
\langle\phi_{1}^{\dagger}\phi_{1}\rangle\ &=\ 
\frac{2\lambda_2 \mu_1^2 - \lambda_{34} \mu_2^2}{4\lambda_1 \lambda_2 
- \lambda_{34}^2} \ \geq\ 0\; ,\label{eq:VEV1 U1} \\
\langle\phi_{2}^{\dagger}\phi_{2}\rangle\ &=\ 
\frac{2\lambda_1 \mu_2^2 - \lambda_{34} \mu_1^2}{4\lambda_1 \lambda_2 
- \lambda_{34}^2} \ \geq\ 0\; .\label{eq:VEV2 U1} 
\end{align}
\end{subequations}
For   a  neutral   vacuum  solution   we  require,   as   before, that
$\mathrm{R}^\mu$    satisfies    $\mathrm{R}_\mu   \mathrm{R}^\mu    =
0$, which leads to the relation:
\begin{equation}
(\mathrm{R}^1)^2 + (\mathrm{R}^2)^2\ =\  \frac{4\left(2\lambda_2
      \mu_1^2 - \lambda_{34} \mu_2^2\right) \left(2\lambda_1 \mu_2^2 -
      \lambda_{34} \mu_1^2 \right)}{\left(4\lambda_1 \lambda_2 -
      \lambda_{34}^2 \right)^2}\ . 
\end{equation}
Employing the  parameterization of ${\rm  R}^\mu$ in~\eqref{eq:r^mu in
  terms  of  vacuum manifold  parameters},  we  find  that the  vacuum
manifold parameters for the  Lagrange multiplier $\zeta_1$ with $v_2^+
= 0$ are:
\begin{subequations}
\begin{align}
  \label{v1PQ}
v_1^0\ &=\ \sqrt{\frac{4\lambda_2 \mu_1^2 - 2\lambda_{34}
    \mu_2^2}{4\lambda_1 \lambda_2 - \lambda_{34}^2}}\ , \\ 
  \label{v2PQ}
v_2^0\ &=\ \sqrt{\frac{4\lambda_1 \mu_2^2 - 2\lambda_{34}
    \mu_1^2}{4\lambda_1 \lambda_2 - \lambda_{34}^2}}\ ,\\ 
\xi\ &\in\ [0, 2\pi)\; .
\end{align}
\end{subequations}
Notice  that  the phase  $\xi$  remains  undetermined, signifying  the
presence   of   a  massless   Goldstone   boson,   the  so-called   PQ
axion~\cite{Weinberg:1977ma,Wilczek:1977pj}.

The  conditions for a  global minimum  are identical  to those  of the
${\rm Z}_2$ case with $\lambda_5 = 0$. Thus, we have a global
minimum with $v^0_{1,2}\neq 0$, provided
\begin{equation}
\frac{\lambda_{34}}{2\lambda_2}\ <\ \frac{\mu_1^2}{\mu_2^2}\ <\
\frac{2 \lambda_1}{\lambda_{34}}\ .  
\end{equation}
The value of the U$(1)_{\mathrm{PQ}}$-invariant 2HDM potential at the
global minimum is given by
\begin{equation}
\mathrm{V}_0\ =\ \frac{\lambda_{34} \mu_1^2 \mu_2^2 - \lambda_1
  \mu_2^4 - \lambda_2 \mu_1^4}{4 \lambda_1 \lambda_2 -
  \lambda_{34}^2}\ .
\label{eq:LM1 U1 Pot Min}
\end{equation}
As before,  we find  that neutral vacua  where $v^0_{1,2} \neq  0$ and
$v^0_1 = 0$ or $v^0_2 = 0$ cannot co-exist.

\subsubsection{U$(1)_{\mathrm{PQ}}$ Topology}

Let us now discuss the topology of the vacuum manifold associated with
the U$(1)_{\mathrm{PQ}}$-invariant 2HDM potential.  The total symmetry
group    of    the   potential    in    the    symmetric   phase    is
$\mathrm{G}_{\mathrm{U}(1)_{\mathrm{PQ}}}                             =
\mathrm{U}(1)_{\mathrm{PQ}}     \otimes    \mathrm{SU}(2)_{\mathrm{L}}
\otimes   \mathrm{U}(1)_{\mathrm{Y}}$.   After   electroweak  symmetry
breaking  [cf.~\eqref{eq:Electroweak Symmetry  Breaking  Scheme}], the
U$(1)_{\mathrm{PQ}}$ symmetry  breaks  into  the  identity ${\bf  I}$,  so  the
unbroken  group  is  $\mathrm{H}_{\mathrm{U}(1)_{\mathrm{PQ}}} =  {\bf
  I}\otimes \mathrm{U}(1)_{\mathrm{em}}$. As a consequence, the vacuum
manifold in the $\Phi$-space is given by the set
\begin{equation}
\mathcal{M}^{\mathrm{U(1)}_{\mathrm{PQ}}}_{\Phi}\ =\
\mathrm{G}_{\mathrm{U}(1)_{\mathrm{PQ}}}/\mathrm{H}_{\mathrm{U}(1)_{\mathrm{PQ}}}\ \simeq\
S^1 \times S^3\; ,
\end{equation}
where we  used the fact  that U$(1)_{\mathrm{PQ}}$ is  homeomorphic to
$S^1$.  We  now observe that the  first homotopy group  of this vacuum
manifold is  non-trivial, i.e.~$\Pi_1 \left[ S^1 \times  S^3 \right] =
\Pi_1\left[S^1  \right] \times \Pi_1\left[S^3\right]  = \Pi_1\left[S^1
  \right]  = \mathbb{Z}  \neq {\bf  I}$,  since $\Pi_1\left[S^3\right]
={\bf I}$.  This  implies that the U$(1)_{\mathrm{PQ}}$-invariant 2HDM
has   a   string   or   vortex   solution,   which   we  analyze  in
Section~\ref{Topological Defects of the 2HDM}.

It is interesting  to discuss the construction of  the vacuum manifold
in the $\Phi$-space. As stated in~\eqref{eq:calM+}, a general point of
the         vacuum         manifold         is        given         by
$\Phi\ =\ \mathcal{M}^{\mathrm{U(1)}_{\mathrm{PQ}}}_+\, \Phi_0$, where
$\Phi_0$ is  defined in terms  of the non-zero VEVs  $v^0_{1,2}$ given
in~\eqref{v1PQ} and \eqref{v2PQ} and  by setting $\xi = 0$.  Moreover,
in  the 8-dimensional  Majorana  $\Phi$-space,  the HF  transformation
matrix  $\mathcal{M}^{\mathrm{U(1)}_{\mathrm{PQ}}}_+$   takes  on  the
form:
\begin{eqnarray}
\mathcal{M}^{\mathrm{U(1)}_{\mathrm{PQ}}}_+ & = & \left(\frac{\sigma^0
  + \sigma^3}{2} \right) \otimes \mathcal{T}_+ \otimes \mathrm{U}_+
(\tilde{\theta}^1,\tilde{\theta}^2,\theta_-) \: +\: \left(\frac{\sigma^0 -
  \sigma^3}{2} \right) \otimes \mathcal{T}^{*}_+ \otimes
\mathrm{U}_-(\tilde{\theta}^1,\tilde{\theta}^2,\theta_-)\nonumber\\[3mm] 
& = & \left(\frac{\sigma^0 + \sigma^3}{2} \right)
\otimes \exp \left[\, 2i\alpha \left( \frac{\sigma^0 -
    \sigma^3}{2}\right) \right] \otimes \mathrm{U}_+
(\tilde{\theta}^1,\tilde{\theta}^2,\theta_- - \alpha)\\ 
&&+\ \left(\frac{\sigma^0 - \sigma^3}{2} \right) \otimes
\exp \left[\, -2i\alpha \left( \frac{\sigma^0 - \sigma^3}{2}\right)
  \right] \otimes
\mathrm{U}_-(\tilde{\theta}^1,\tilde{\theta}^2,\theta_- 
- \alpha)\; .\nonumber
\end{eqnarray}
Here, we  have explicitly displayed the dependence  of the gauge-group
factors ${\rm  U}_\pm$ on their group  parameters and made  use of the
fact   that   the    HF   transformation   matrix   $\mathcal{T}_+   =
e^{-i\alpha\sigma_3}$  for   the  PQ   symmetry  may  be   written  as
$\mathcal{T}_+ =  e^{-i\alpha} e^{i\alpha (\sigma^0  - \sigma^3)}$. 
We  may now re-define the  group parameter
$\theta_-$ as $\tilde{\theta}_- = \theta_- - \alpha \in [0,2\pi)$, and
  so having the group parameter $2\alpha \in [0,2\pi)$ to span now the
    complete  space of  the  U(1)$_{\rm PQ}$  group.   Note that  this
    result is identical to the one that would be obtained in the ${\rm
      R}^\mu$    space,    as    can    be   easily    deduced    from
    Table~\ref{table:Lambda Matrix Table}.

\subsection{SO$(3)_{\mathrm{HF}}$ Symmetry}
\label{SU2 Symmetry}

An interesting  HF symmetry  emerges from the  invariance of  the 2HDM
potential under a naive ${\rm SU(2)_{HF}}$ transformation of the Higgs
fields, i.e.
\begin{eqnarray*}
\phi_1 &\to& \phi'_1\ =\ e^{-i \alpha} \cos \theta\, \phi_1\: +\: e^{-i
  \beta} \sin \theta\, \phi_2\; , \\ 
\phi_2 &\to& \phi'_2\ =\ -e^{i \beta} \sin \theta\, \phi_1\: +\: e^{i
  \alpha} \cos \theta\, \phi_2 \; .
\end{eqnarray*}
To avoid a double cover of the ${\rm SU(2)_{HF}}$ group because of the
presence of ${\rm U(1)_Y}$  hypercharge rotations, we have to restrict
the  group  parameters  $\theta,\,  \alpha,\,  \beta$ to  lie  in  the
interval $[0,\pi)$.  Hence, the actual  HF symmetry is the coset group
  ${\rm   SU(2)_{HF}}/{\rm   Z_2}$~\cite{Ferreira:2009wh},  which   is
  isomorphic to ${\rm SO(3)_{HF}}$ in the field-bilinear ${\rm R}^\mu$
  space.   For  this  reason,  this  symmetry  was  called  the  ${\rm
    SO(3)_{HF}}$ symmetry.

The  parameters of  the 2HDM  potential under  the  ${\rm SO(3)_{HF}}$
symmetry  are  restricted,  as  shown in  Table~\ref{table:HF  Parameter
  Conditions}. In  fact, most of  the results can be  easily recovered
from the ${\rm Z}_2$ case  in Section~\ref{Z2 Symmetry}, by making the
replacements: $\lambda_2  \to \lambda_1$, $\lambda_4  \to 2\lambda_1 -
\lambda_3$,   $\mu^2_2   \to  \mu^2_1$   and   putting  $\lambda_5   =
0$. Therefore,  we will only  report key intermediate results  in this
section.    As  before,   we  first   assume  that   the   inverse  of
$\mathrm{N}_{\mu    \nu}$   exists,    with    the   determinant    of
$\mathrm{N}_{\mu \nu}$ given by
\begin{equation}
\mathrm{det} [\mathrm{N}_{\mu \nu}]\ =\ \left( 2\lambda_1 + \lambda_3 -
  \zeta \right)\, \left( 2\lambda_1 - \lambda_3 + \zeta \right)^3\; . 
\label{eq:SU2 Det}
\end{equation}
Because of  the more  restrictive nature of  the SO$(3)_{\mathrm{HF}}$
symmetry, the  matrix equation $\mathrm{N}_{\mu  \nu} \mathrm{R}^\nu =
\mathrm{M}_\mu$ splits into four separate equations:
\begin{subequations}
\begin{align}
  \label{eq:r0 O3 equation} 
(2\lambda_1 + \lambda_3 - \zeta)\, \mathrm{R}^0\ &=\ 2\mu_1^2\; ,\\
  \label{eq:r1 O3 equation} 
(2\lambda_1 - \lambda_3 + \zeta)\, \mathrm{R}^1\ &=\ 0\; ,\\
  \label{eq:r2 O3 equation} 
(2\lambda_1 - \lambda_3 + \zeta)\, \mathrm{R}^2\ &=\ 0\; ,\\
  \label{eq:r3 O3 equation} 
(2\lambda_1 - \lambda_3 + \zeta)\, \mathrm{R}^3\ &=\ 0\; . 
\end{align}
\end{subequations}
On the basis of the above assumption that $\mathrm{N}_{\mu \nu}$ is
invertible, the components of the 4-vector $\mathrm{R}^\mu$ are 
easily found to be
\begin{subequations}
\begin{align}
&\mathrm{R}^0\ =\ \frac{2\mu_1^2}{2\lambda_1 + \lambda_3 - \zeta}\ , \\
&\mathrm{R}^1\ =\ \mathrm{R}^2\ =\ \mathrm{R}^3\ =\ 0\; .
\end{align}
\end{subequations}
On  the other  hand,  the  constraint for  a  neutral vacuum  solution
requires that ${\rm R}^\mu$ is  a null vector, satisfying ${\rm R}_\mu
{\rm  R}^\mu  =  0$.   Since  all  the  ``spatial''  components  ${\rm
  R}^{1,2,3}$ vanish, so should  the ``time'' component, i.e.~${\rm
  R}^0 = 0$. This last result  tells us that the Higgs doublets should
have  vanishing VEVs,  i.e.~$v^0_{1,2} =  0$, leaving  the electroweak
gauge  group unbroken.   This is  an unrealistic  scenario and  can only be
obtained in the limit $\mu^2_1 \to  0$, or $\zeta \to \pm \infty$.  We
will therefore  investigate neutral vacuum solutions that result
from a singular matrix~${\rm N}_{\mu\nu}$.

\subsubsection{Neutral Vacuum Solutions from a Singular Matrix N}
\label{sec:SO3HFLMsingular}

From~\eqref{eq:SU2 Det}, we readily  see that the following choices of
the Lagrange multiplier render $\mathrm{N}_{\mu \nu}$ singular:
\begin{subequations}
\begin{align}
\zeta_1\ &=\ -2\lambda_1 + \lambda_3\; , \\
\zeta_2\ &=\ 2\lambda_1 + \lambda_3\; .
\end{align}
\end{subequations}
However, from~\eqref{eq:r0  O3 equation}, we notice  that the solution
$\zeta_2$ implies  either $\mathrm{R}^0  \to \infty$, or  $\mu^2_1 \to
0$,  both  of  which  lead  to unrealistic  scenarios  of  electroweak
symmetry   breaking.  Therefore,  we   concentrate  on   the  Lagrange
multiplier solution $\zeta_1$.

Considering the Lagrange multiplier solution $\zeta_1$, we obtain
from~\eqref{eq:r0  O3 equation} that 
\begin{equation}
  \label{SO3:R0}
\mathrm{R}^0\ =\ \frac{\mu_1^2}{2 \lambda_1}\ . 
\end{equation}
Instead, from \eqref{eq:r1 O3 equation}--\eqref{eq:r3 O3 equation}, we
see  that  all the  ``spatial  components''  ${\rm R}^{1,2,3}$  remain
undetermined.  The only  constraint that can be placed  upon the three
``spatial''  components of  $\mathrm{R}^\mu$ is  the requirement  of a
neutral  vacuum solution, $\mathrm{R}_\mu  \mathrm{R}^\mu =  0$, which
implies that
\begin{equation}
  \label{SO3:R123}
(\mathrm{R}^1)^2 + (\mathrm{R}^2)^2 + (\mathrm{R}^3)^2\ =\
  \frac{\mu_1^4}{4 \lambda_1^2} \ .
\end{equation}
In  terms of  the vacuum  manifold parameters  $v^0_{1,2}$  and $\xi$,
\eqref{SO3:R0} and \eqref{SO3:R123} are translated into
\begin{equation}
  \label{VEVs:SO3}
v_1^0\ =\ \frac{\mu_1}{\sqrt{\lambda_1}}\ \sin\theta\; ,\qquad  
v_2^0\ =\ \frac{\mu_1}{\sqrt{\lambda_1}}\ \cos\theta\; ,
\end{equation}
where  $\xi  \in  [0, 2\pi  )$  and  $\theta  \in  [0, \pi  )$  remain
    undetermined.  The latter signifies  the presence of two Goldstone
    bosons. Specifically, the one associated with the phase $\xi$ is a
    CP-odd scalar, whereas the one related to the polar angle $\theta$
    is  a   `CP-even'  boson.    This  result  can   be  cross-checked
    independently from  the explicit analytical  expressions presented
    in~\cite{Pilaftsis:1999qt} for the  general Higgs-boson mass matrices.
    The  global minimum  of  the SO$(3)_{\mathrm{HF}}$-symmetric  2HDM
    potential is given by
\begin{equation}
\mathrm{V}_0\ =\ -\; \frac{\mu_1^4}{4 \lambda_1}\ . 
\end{equation}
Such a  global minimum is always  guaranteed, as long  as $\mu^2_1$ is
positive   and  the   bounded-from-below   condition,  $2\lambda_1   >
|\lambda_3|$  given  in  Table~\ref{table:HF Convexity  Conditions}  is
satisfied.

\subsubsection{SO$(3)_{\mathrm{HF}}$ Topology}

It  is interesting  to analyze  the  topology of  the vacuum  manifold
arising    from   the    spontaneous   symmetry    breaking    of   an
SO$(3)_{\mathrm{HF}}$-invariant  2HDM  potential.   In  the  symmetric
phase   of  the   theory,  the   SO$(3)_{\mathrm{HF}}$-invariant  2HDM
potential   has   the   symmetry,    which   is   described   by   the
group~\cite{Ferreira:2009wh}
\begin{equation}
  \label{GSO3}
\mathrm{G}_{\mathrm{SO(3)}_{\mathrm{HF}}}\ =\ (\mathrm{SU(2)}_{\mathrm{HF}}/{\rm
  Z}_2) \otimes \mathrm{SU}(2)_{\mathrm{L}} \otimes
\mathrm{U}(1)_{\mathrm{Y}}\ \cong\ \mathrm{SO(3)}_{\mathrm{HF}}
\otimes \mathrm{SU}(2)_{\mathrm{L}} \otimes \mathrm{U}(1)_{\mathrm{Y}}\; .
\end{equation}
Using the  results of  the previous  section, we see  that out  of the
three  generators of  the SU$(2)_{\mathrm{HF}}/{\rm  Z}_2$  group, one
linear combination of  generators, $(\sigma^0 +\sigma^3)/2$ related to
a residual HF  symmetry, which we call ${\rm  U(1)}_{\rm HF}$, remains
unbroken  after the  electroweak symmetry  breaking, resulting  in the
little group
\begin{equation}
  \label{HSO3}
\mathrm{H}_{\mathrm{SO(3)}_{\mathrm{HF}}}\ =\ \mathrm{U(1)}_{\mathrm{HF}}
\otimes \mathrm{SU}(2)_{\mathrm{L}} \otimes \mathrm{U}(1)_{\mathrm{Y}}
\ \cong\ \mathrm{SO(2)}_{\mathrm{HF}}
\otimes \mathrm{SU}(2)_{\mathrm{L}} \otimes \mathrm{U}(1)_{\mathrm{Y}}\; .
\end{equation}
Then,                the                vacuum                manifold
$\mathcal{M}^{\mathrm{SO(3)}_{\mathrm{HF}}}_{\Phi}$  may  be described
by the product of spaces:
\begin{equation}
\mathcal{M}^{\mathrm{SO(3)}_{\mathrm{HF}}}_{\Phi}\ =\  
\mathrm{G}_{\mathrm{SO(3)}_{\mathrm{HF}}} 
/\mathrm{H}_{\mathrm{SO(3)}_{\mathrm{HF}}}\ \simeq\ S^2 \times S^3 \; , 
\end{equation}
where the first factor $S^2$ is obtained using the known homeomorphism
${\rm SO(3)_{HF}  /SO(2)_{HF}}$ $\simeq S^2$ and the  second one $S^3$
is due  to the breaking of  the electroweak group  to ${\rm U(1)}_{\rm
  em}$.    We   observe   that    the   second   homotopy   group   of
$\mathcal{M}^{\mathrm{SO(3)}_{\mathrm{HF}}}_{\Phi}$                  is
non-trivial. More  explicitly, $\Pi_2 \left[ S^2 \times  S^3 \right] =
\Pi_2  \left[ S^2\right]  \times  \Pi_2 \left[  S^3  \right] =  \Pi_2
\left[ S^2\right]  \neq {\bf  I}$, since $\Pi_2  \left[ S^3  \right] =
     {\bf  I}$.  Consequently,  spontaneous symmetry  breaking  of the
     $\mathrm{SO(3)}_{\mathrm{HF}}$-symmetric~2HDM  can  give rise  to
     global monopoles.

As  with the  previous HF  symmetries, we  are able  to  construct the
entire vacuum  manifold by  the transitive action  of the  total group
$\mathrm{G}_{\mathrm{SO(3)}_{\mathrm{HF}}}$  stated in~\eqref{GSO3} on
the vacuum  point $\Phi_0$, which  remains invariant under  the little
group $\mathrm{H}_{\mathrm{SO(3)}_{\mathrm{HF}}}$ given in~\eqref{HSO3}.
An appropriate  representation of $\Phi_0$ consistent  with the latter
property is given by the VEVs
\begin{equation}
{\bf V}_1\ =\ \left( \begin {array}{c} 0\\\noalign{\medskip} 0 \end 
{array}\right) \;, \qquad 
{\bf V}_2\ =\ \frac{1}{\sqrt{2}}
\left( \begin {array}{c} 0\\\noalign{\medskip}
  \frac{\mu_1}{\sqrt{\lambda_1}} \end {array} \right)\; , 
\end{equation}
where  we set  $\theta =  \xi =  0$ in~\eqref{VEVs:SO3}.   The general
point $\Phi$ on the vacuum manifold is then given by the action of the
coset       set        of       HF       transformation       matrices
$\mathcal{M}^{\mathrm{SO(3)}_{\mathrm{HF}}}_+$   on   $\Phi_0$,   i.e.
$\Phi = \mathcal{M}^{\mathrm{SO(3)}_{\mathrm{HF}}}_+\, \Phi_0$, where
\begin{equation}
\mathcal{M}^{\mathrm{SO(3)}_{\mathrm{HF}}}_+\ =\ \left(\frac{\sigma^0
  + \sigma^3}{2} \right) \otimes \mathcal{T}_+ \otimes \mathrm{U}_+
(\tilde{\theta}^1,\tilde{\theta}^2,\theta_-) \: +\: \left(\frac{\sigma^0 -
  \sigma^3}{2} \right) \otimes \mathcal{T}^{*}_+ \otimes
\mathrm{U}_-(\tilde{\theta}^1,\tilde{\theta}^2,\theta_-)\; .
\end{equation}
Here, the  $2\times 2$ HF transformation matrices  ${\cal T}_+$ belong
to the  coset space  of the $\rm  SO(3)_{HF}$ symmetry in  the adjoint
representation,  i.e.  ${\cal  T}_+  \in  ({\rm  SU(2)}_{\rm  HF}/{\rm
  Z_2})/{\rm U}(1)_{\rm HF}$, and can be represented as
\begin{equation}
{\cal T}_+ \ =\ \left( \begin {array}{cc} e^{-i\alpha} \cos \theta & e^{-i
 \beta}\sin\theta \\\noalign{\medskip} - e^{i\beta}\sin\theta&
e^{i\alpha}\cos \theta \end {array} \right)\,
\left( \begin{array}{cc} e^{i\chi} & 0 \\
\noalign{\medskip} 0 & 1 \end {array} \right)\ =\ e^{-i\beta}\, 
\left( \begin {array}{cc} \cos \theta & \sin\theta \\\noalign{\medskip} 
- e^{i(\alpha + \beta)}\sin\theta & e^{i(\alpha + \beta)}
\cos\theta \end {array} \right) \; , 
\end{equation}
where we set the free ${\rm U}(1)_{\rm HF}$ phase $\chi$ to be $\chi =
\alpha  - \beta$,  in obtaining  the second  equation.  As  in  the PQ
symmetry case,  the overall factor $e^{-i\beta}$ can  be absorbed into
the  definition  of  the  gauge-group  parameter  $\theta_-$,  i.e.~by
defining $\tilde{\theta}_-  = \theta_- - \beta \in  [0,2\pi)$.  The HF
  transformation                                               matrices
  $\mathcal{M}^{\mathrm{SO(3)}_{\mathrm{HF}}}_+$  can then  be written
  down as
\begin{eqnarray}
\mathcal{M}^{\mathrm{SO(3)}_{\mathrm{HF}}}_+\! &=&\! \left(\frac{\sigma^0
  + \sigma^3}{2} \right) \otimes \left( \begin {array}{cc} \cos \theta
  & \sin\theta \\\noalign{\medskip}  
- e^{i(\alpha + \beta)}\sin\theta & e^{i(\alpha +\beta)}
\cos\theta \end {array} \right) \otimes \mathrm{U}_+
(\tilde{\theta}^1,\tilde{\theta}^2,\tilde{\theta}_-) \nonumber\\ 
\!&+&\!\! \left(\frac{\sigma^0 -
  \sigma^3}{2} \right) \otimes \left( \begin {array}{cc} \cos \theta
  & \sin\theta \\\noalign{\medskip}  
- e^{-i(\alpha + \beta)}\sin\theta & e^{-i(\alpha+\beta)}
\cos\theta \end {array} \right) \otimes
\mathrm{U}_-(\tilde{\theta}^1,\tilde{\theta}^2,\tilde{\theta}_-)\; .\qquad
\end{eqnarray}
If  we ignore  the $S^3$  gauge rotations  by setting  ${\rm  U}_\pm =
\sigma^0$,                 the                action                of
$\mathcal{M}^{\mathrm{SO(3)}_{\mathrm{HF}}}_+$    on   $\Phi_0$   then
generates the general vacuum manifold point given in~\eqref{VEVs:SO3},
with  $\xi   =  \alpha   +  \beta  \in   [0,2\pi)$  and   $\theta  \in
  [0,\pi)$. Thus,  the vacuum manifold of  the SO(3)$_{\rm HF}$-broken
    2HDM  is homeomorphic  to  $S^2$, parameterized  by the  azimuthal
    angle $\theta$ and  the polar angle $\xi =  \alpha + \beta$.  This
    parameterization will be  used in Section~\ref{Topological Defects
      of the 2HDM} to analyze the monopole solution in this model.

\bigskip

\section{Neutral Vacuum Solutions of the CP Symmetries}\label{CP Symmetries}

In this section, we will study the three generic CP symmetries, termed
CP1,  CP2   and~CP3.   These  three  CP   symmetries  impose  specific
relations~\cite{Ferreira:2009wh}  among  the  parameters of  the  2HDM
potential,   which  are   presented  in   Table~\ref{table:CP  Parameter
  Conditions}.

\begin{table}[t!]
\begin{center}
\begin{tabular}{c|cccccccccc}
\hline
Symmetry & 	$\mu_1^2$	&	$\mu_2^2$	&	$m_{12}^2$	&	$\lambda_1$	&	$\lambda_2$	&	$\lambda_3$	&	$\lambda_4$	&	$\lambda_5$	&	$\lambda_6$	&	$\lambda_7$ \\
\hline
\hline

CP1	&	--	&	--	&	Real	&	--	&	--	&	--	&	--	&	Real	&	Real	&	Real	 \\

\hline

CP2	&	--	&	$\mu_1^2$	&	0	&	--	&	$\lambda_1$	&	--	&	--	&	--	&	--	&	$-\lambda_6$ \\

\hline

CP3	&	--	&	$\mu_1^2$	&	0	&	--	&	$\lambda_1$	&	--	&	--	&	$2\lambda_1 - \lambda_3 - \lambda_4$	&	0	&	0\\

\hline
\end{tabular}
\caption{\it  Parameter relations  in the  2HDM potential  that result
  from  the imposition  of the  three  generic CP  symmetries. A  dash
  indicates  the   absence  of  a   constraint.\label{table:CP  Parameter
    Conditions}}
\end{center}
\end{table}

Implementing the constraints on the potential parameters due to the CP
symmetries,  the  four  general  convexity  conditions  \eqref{eq:2HDM
  Convexity 1}--\eqref{eq:2HDM Convexity 3} and \eqref{detL} take on a
simpler    form.    These   four    conditions   are    displayed   in
Table~\ref{table:CP  Convexity Conditions}. In  particular, for  the CP3
case, the  four convexity conditions  are not all independent  of each
other, so only the two distinct conditions are presented.

\begin{table}[t!]
\begin{center}
\begin{tabular}{c||c|c|c}
\hline
Convexity	& 		&		&		\\
Condition	&	CP1	&	CP2	&	CP3	\\
\hline
\hline

1	&	$\lambda_1 + \lambda_2 + \lambda_3 > 0$	&	$2\lambda_1 > - \lambda_3$	&	$2\lambda_1 > |\lambda_3|$	\\

\hline

2	&	$\lambda_4 > - \lambda_5 + \frac{(\lambda_6 + \lambda_7)^2}{\lambda_1 + \lambda_2 + \lambda_3}$	&	$\lambda_4 > - R_5$	&	$2\lambda_4 > 2\lambda_1 - \lambda_3$	\\

\hline

3	&	$\lambda_4 > \lambda_5$	&	$\lambda_4^2 > |\lambda_5|^2$	&	--	\\

\hline

4	&	$\lambda_1 \lambda_2 - \frac{1}{4}\lambda_3^2 >$	&	$2\lambda_1 - \lambda_3 >$	&	--	\\
& $\frac{\lambda_1 \lambda_6^2 + \lambda_2 \lambda_7^2 - \lambda_3 \lambda_6 \lambda_7}{\lambda_4 + \lambda_5}$ 	&	$\frac{4|\lambda_6|^2 (\lambda_4 - R_5) - 8 I_6 (I_5 R_6 - R_5 I_6)}{\lambda_4^2 - |\lambda_5|^2}$	& \\

\hline
\end{tabular}
\caption{\it The four convexity conditions for a bounded-from-below 2HDM
  potential for each of the three CP symmetries. A dash signifies 
the absence of any additional constraint. \label{table:CP Convexity Conditions} }
\end{center}
\end{table}

As  in the  previous section,  our aim  is to  derive analytical
expressions for the neutral VEVs  of $\phi_{1,2}$ in terms of the 2HDM
potential parameters  for each of  the three CP symmetries,  by making
use of the Lagrange multiplier  method.  These results will be used to
determine the existence and the nature of possible topological defects
which will be discussed  in detail in Section~\ref{Topological Defects
  of the 2HDM}.

\subsection{CP1 Symmetry} \label{CP1 Symmetry}

The  discrete CP1  symmetry of  the  2HDM represents  the standard  CP
transformation of the two Higgs doublets $\phi_{1,2}$, given by
\begin{eqnarray}
\phi_1 &\to& \phi'_1\ =\ \phi_1^* \; , \nonumber \\
\phi_2 &\to& \phi'_2\ =\ \phi_2^* \; . \nonumber
\end{eqnarray}
Taking  into   account  the   CP1  parameter  restrictions   of  Table
\ref{table:CP   Parameter  Conditions},   we  calculate   the   VEVs  of
$\phi_{1,2}$     by    imposing     the     extremization    condition
\eqref{eq:Extremization        Condition       1}        and       the
condition~\eqref{eq:Neutral  Vacuum Condition  2} for  an electrically
neutral vacuum.   As before, we consider two  cases: (i) $\mathrm{det}
[\mathrm{N}_{\mu \nu}] \neq 0$ and (ii)~$\mathrm{det} [\mathrm{N}_{\mu
    \nu}] = 0$.

The   determinant   of  $\mathrm{N}_{\mu   \nu}$   resulting  from   a
CP1-invariant   2HDM   potential   follows  from   Appendix   \ref{The
  Determinant of N} and can be expressed in the factorized form:
\begin{equation}
\mathrm{det} [\mathrm{N}_{\mu \nu}]\ =\ (\bar{\lambda}_{45} + \zeta )
\left[\left(\lambda_{45} + \zeta \right) \left(4\lambda_1 \lambda_2 -
  (\lambda_3-\zeta)^2 \right) -4\lambda_1 \lambda_6^2 - 4\lambda_2
  \lambda_7^2 + 4 \lambda_6 \lambda_7 (\lambda_3 - \zeta) \right]\; . 
\label{eq:CP1 det} 
\end{equation}
Moreover,   the    extremization   condition   $\mathrm{N}_{\mu   \nu}
\mathrm{R}^\nu = \mathrm{M}_\mu$ decomposes into two equations:
\begin{subequations}
\begin{align}
\left( \begin {array}{ccc} \lambda_1 \!+\! \lambda_2 \!+\! \lambda_3
  \!-\! \zeta&\lambda_6 \!+\! \lambda_7&\lambda_1 \!-\!
  \lambda_2\\\noalign{\medskip}\lambda_6 \!+\! \lambda_7 &\lambda_4
  \!+\! \lambda_5 \!+\! \zeta&\lambda_6 \!-\!
  \lambda_7\\\noalign{\medskip} \lambda_1 \!-\! \lambda_2&\lambda_6
  \!-\! \lambda_7&\lambda_1 \!+\! \lambda_2 \!-\! \lambda_3 \!+\!
  \zeta\end {array} \right) \!\! \left( \begin {array}{c}
    \mathrm{R}^0\\\noalign{\medskip}\mathrm{R}^1\\
\noalign{\medskip}\mathrm{R}^3\end 
           {array} \right)\ &=\ \left( \begin {array}{c} \mu_1^2 +
             \mu_2^2\\\noalign{\medskip} 2m_{12}^2\\\noalign{\medskip}
             \mu_1^2 - \mu_2^2\end {array} \right) \; , \label{eq:r0
               r1 r3 CP1 equation} \\[2mm] 
(\lambda_4 - \lambda_5 + \zeta)\, \mathrm{R}^2\ &=\ 0\; . 
\label{eq:r2 CP1 equation}
\end{align}
\end{subequations}
Assuming  that the  matrix  $\mathrm{N}_{\mu \nu}$  is invertible,  we
observe  that   $\mathrm{R}^2  =  0$,  which   implies  that  $\langle
\phi_1^{\dagger}  \phi_2  \rangle  = \langle  \phi_2^{\dagger}  \phi_1
\rangle$.   This latter  condition can  be satisfied  in two  ways, if
$v^0_2 \neq  0$. The  first possibility  is to have  $\xi =  0$, which
amounts to  the non-breaking of the  CP1 symmetry by  the vacuum.  The
second  possibility is  to  have $v^0_1  =0$,  with $\xi  \neq 0$  and
possibly $v^+_2 \neq 0$. However, $\xi$ and $v^+_2$ can be set to zero
by an SU$(2)_{\rm  L}$ gauge rotation, giving rise  to a CP1-invariant
vacuum.   Hence,   the  neutral  vacuum  solutions   arising  from  an
invertible matrix $\mathrm{N}_{\mu \nu}$ do not break the discrete CP1
symmetry and  so do  not lead to  topological defects, such  as domain
walls. We therefore turn our attention to situations where
the  determinant of  the  matrix $\mathrm{N}_{\mu  \nu}$ is  singular,
thanks to specific choices of the Lagrange multiplier $\zeta$.

\subsubsection{Neutral Vacuum Solutions from a Singular Matrix N}
\label{Utilizing CP1 zeta}

In  order for  the matrix  $\mathrm{N}_{\mu \nu}$  to have  no inverse
under the CP1 symmetry, we require that the determinant of $\rm N_{\mu
  \nu}$  vanishes. This is  guaranteed by  setting the  the expression
in~\eqref{eq:CP1 det} to zero. We find four possible solutions for the
Lagrange  multiplier,  three  attributed  to~\eqref{eq:r0  r1  r3  CP1
  equation}   and  one   attributed  to~\eqref{eq:r2   CP1  equation}.
However, as we  have previously seen for the  other symmetries studied
so  far, since  the  RHS of~\eqref{eq:r0  r1  r3 CP1  equation} is  in
general a non-zero vector in this case, unless $\mu^2_1 = \mu^2_2 = 0$
and   $\mathrm{Re}(m_{12}^2)   =   0$,   this   matrix   equation   is
overdetermined.      Unless      the     parameters     $\mu^2_{1,2}$,
$\mathrm{Re}(m_{12}^2)$      and       the      quartic      couplings
$\lambda_{1,2,\dots,7}$ satisfy an unnatural fine-tuning relation, the
matrix equation~\eqref{eq:r0 r1  r3 CP1 equation} becomes incompatible
for  the Lagrange  multipliers  that result  from  requiring that  the
matrix of~\eqref{eq:r0 r1 r3  CP1 equation} is singular.  We therefore
reject  these three  possible Lagrange  multipliers and  focus  on the
single Lagrange multiplier solution to~\eqref{eq:r2 CP1 equation}:
\begin{equation}
  \label{eq:zetaCP1}
\zeta\ =\ - \bar{\lambda}_{45}\; .
\end{equation}
This choice of $\zeta$ lifts  the constraint $\mathrm{R}^2 = 0$, which
resulted  from  a   non-singular  matrix~$\mathrm{N}_{\mu  \nu}$.   As
consequence, the CP-odd  phase $\xi$ can be non-zero  in general, thus
triggering  spontaneous  breakdown  of  the  CP1  symmetry  after  the
electroweak symmetry breaking.   This phenomenon is called spontaneous
CP violation in the literature~\cite{Lee:1973iz,Branco:1980sz}.

Substituting  the   value  of  the  Lagrange   multiplier  $\zeta$  in
\eqref{eq:zetaCP1}  into the  matrix equation~\eqref{eq:r0  r1  r3 CP1
  equation},  we  can calculate  the  individual  components  of  the
4-vector $\mathrm{R}^\mu$. These are given by
\begin{subequations}
\begin{align}
\mathrm{R}^0 \ &=\  \frac{1}{\rm A}\, 
\Big\{ \left[\lambda_5(2\lambda_2 -
    \bar{\lambda}_{345}) + \bar{\lambda}_{67}\lambda_7 \right]
  \mu_1^2 + \left[\lambda_5 (2\lambda_1 -
    \bar{\lambda}_{345}) - \lambda_6\bar{\lambda}_{67} \right]
  \mu_2^2 \nonumber\\
\ & \hspace{1.5cm} +\, \left[\bar{\lambda}_{12}\bar{\lambda}_{67} -
    (\lambda_{12} \!-\! \bar{\lambda}_{345})\lambda_{67} \right] 
  m_{12}^2\Big\} \; , \label{eq:R0 Zeta1 CP1} \\ 
\mathrm{R}^1 \ &=\ \frac{1}{\rm A}\, 
\Big\{ \left(\bar{\lambda}_{345}\lambda_7 -
    2\lambda_2\lambda_6 \right) \mu_1^2 +
  \left(\bar{\lambda}_{345}\lambda_6 - 2\lambda_1\lambda_7 \right)
  \mu_2^2 + \left(4\lambda_1\lambda_2 - \bar{\lambda}_{345}^2 \right)
  m_{12}^2\,\Big\} \; , \label{eq:R1 Zeta1 CP1} \\[3mm] 
\mathrm{R}^3 \ &=\ \frac{1}{\rm A}\, 
\Big\{ \left[\lambda_5(2\lambda_2 +
    \bar{\lambda}_{345}) - \lambda_{67} \lambda_7 \right]
  \mu_1^2 + \left[ \lambda_6 \lambda_{67} - \lambda_5(2\lambda_1 + 
\bar{\lambda}_{345})\right] \mu_2^2\nonumber\\
 \ & \hspace{1.5cm} +\, \left[\bar{\lambda}_{12} \lambda_{67} - (\lambda_{12}
    + \bar{\lambda}_{345})\bar{\lambda}_{67}) \right]
  m_{12}^2\, \Big\} \, , \label{eq:R3 Zeta1 CP1}
\end{align}
\end{subequations}
with 
\begin{equation}
\mathrm{A} \ =\ 
\lambda_5 \left(4\lambda_1 \lambda_2 - \bar{\lambda}_{345}^2 \right) -
2\lambda_1 \lambda_7^2 - 2\lambda_2 \lambda_6^2 + 2
\bar{\lambda}_{345} \lambda_6 \lambda_7 \; . \label{eq:A Parameter for
  Zeta1 CP1} 
\end{equation}
From \eqref{eq:R0 Zeta1  CP1} and \eqref{eq:R3 Zeta1 CP1},  we can now
calculate, by  means of~\eqref{eq:r^mu  definition}, the VEVs  for the
bilinear field expressions:
\begin{subequations}
\begin{align}
\langle\phi_1^{\dagger} \phi_1\rangle\ &=\ \frac{\left(2\lambda_2 \lambda_5 - \lambda_7^2\right) \mu_1^2 + \left( \lambda_6\lambda_7 -\bar{\lambda}_{345} \lambda_5\right) \mu_2^2 + \left(\bar{\lambda}_{345} \lambda_7 - 2\lambda_2 \lambda_6 \right) m_{12}^2}{\lambda_5 \left(4\lambda_1 \lambda_2 - \bar{\lambda}_{345}^2 \right) - 2\lambda_1 \lambda_7^2 - 2\lambda_2 \lambda_6^2 + 2 \bar{\lambda}_{345} \lambda_6 \lambda_7}\ > \ 0 \; , \label{eq:Phi1 VEV Zeta1 CP1} \\[3mm]
\langle\phi_2^{\dagger} \phi_2 \rangle\ &=\ \frac{\left(\lambda_6 \lambda_7 -\bar{\lambda}_{345}\lambda_5 \right) \mu_1^2 + \left(2\lambda_1 \lambda_5 - \lambda_6^2 \right) \mu_2^2 + \left(\bar{\lambda}_{345} \lambda_6 - 2\lambda_1 \lambda_7 \right) m_{12}^2}{\lambda_5 \left(4\lambda_1 \lambda_2 - \bar{\lambda}_{345}^2 \right) - 2\lambda_1 \lambda_7^2 - 2\lambda_2 \lambda_6^2 + 2 \bar{\lambda}_{345} \lambda_6 \lambda_7}\ > \ 0 \; . \label{eq:Phi2 VEV Zeta1 CP1} 
\end{align}
\end{subequations}
In order  to fix the remaining  undetermined component $\mathrm{R}^2$,
we  impose  the   neutral  vacuum  condition~\eqref{eq:Neutral  Vacuum
  Condition 2} on the 4-vector $\mathrm{R}^\mu$, i.e.~$\mathrm{R}^\mu$
has to be a null vector. In this way, we find for the second component
$\mathrm{R}^2$ that
\begin{eqnarray}
\label{eq:R2 Zeta1 CP1} 
\mathrm{R}^2 \!\!&=&\!\! \pm \frac{1}{\mathrm{A}}\; \bigg\{\, 4\Big[
  \left(2\lambda_2 \lambda_5 - \lambda_7^2\right) \mu_1^2 +
  \left(\lambda_6 \lambda_7 -\bar{\lambda}_{345} \lambda_5\right)
  \mu_2^2 + \left(\bar{\lambda}_{345} \lambda_7 - 2 \lambda_2
  \lambda_6 \right) m_{12}^2 \Big] \nonumber \\  
\!&& \!\times\, \Big[\left(\lambda_6 \lambda_7
  -\bar{\lambda}_{345}\lambda_5 \right) \mu_1^2 + \left(2\lambda_1
  \lambda_5 - \lambda_6^2 \right) \mu_2^2 + \left(\bar{\lambda}_{345}
  \lambda_6 - 2\lambda_1 \lambda_7 \right) m_{12}^2 \Big] 
\\  
\!&& \!- \Big[\left(\bar{\lambda}_{345}\lambda_7 - 2\lambda_2\lambda_6
  \right) \mu_1^2 + \left(\bar{\lambda}_{345}\lambda_6 -
  2\lambda_1\lambda_7 \right) \mu_2^2 + \left(4\lambda_1\lambda_2 -
  \bar{\lambda}_{345}^2 \right) m_{12}^2 \Big]^2\, \bigg\}^{1/2} \,.\qquad \nonumber
\end{eqnarray}
After determining  all the components  of ${\rm R}^\mu$  and comparing
them  with  \eqref{eq:V1  Ansatz}  and  \eqref{eq:V2  Ansatz},  it  is
straightforward  to  find  the  vacuum  manifold  parameters  for  the
Lagrange multiplier solution $\zeta$ given in~\eqref{eq:zetaCP1}, with
$v_2^+ = 0$. These are given by
\begin{subequations}
\begin{align}
v_1^0\ &=\ \sqrt{ \frac{2\left(2\lambda_2 \lambda_5 -
    \lambda_7^2\right) \mu_1^2 + 2 \left( \lambda_6 \lambda_7
    -\bar{\lambda}_{345} \lambda_5\right) \mu_2^2 +
    2\left(\bar{\lambda}_{345} \lambda_7 - 2\lambda_2 \lambda_6
    \right) m_{12}^2}{\lambda_5 \left(4\lambda_1 \lambda_2 -
    \bar{\lambda}_{345}^2 \right) - 2\lambda_1 \lambda_7^2 -
    2\lambda_2 \lambda_6^2 + 2 \bar{\lambda}_{345} \lambda_6
    \lambda_7} }\; , \label{eq:CP1 v1} \\[2mm]   
v_2^0\ &=\ \sqrt{\frac{2\left(\lambda_6 \lambda_7
    -\bar{\lambda}_{345}\lambda_5 \right) \mu_1^2 + 2 \left(2\lambda_1
    \lambda_5 - \lambda_6^2 \right) \mu_2^2 +
    2\left(\bar{\lambda}_{345} \lambda_6 - 2\lambda_1 \lambda_7
    \right) m_{12}^2}{\lambda_5 \left(4\lambda_1 \lambda_2 -
    \bar{\lambda}_{345}^2 \right) - 2\lambda_1 \lambda_7^2 -
    2\lambda_2 \lambda_6^2 + 2 \bar{\lambda}_{345} \lambda_6
    \lambda_7} } \; , \label{eq:CP1 v2} \\[2mm]  
\cos \xi\ &=\ \frac{2m_{12}^2 - \lambda_6 (v_1^0)^2 - \lambda_7
  (v_2^0)^2}{2 \lambda_5 v_1^0 v_2^0} \ . \label{eq:CP1 xi value}  
\end{align}
\end{subequations}
We note  the necessary condition $0  < |\cos \xi| <  1$, for obtaining
spontaneous   electroweak  breaking   of  the   CP  symmetry   in  the
CP1-invariant 2HDM.

In order for the above extremal solutions to  represent local minima,
we require that the Hessian of the CP1-invariant potential be positive
definite  when evaluated  at  the extremal  points.  The Hessian  with
respect to $v^0_1$, $v^0_2$  and $\xi$ for the CP1-invariant potential
has the elements:
\begin{subequations}
\begin{align}
\mathrm{H}_{11}\ &=\ -\mu_1^2 + 3 \lambda_1 (v_1^0)^2 + \frac{1}{2}
\left[\bar{\lambda}_{345} + 2\lambda_5 \cos^2 \xi \right] (v_2^0)^2 +
3 \lambda_6 v_1^0 v_2^0 \cos \xi \; , \\ 
\mathrm{H}_{12}\ &=\ \left[\bar{\lambda}_{345} + 2\lambda_5 \cos^2 \xi
  \right] v_1^0 v_2^0 + \cos \xi \left[-m_{12}^2 + \frac{3}{2}
  \lambda_6 (v_1^0)^2 + \frac{3}{2} \lambda_7 (v_2^0)^2 \right] \; ,\\ 
\mathrm{H}_{13}\ &=\ - v_2^0 \sin \xi \left[2 \lambda_5 v_1^0 v_2^0
  \cos \xi - m_{12}^2 + \frac{3}{2} \lambda_6 (v_1^0)^2 + \frac{1}{2}
  \lambda_7 (v_2^0)^2 \right] \; , \\ 
\mathrm{H}_{22}\ &=\ -\mu_2^2 + 3 \lambda_2 (v_2^0)^2 +
\frac{1}{2}\left[\bar{\lambda}_{345} + 2\lambda_5 \cos^2 \xi \right]
(v_1^0)^2 + 3 \lambda_7 v_1^0 v_2^0 \cos \xi \; , \\ 
\mathrm{H}_{23}\ &=\ - v_1^0 \sin \xi \left[2 \lambda_5 v_1^0 v_2^0
  \cos \xi - m_{12}^2 + \frac{1}{2} \lambda_6 (v_1^0)^2 + \frac{3}{2}
  \lambda_7 (v_2^0)^2 \right] \; , \\ 
\mathrm{H}_{33}\ &=\ -\lambda_5 (v_1^0)^2 (v_2^0)^2 \cos 2\xi -
\left[-m_{12}^2 + \frac{1}{2} \lambda_6 (v_1^0)^2 + \frac{1}{2}
  \lambda_7 (v_2^0)^2 \right] v_1^0 v_2^0 \cos \xi \; . 
\end{align}
\end{subequations}
It is difficult  to obtain compact analytical expressions  in terms of
the set of  potential parameters  $\{\mu_1^2,  \mu_2^2, m_{12}^2,  \lambda_1,
\lambda_2,  \lambda_3, \lambda_4, \lambda_5,  \lambda_6, \lambda_7\}$,
so  the positivity  of  the symmetric  ${\rm  H}$ matrix  can only  be
checked  numerically  for  a  given  set of  input  parameters.   This
procedure  forms part  of our  numerical analysis  in Section~\ref{CP1
  Domain Walls}.

\subsubsection{CP1 Topology}
\label{CP1 Topology}

The topology  of the CP1-invariant  2HDM potential is very  similar to
the   ${\rm   Z}_2$-symmetric   case  discussed   in   Section~\ref{HF
  Symmetries}.   In  the symmetric  phase  of  the  theory, the  total
symmetry  group   of  the  potential   is  $\mathrm{G}_\mathrm{CP1}  =
\mathrm{CP1} \otimes \mathrm{SU}(2)_\mathrm{L} \otimes
\mathrm{U}(1)_\mathrm{Y} \simeq \mathrm{Z}_2 \times S^3 \times
S^1$.  Here  we  have  used  the  fact that  CP1  is  homeomorphic  to
$\mathrm{Z}_2$. After electroweak symmetry breaking
[cf.~\eqref{eq:Electroweak Symmetry Breaking Scheme}], the CP1
symmetry breaks into the identity ${\bf I}$, so the unbroken group is
$\mathrm{H}_\mathrm{CP1} = {\bf I} \otimes \mathrm{U}(1)_{\mathrm{em}}
\simeq S^1$. In the $\Phi$-space, the vacuum manifold is then given by
the set
\begin{equation}
\mathcal{M}^\mathrm{CP1}_{\Phi}\ =\ \mathrm{G}_\mathrm{CP1}/ 
\mathrm{H}_\mathrm{CP1}\ \simeq\ \mathrm{Z}_2 \times S^3 \; .  
\end{equation}
This vacuum manifold is homeomorphic  to that of the $\mathrm{Z}_2$ HF
symmetry          and         we          conclude         that~$\Pi_0
\left[\mathcal{M}^\mathrm{CP1}_{\Phi}  \right]  \neq  {\bf  I}$.  This
implies that the CP1-invariant 2HDM  has a domain wall solution, which
is studied in Section~\ref{Topological Defects of the 2HDM}.

The construction  of the vacuum manifold in  the $\Phi$-space proceeds
in  a  rather analogous  manner.   As  stated in~\eqref{eq:calM+}  and
\eqref{eq:calM-}, a general point of the vacuum manifold due a CP1
symmetry is given by $\Phi = \mathcal{M}^\mathrm{CP1}_\pm\, \Phi_0$,
where $\Phi_0$ is defined in terms of the non-zero VEVs $v^0_{1,2}$
and the  CP-odd phase $\xi$ given  in~\eqref{eq:CP1 v1}, \eqref{eq:CP1
  v2} and \eqref{eq:CP1 xi  value}, respectively. In the 8-dimensional
Majorana  $\Phi$   space,  the  HF  and   CP  transformation  matrices
$\mathcal{M}^\mathrm{CP1}_\pm$         of~\eqref{eq:calM+}         and
\eqref{eq:calM-} have  $\mathcal{T}_\pm = \mathrm{T}_\pm  = \sigma^0$.
Ignoring gauge transformations, there  are two distinct neutral vacuum
solutions:
\begin{equation}
\phi_{1} = \frac{1}{ \sqrt{2} } \left( \begin {array}{c}
  0\\\noalign{\medskip}v^0_{1}\end {array} \right) \, ,\
\phi_{2} = \frac{1}{ \sqrt{2} } \left( \begin {array}{c}
  0\\\noalign{\medskip}v^0_{2} e^{i \xi}\end {array} \right)\ \mbox{and}\
\;\; \phi_{1} = \frac{1}{ \sqrt{2} } \left( \begin {array}{c}
  0\\\noalign{\medskip}v^0_{1}\end {array} \right) \, ,\
\phi_{2} = \frac{1}{ \sqrt{2} } \left( \begin {array}{c}
  0\\\noalign{\medskip}v^0_{2} e^{-i \xi}\end {array} \right)\, ,
\end{equation}
in the gauge basis, where $v^0_1 > 0$. Finally, it is worth mentioning
that under the additional parameter restrictions $m_{12}^2 = \lambda_6
= \lambda_7  = 0$, the phase $\xi$  takes on the special  value $\xi =
\frac{\pi}{2}$. Given the freedom  of reparameterization $\phi_2 \to i
\phi_2$~\cite{Branco:1985pf}, the  CP1 vacuum manifold  coincides with
the one of the $\mathrm{Z}_2$ vacuum manifold in this case.

\subsection{CP2 Symmetry}\label{CP2 Symmetry}

The discrete CP2 symmetry of the 2HDM is defined as follows: 
\begin{eqnarray*}
\phi_1 &\to& \phi'_1\ =\ \phi_2^{*} \; , \\
\phi_2 &\to& \phi'_2\ =\ -\phi_1^{*} \; .
\end{eqnarray*}
Using the  CP2 parameter restrictions of  Table \ref{table:CP Parameter
  Conditions}, we  derive the VEVs of $\phi_{1,2}$  by considering the
two        conditions~\eqref{eq:Extremization       Condition       1}
and~\eqref{eq:Neutral Vacuum Condition 2}.   As before, we examine the
two distinct  cases: (i)~$\mathrm{det} [\mathrm{N}_{\mu  \nu}] \neq 0$
and (ii)~$\mathrm{det} [\mathrm{N}_{\mu \nu}] = 0$.

To start with, we  first calculate the determinant of $\mathrm{N}_{\mu
  \nu}$, which may be conveniently expressed as follows:
\begin{eqnarray}
\mathrm{det} [\mathrm{N}_{\mu \nu}]\ =\ (2\lambda_1 + \lambda_3 -
\zeta)  \left[ (2\lambda_1 - \lambda_3 + \zeta)((\lambda_4 + \zeta)^2
  - |\lambda_5|^2) \right.  \nonumber \\ 
\left. - 4|\lambda_6|^2 (\lambda_4 - R_5 + \zeta ) + 8 I_6 (I_5 R_6 -
R_5 I_6) \right] \; . 
\label{eq:CP2 Det}
\end{eqnarray}
Then,  for   the  CP2-invariant  2HDM   potential,  the  extremization
condition $\mathrm{N}_{\mu \nu} \mathrm{R}^\nu = \mathrm{M}_\mu$ gives
the two equations:
\begin{subequations}
\begin{align}
(2\lambda_1 + \lambda_3 - \zeta)\, \mathrm{R}^0\ &=\ 2\mu_1^2 \;
  ,\label{eq:r0 CP2 equation} \\[3mm]
\left( \begin {array}{ccc} \lambda_4+\mathrm{Re}(\lambda_5)+\zeta&
  -\mathrm{Im}(\lambda_5)&2\mathrm{Re}(\lambda_6)\\
\noalign{\medskip}-\mathrm{Im}(\lambda_5)  
  &\lambda_4-\mathrm{Re}(\lambda_5)+\zeta&-2\mathrm{Im}(\lambda_6)\\
\noalign{\medskip} 
  2\mathrm{Re}(\lambda_6)&-2\mathrm{Im}(\lambda_6)&2\lambda_1 -
  \lambda_3 + \zeta\end {array} \right) \left( \begin {array}{c}
    \mathrm{R}^1\\
\noalign{\medskip}\mathrm{R}^2\\\noalign{\medskip}\mathrm{R}^3 \end{array} 
  \right)\ &=\ \left( \begin {array}{c} 0\\\noalign{\medskip}
    0\\\noalign{\medskip} 0\end {array} \right) \label{eq:r1 r2 r3 CP2
      equation} \; . \quad 
\end{align}
\end{subequations}
Now,  if   the  matrix  $\mathrm{N}_{\mu  \nu}$   is  invertible,  the
components of $\mathrm{R}^\mu$ are found to be
\begin{subequations}
\begin{align}
\mathrm{R}^0\ &=\ \frac{2\mu_1^2}{2\lambda_1 + \lambda_3 - \zeta} \;
, \label{eq:CP2 R0} \\ 
\mathrm{R}^1\ &=\ \mathrm{R}^2\ =\ \mathrm{R}^3\ =\ 0 \;
. \label{eq:CP2 R1 R2 R3} 
\end{align}
\end{subequations}
Since only the  component ${\rm R}^0$ is non-zero,  this result is not
compatible with the  neutral vacuum condition~\eqref{eq:Neutral Vacuum
  Condition 2}, with ${\rm R}^\mu  {\rm R}_\mu = 0$, unless $\mu^2_1 =
0$.  This  is not  a viable scenario,  since $v^0_{1,2} =  0$, without
electroweak symmetry  breaking. For this  reason, we now  consider the
second possibility of a  singular matrix ${\rm N}_{\mu\nu}$, with $\rm
det [N_{\mu \nu}] = 0$.

\subsubsection{Neutral Vacuum Solutions from a Singular Matrix N}
\label{Utilizing CP2 zeta}

We now analyze the neutral vacuum solutions, for which the determinant
of  $\rm N_{\mu  \nu}$  vanishes due  to  a particular  choice of  the
Lagrange multiplier  $\zeta$. From~\eqref{eq:r0 CP2  equation}, we see
that the  singular solution  $\zeta = 2\lambda_1  + \lambda_3$  is not
compatible,  unless $\mu^2_1 =  0$. Therefore,  we concentrate  on the
other three  possible solutions obtained from  requiring the vanishing
of the determinant of the matrix  on the LHS of~\eqref{eq:r1 r2 r3 CP2
  equation}.

Employing standard methods for  solving cubic equations, we obtain the
three roots:
\begin{subequations}
\begin{align}
\zeta_1\ &=\ \frac{d}{6} - \frac{6b - 2a^2}{3d} - \frac{a}{3} \; , \\
\zeta_2\ &=\ -\frac{(1+i\sqrt{3})d}{12} + \frac{(1-i\sqrt{3})(3b -
  a^2)}{3d} - \frac{a}{3} \; , \\ 
\zeta_3\ &=\ -\frac{(1-i\sqrt{3})d}{12} + \frac{(1+i\sqrt{3})(3b -
  a^2)}{3d} - \frac{a}{3} \; , 
\end{align}
\end{subequations}
where $a,\,b,\,c$ and $d$ are defined as
\begin{subequations}
\begin{align}
a\ &=\ 2\lambda_1 - \lambda_3 + 2\lambda_4 \; , \\
b\ &=\ 2\lambda_4 ( 2\lambda_1 - \lambda_3) + \lambda_4^2 -
|\lambda_5|^2 - 4|\lambda_6|^2 \; , \\ 
c\ &=\ (2\lambda_1 - \lambda_3)(\lambda_4^2 - |\lambda_5|^2) -
4|\lambda_6|^2 (\lambda_4 - R_5) + 8 I_6(I_5 R_6 - R_5 I_6) \; , \\ 
d\ &=\ \left(36ab -108c - 8a^3 + 12\sqrt{12b^3 - 3a^2 b^2 - 54abc +
    81c^2 + 12a^3 c} \right)^{1/3} \; . 
\end{align}
\end{subequations}

Since  the  matrix  equation~\eqref{eq:r1   r2  r3  CP2  equation}  is
underdetermined, we may exploit this fact to express the components
${\rm R}^2$ and ${\rm R}^3$ in terms of ${\rm R}^1$ as
\begin{subequations}
\begin{align}
\mathrm{R}^2\ &=\ \frac{I_5 R_6 - I_6(\lambda_4 + R_5 + \zeta)}{R_6
  (\lambda_4 - R_5 + \zeta) - I_5 I_6}\ 
\mathrm{R}^1 \; , \label{eq:CP2 Singular R2}\\ 
\mathrm{R}^3\ &=\ \frac{2 I_5 R_6 - 2 I_6 (\lambda_4 + R_5 + \zeta)}{4
  I_6 R_6 - I_5(2 \lambda_1 - \lambda_3 + \zeta)} \ 
\mathrm{R}^1 \; . \label{eq:CP2 Singular R3} 
\end{align}
\end{subequations}
To determine the  component ${\rm R}^1$, we impose  the neutral vacuum
condition $\rm R^\mu R_\mu = 0$. In this way, we obtain that
\begin{equation}
\mathrm{R}^1\ =\ \pm\; \frac{1}{\mathrm{B}}\
  \frac{2\mu_1^2}{2\lambda_1 + \lambda_3 - \zeta} \ ,
\label{eq:CP2 Singular R1}
\end{equation}
where the parameter $\mathrm{B}$ is given by 
\begin{equation}
\mathrm{B}\ =\ \sqrt{\left[\frac{I_5 R_6 - I_6(\lambda_4 + R_5 +
      \zeta)}{R_6 (\lambda_4 - R_5 + \zeta) - I_5 I_6} \right]^2
  +\ \left[\frac{2 I_5 R_6 - 2 I_6 (\lambda_4 + R_5 + \zeta)}{4 I_6
      R_6 - I_5(2 \lambda_1 - \lambda_3 + \zeta)} \right]^2 +\ 1} \ .  
\end{equation}
We observe  that there are two possible  solutions for $\mathrm{R}^1$,
and therefore  for ${\rm R}^2$ and  ${\rm R}^3$, through~\eqref{eq:CP2
  Singular  R2}  and~\eqref{eq:CP2  Singular  R3}. The  two  solutions
differ by a  common overall sign and they  are topologically connected
via   the   CP2  transformation   ${\cal   O}_{\rm   CP2}$  given   in
Table~\ref{table:Lambda Matrix  Table} (see also  our discussion below
in  Section~\ref{sec:CP2 topology}).   Considering  only the  positive
solution of ${\rm  R}^1$, the vacuum manifold parameters  for $v_2^+ =
0$ are calculated to be
\begin{subequations}
\begin{align}
  \label{eq:CP2 v1}
v_{1}^0\ &=\ \sqrt{\left(\frac{2\mu_1^2}{2\lambda_1 + \lambda_3 -
    \zeta} \right) \!\! \left(1 + \frac{1}{\mathrm{B}} \, \frac{2 I_5
    R_6 - 2 I_6 (\lambda_4 + R_5 + \zeta)}{4 I_6 R_6 - I_5(2 \lambda_1
    - \lambda_3 + \zeta)} \right)} \; ,  \\ \label{eq:CP2 v2}   
v_{2}^0\ &=\ \sqrt{\left(\frac{2\mu_1^2}{2\lambda_1 + \lambda_3 -
    \zeta} \right) \!\! \left(1 - \frac{1}{\mathrm{B}} \, \frac{2 I_5
    R_6 - 2 I_6 (\lambda_4 + R_5 + \zeta)}{4 I_6 R_6 - I_5(2 \lambda_1
    - \lambda_3 + \zeta)} \right)} \; , \\  
  \label{eq:CP2 xi} 
\tan \xi\ &=\ \frac{(\lambda_4 + R_5 + \zeta) I_6 - I_5 R_6}{I_5 I_6 -
  (\lambda_4 - R_5 + \zeta) R_6} \; .  
\end{align}
\end{subequations}
Note  that  the  negative  solution  of ${\rm  R}^1$  is  obtained  by
interchanging $v^0_1 \leftrightarrow v^0_2$  and shifting $\xi \to \xi
+ \pi$. 

It is important  to remark here that the  phase $\xi$ in~\eqref{eq:CP2
  xi}   does    not   signal   spontaneous   breaking    of   the   CP
symmetry~\cite{Davidson:2005cw}.   Within  the  bilinear  scalar-field
formalism, it is not difficult to see that under a unitary rotation of
the Higgs doublets $\phi_{1,2}$,  which induces an orthogonal rotation
to   the   `spatial'   components   ${\rm  R}^{1,2,3}$,   the   matrix
equation~\eqref{eq:r1 r2 r3 CP2  equation} remains form invariant.  In
particular, one  can always find an induced  orthogonal rotation, such
that the matrix on the LHS of~\eqref{eq:r1 r2 r3 CP2 equation} becomes
diagonal~\cite{Maniatis:2011qu}.  It is  obvious that in this diagonal
basis,  the transformed quartic  couplings $\lambda_{6,7}$  vanish and
${\rm Im}\, \lambda_5 = 0$.  This result is identical to the one found
previously in~\cite{Gunion:2005ja}, which is based on the construction
of all possible Jarlskog-like~\cite{Jarlskog:1985ht,Bernabeu:1986fc},
Higgs-basis                     independent                     CP-odd
invariants~\cite{Lavoura:1994fv,Botella:1994cs} (for  a recent review,
see~\cite{Branco:2011iw}).

For  illustration,  we  display  in Table~\ref{table:CP2  VEV  parameter
  values} the  numerical values of the vacuum  manifold parameters for
$\zeta_{1,2,3}$ in a  CP2-invariant 2HDM, where 
\begin{equation}
  \label{CP2inputset}
\{\mu_1^2, \lambda_1, \lambda_3, \lambda_4, \lambda_5, \lambda_6\}\ =\ 
\{1, 8, 1, 3, 1 - 2i, 1 - 2i\}\; ,
\end{equation}
in  arbitrary  mass units.   This  particular  set  of parameters  is
chosen  so  as to  satisfy  the  CP2  convexity conditions  of  Table
\ref{table:CP Convexity  Conditions}. The  values of the  three Lagrange
multipliers are: $\zeta_1 = -0.295$,  $\zeta_2 = -4.09$ and $\zeta_3 =
-16.6$.   In order  to  determine whether  the  three extremal  points
presented  in  Table~\ref{table:CP2  VEV  parameter  values}  are  local
minima, we need to analyze  the positivity of the Hessian matrix ${\rm
  H}$.

\begin{table}[t!]
\begin{center}
\begin{tabular}{c|c|c|c}
\hline
Quantity	& 	$\zeta_1$	&	$\zeta_2$	&
$\zeta_3$ \\ 
\hline 

$v_1^0$	&	0.372	&	0.349	&	0.340	\\

$v_2^0$	&	0.305	&	0.261	&	0.060	\\

$\xi$	&	-1.17	&	0.343	&	0.971	\\

\hline

$\mathrm{V}(v_1^0,v_2^0, \xi)$	&	-0.0578	&	-0.0474	&
-0.0297	\\ 

\hline
\end{tabular}
\caption{\it The  numerical values for the  vacuum manifold parameters
  and potential  value at  the extremal points  for the  parameter set
  $\{\mu_1^2, \lambda_1,  \lambda_3, \lambda_4, \lambda_5, \lambda_6\}
  =  \{1,   8,  1,  3,  1  -   2i,  1  -  2i\}$,   in  arbitrary  mass
  units. \label{table:CP2 VEV parameter values}}
\end{center}
\end{table}

The Hessian  for the  CP2-invariant 2HDM potential  is a $3  \times 3$
symmetric matrix, with the elements
\begin{subequations}
\begin{align}
\mathrm{H}_{11}\ &=\ -\mu_1^2 + 3 \lambda_1 (v_1^0)^2 + 
\frac{1}{2} \Big(\lambda_{34} + R_5 \cos 2 \xi - I_5 \sin 2 \xi
  \Big) (v_2^0)^2\nonumber\\
\ & \hspace{6mm}  + 3 v_1^0 v_2^0 \Big(R_6 \cos \xi - I_6 \sin \xi
  \Big) \; , \label{eq:CP2 Hessian H11} \\[3mm] 
\mathrm{H}_{12}\ &=\ \Big(\lambda_{34} + R_5 \cos 2 \xi - I_5 \sin 2
  \xi \Big) v_1^0 v_2^0 + \frac{3}{2}\Big[(v_1^0)^2 - (v_2^0)^2
  \Big] \Big(R_6 \cos \xi - I_6 \sin \xi \Big) \; , \nonumber \\ 
& \\
\mathrm{H}_{13}\ &=\ - \Big( R_5 \sin 2\xi + I_5 \cos 2\xi \Big) v_1^0
(v_2^0)^2 - \frac{1}{2} v_2^0 \Big[ 3(v_1^0)^2 - (v_2^0)^2 \Big]
\Big( R_6 \sin \xi + I_6 \cos \xi \Big) \; , \\ 
\mathrm{H}_{22}\ &=\ -\mu_1^2 + 3 \lambda_1 (v_2^0)^2 + \frac{1}{2}
\Big(\lambda_{34} + R_5 \cos 2 \xi - I_5 \sin 2 \xi \Big) (v_1^0)^2\nonumber\\
\ & \hspace{6mm} - 3 v_1^0 v_2^0 \Big(R_6 \cos \xi - I_6 \sin \xi
\Big) \; , \\[3mm]  
\mathrm{H}_{23}\ &=\ - \Big(R_5 \sin 2\xi + I_5 \cos 2\xi \Big)
(v_1^0)^2 v_2^0 - \frac{1}{2} v_1^0 \Big[(v_1^0)^2 - 3(v_2^0)^2\Big] 
\Big( R_6 \sin \xi + I_6 \cos \xi \Big) \; , \\ 
\mathrm{H}_{33}\ &=\ - \Big( R_5 \cos 2\xi - I_5 \sin 2\xi \Big)
(v_1^0)^2 (v_2^0)^2 - \frac{1}{2} v_1^0 v_2^0 \Big[(v_1^0)^2 -
  (v_2^0)^2 \Big] \Big( R_6 \sin \xi - I_6 \cos \xi \Big) \; . 
\label{eq:CP2 Hessian H33} 
\end{align}
\end{subequations}
We can numerically  check the positivity of the  matrix ${\rm H}$.  In
this way, we find that for a convex CP2-invariant potential with input
parameters   as  given   in~\eqref{CP2inputset},  only   the  Lagrange
multiplier  $\zeta_1$ represents a  local minimum,  which is  a global
minimum.   As we will  see below,  this global  minimum has  a twofold
degeneracy, as a consequence of the CP2 symmetry.

\subsubsection{CP2 Topology}\label{sec:CP2 topology}

In the symmetric phase of the  theory, the total symmetry group of the
CP2-invariant    2HDM   potential   is    $\mathrm{G}_\mathrm{CP2}   =
\mathrm{CP2}       \otimes      \mathrm{SU}(2)_\mathrm{L}      \otimes
\mathrm{U}(1)_\mathrm{Y}  \cong  \mathrm{Z}_2  \otimes  \Pi_2  \otimes
\mathrm{SU}(2)_\mathrm{L}   \otimes  \mathrm{U}(1)_\mathrm{Y}$,  where
$\Pi_2$ is  the permutation symmetry  $\phi_1 \leftrightarrow \phi_2$.
To      be      specific,     we      have      used     here      the
isomorphism~\cite{Ferreira:2009wh}:  ${\rm   CP2}  \cong  \mathrm{Z}_2
\otimes \Pi_2$, which is  evident in the ${\rm Z}_2$-constrained Higgs
basis~\cite{Ferreira:2009wh}, where $\lambda_6 = \lambda_7 = I_5 = 0$.
After   electroweak   symmetry   breaking   [cf.~\eqref{eq:Electroweak
    Symmetry  Breaking  Scheme}],  the  permutation  symmetry  $\Pi_2$
remains   intact,  so   the  residual   unbroken  group   of   CP2  is
$\mathrm{H}_\mathrm{CP2}       =       {\rm       \Pi}_2       \otimes
\mathrm{U}(1)_{\mathrm{em}}$.  As  a consequence, the  vacuum manifold
$\mathcal{M}^\mathrm{CP2}_{\Phi}$ in the $\Phi$-space has the topology
of the coset space:
\begin{equation}
\mathcal{M}^\mathrm{CP2}_{\Phi}\ =\ 
\mathrm{G}_\mathrm{CP2}/\mathrm{H}_\mathrm{CP2}\
\simeq\ \mathrm{Z}_2 \times S^3 \; .
\end{equation}
The vacuum  manifold $\mathcal{M}^\mathrm{CP2}_{\Phi}$ is homeomorphic
to that of  the $\mathrm{Z}_2$ HF symmetry, thus  having a non-trivial
zeroth               homotopy               group               $\Pi_0
\left[\mathcal{M}^\mathrm{CP2}_{\Phi}\right] =  \Pi_0 \left[ {\rm Z}_2
  \right] \neq {\bf I}$. This  implies that the CP2-invariant 2HDM has
a domain  wall solution, which we  analyze in Section~\ref{Topological
  Defects of the 2HDM}.

An arbitrary point $\Phi$ of the vacuum manifold due to a CP2 symmetry
may    be   obtained   with    the   help    of~\eqref{eq:calM+}   and
\eqref{eq:calM-}, i.e.~$\Phi = \mathcal{M}^\mathrm{CP2}_\pm\, \Phi_0$,
where $\Phi_0$ is  defined in terms of the non-zero VEVs $v^0_{1,2}$
given in~\eqref{eq:CP2 v1}, \eqref{eq:CP2 v2} and $\xi$
in~\eqref{eq:CP2   xi}.   The  HF   and  CP   transformation  matrices
$\mathcal{M}^\mathrm{CP2}_\pm$         of~\eqref{eq:calM+}         and
\eqref{eq:calM-}  are $\mathcal{T}_+  = \mathrm{T}_+  =  \sigma^0$ and
$\mathcal{T}_-  = \mathrm{T}_- =  i\sigma^2$, respectively.   From the
action of these transformation matrices  on $\Phi_0$, we find that the
vacuum manifold  is comprised of two disconnected  sets.  The elements
within  each   set  are  related   by  $S^3$  gauge   rotations  ${\rm
  U}_\pm$. Two representative vacuum manifold points from each set are
\begin{equation}
\phi_{1} = \frac{1}{ \sqrt{2} } \left( \begin {array}{c}
  0\\\noalign{\medskip}v^0_{1}\end {array} \right) \; ,\qquad
\phi_{2} = \frac{1}{ \sqrt{2} } \left( \begin {array}{c}
  0\\\noalign{\medskip}v^0_{2} e^{i \xi}\end {array} \right)
\end{equation}
and
\begin{equation}
\phi_{1} = \frac{1}{ \sqrt{2} } \left( \begin {array}{c}
  0\\\noalign{\medskip}v^0_{2}\end {array} \right) \; ,\qquad
\phi_{2} = \frac{1}{ \sqrt{2} } \left( \begin {array}{c}
  0\\\noalign{\medskip} -\,v^0_{1} e^{i \xi}\end {array} \right)\; ,
\end{equation}
where we  used the  freedom of the  gauge rotations ${\rm  U}_\pm$, in
order to adjust the neutral component of $\phi_1$ to be positive.

\subsection{CP3 Symmetry}\label{CP3 Symmetry}

The CP3  symmetry is a  continuous CP symmetry  and is defined  by the
transformations
\begin{eqnarray*}
\phi_1 &\to& \phi'_1\ =\ \cos \theta \, \phi_1^{*} + \sin \theta \,
\phi_2^{*} \; , \\ 
\phi_2 &\to& \phi'_2\ =\ - \sin \theta \, \phi_1^{*} + \cos \theta \,
\phi_2^{*} \; , 
\end{eqnarray*}
where $\theta  \in [0,\pi)$. As before,  we first consider the  case $\mathrm{det} [\mathrm{N}_{\mu
    \nu}] \neq 0$. The  determinant of $\mathrm{N}_{\mu \nu}$ is given
by
\begin{equation}
\mathrm{det} [\mathrm{N}_{\mu \nu}]\ =\ \left(2\lambda_1 + \lambda_3 +
\zeta \right) \left(2\lambda_1 - \lambda_3 - \zeta \right)^2
\left(2\lambda_4 + \lambda_3 - 2\lambda_1 -\zeta \right) \; . 
\label{eq:CP3 Det}
\end{equation}
For  the  CP3-invariant 2HDM  potential,  the extremization  condition
$\mathrm{N}_{\mu \nu}  \mathrm{R}^\nu = \mathrm{M}_\mu$  leads to four
separate equations:
\begin{subequations}
\begin{align}
(2\lambda_1 + \lambda_3 - \zeta) \mathrm{R}^0\ &=\ 2\mu_1^2 \;
  , \label{eq:r0 CP3 equation} \\ 
(2\lambda_1 - \lambda_3 + \zeta) \mathrm{R}^1\ &=\ 0 \; , \label{eq:r1
    CP3 equation} \\ 
(2\lambda_4 - 2\lambda_1 + \lambda_3 + \zeta) \mathrm{R}^2\ &=\ 0 \;
  , \label{eq:r2 CP3 equation} \\ 
(2\lambda_1 - \lambda_3 + \zeta) \mathrm{R}^3\ &=\ 0 \; .\label{eq:r3
    CP3 equation} 
\end{align}
\end{subequations}
Based on  the assumption that  ${\rm N}_{\mu\nu}$ can be  inverted, we
find that all ``spatial'' components ${\rm R}^{1,2,3}$ vanish. Like in
the CP2 case, the condition  for having a neutral vacuum restricts the
remaining  component ${\rm  R}^0$ to  be zero  as well,  which  can be
naturally fulfilled, only  if $\mu^2_1 = 0$.  In  such a scenario, one
has $v^0_{1,2} =  0$ and so absence of  electroweak symmetry breaking.
Therefore, we now investigate the  case where $\rm det [N_{\mu \nu}] =
0$.

\subsubsection{Neutral Vacuum Solutions from a Singular Matrix N}
\label{CP3 Singular Section}

From the system  of equations~\eqref{eq:r0 CP3 equation}--\eqref{eq:r3
  CP3 equation}, it is easy to  see that there are only two compatible
singular  solutions   of  $\mathrm{N}_{\mu  \nu}$   for  the  Lagrange
multipliers:
\begin{subequations}
\begin{align}
  \label{CP3zeta1}
\zeta_1\ &=\ -2\lambda_1 + \lambda_3 \; , \\
  \label{CP3zeta2}
\zeta_2\ &=\ 2\lambda_1 - \lambda_3 - 2\lambda_4 \; .
\end{align}
\end{subequations}

Let  us   first  consider  the  solution  $\zeta_1   =  -2\lambda_1  +
\lambda_3$.   In this  case,  only the  components $\mathrm{R}^0$  and
$\mathrm{R}^2$ of the 4-vector $\mathrm{R}^\mu$ are determined as
\begin{subequations}
\begin{align}
\mathrm{R}^0\ &=\ \frac{\mu_1^2}{2 \lambda_1} \label{eq:CP3 zeta1 R^0} \; , \\
\mathrm{R}^2\ &=\ 0 \; .
\end{align}
\end{subequations}
Instead, $\mathrm{R}^1$ and  $\mathrm{R}^3$ are free parameters, which
are  constrained by the  neutral vacuum  condition: ${\rm  R}^\mu {\rm
  R}_\mu =  0$. Specifically, the  latter condition gives rise  to the
constraint:
\begin{equation}
  \label{R1R3}
({\rm R}^1)^2\ +\ ({\rm R}^3)^2\ =\ \left( 
\frac{\mu_1^2}{2 \lambda_1}\right)^2 \; .
\end{equation}
The constraint $\mathrm{R}^2 = 0$ implies that $\xi = n\,\pi$, where $n$
is an integer. In terms of the vacuum manifold parameters $v^0_{1,2}$,
we have the general solution
\begin{equation}
  \label{eq:CP3 zeta1 VM Point}
v_1^0\ =\ \frac{\mu_1}{\sqrt{\lambda_1}}\ \sin\theta\; ,\qquad  
v_2^0\ =\ \frac{\mu_1}{\sqrt{\lambda_1}}\ \cos\theta\; ,
\end{equation} 
where $\xi  = n\,\pi$  and $\theta \in  [ 0,  \pi )$.  The  free angle
  $\theta$ is associated with a massless `CP-even' Goldstone boson, as
  can be verified independently  from the analytical results presented
  in~\cite{Pilaftsis:1999qt}.

In order for  the extremal point  given in~\eqref{eq:CP3  zeta1 VM
  Point} to  be a local minimum,  we require that the  elements of the
Hessian  matrix ${\rm  H}$  for the  CP3-invariant  2HDM potential  be
positive. The elements of the symmetric matrix ${\rm H}$ read:
\begin{subequations}
\begin{align}
\mathrm{H}_{11} &= -\mu_1^2 + 3\lambda_1 (v_1^0)^2 + \frac{1}{2}
\left[\lambda_{34} + (2 \lambda_1 - \lambda_{34}) \cos 2 \xi \right]
(v_2^0)^2 \; , \label{eq:CP3 Hessian H11} \\ 
\mathrm{H}_{12} &= \left[\lambda_{34} + (2 \lambda_1 - \lambda_{34})
  \cos 2 \xi \right] v_1^0 v_2^0 \; , \\ 
\mathrm{H}_{13} &= -\left(2\lambda_1 - \lambda_{34} \right) v_1^0
(v_2^0)^2 \sin 2 \xi \; , \\ 
\mathrm{H}_{22} &= -\mu_1^2 + 3\lambda_1 (v_2^0)^2 + \frac{1}{2}
\left[\lambda_{34} + (2 \lambda_1 - \lambda_{34}) \cos 2 \xi \right]
(v_1^0)^2 \; , \\ 
\mathrm{H}_{23} &= - \left(2\lambda_1 - \lambda_{34} \right) (v_1^0)^2
v_2^0 \sin 2 \xi \; , \\ 
\mathrm{H}_{33} &= - \left(2\lambda_1 - \lambda_{34} \right) (v_1^0)^2
(v_2^0)^2 \cos 2 \xi \; . \label{eq:CP3 Hessian H33} 
\end{align}
\end{subequations}
Then, the conditions for positivity of ${\rm H}$ are simply given by
\begin{equation}
  \label{eq:CP3 Zeta1 Minima Condition}
\mu_1^2 \ >\ 0 \; ,\qquad \lambda_{34} \ >\ 2 \lambda_1 \; .
\end{equation}
Note   that  the  second   condition  in~\eqref{eq:CP3   Zeta1  Minima
  Condition}  is supplementary to  the two  conditions given  in Table
\ref{table:CP Convexity Conditions}  for ensuring a convex CP3-invariant
2HDM potential.   This local minimum has the potential value
\begin{equation}
\mathrm{V}_0\ =\ - \frac{\mu_1^4}{4 \lambda_1}\ .
\label{eq:CP3 Zeta1 Potential Value}
\end{equation}

Let  us  now  investigate  the  second singular  solution  $\zeta_2  =
2\lambda_1 -  \lambda_3 - 2\lambda_4$.  In this case, we obtain
\begin{subequations}
\begin{align}
  \label{eq:CP3 zeta2 R^0} 
\mathrm{R}^0\ &=\ \frac{\mu_1^2}{2 (\lambda_3 + \lambda_4)} \; , \\
  \label{eq:CP3 zeta2 R^1} 
\mathrm{R}^1\ &=\ \mathrm{R}^3\ =\ 0 \; .
\end{align}
\end{subequations}
The  component  ${\rm  R}^2$  is  constrained by  the  neutral  vacuum
condition~\eqref{eq:Neutral  Vacuum  Condition  2}  imposed  on  ${\rm
  R}^\mu$, i.e.~${\rm R}^\mu {\rm R}_\mu = 0$, from which we find that
\begin{equation}
  \label{eq:CP3 zeta2 R^2}
{\rm R}^2\ =\ \pm\, {\rm R}^0\; .
\end{equation}
Taking the  constraints~\eqref{eq:CP3 zeta2 R^0},  \eqref{eq:CP3 zeta2
  R^1} and~\eqref{eq:CP3 zeta2 R^2} into account, we derive the vacuum
manifold parameters
\begin{equation}
  \label{eq:CP3 zeta2 VM Point}
v^0_1\ =\ \frac{v'}{\sqrt{2}}\ ,\qquad 
v^0_2\ =\ \pm\; \frac{i v'}{\sqrt{2}} \ ,
\end{equation} 
with $v^\prime =  \mu_1/\sqrt{\lambda_{34}}$.  The conditions for this
neutral  vacuum  solution  to  be  a local  minimum  result  from  the
positivity  of the  Hessian matrix  ${\rm H}$  given  in \eqref{eq:CP3
  Hessian H11}--\eqref{eq:CP3 Hessian H33}.  These conditions are
\begin{equation}
  \label{eq:CP3 Zeta2 Minima Condition}
\mu_1^2\ >\ 0 \; ,\qquad
2 \lambda_1 \ >\ \lambda_{34} \; .
\end{equation}
These  conditions  are,  in  general,  not guaranteed  solely  by  the
convexity conditions  for the CP3-invariant potential  stated in Table
\ref{table:CP Convexity  Conditions}. Direct comparison  of these minima
conditions  with those  for  the $\zeta_1$  solution in  \eqref{eq:CP3
  Zeta1  Minima  Condition}  shows   that  both  local  minima  cannot
co-exist.   It depends  on the  relative values  of $2  \lambda_1$ and
$\lambda_{34}$ which solution becomes the local minimum, and this will
then be the global minimum as well. The value of the potential arising
from the second solution $\zeta_2$ is easily evaluated to be
\begin{equation}
  \label{eq:CP3 Zeta2 Potential Value}
\mathrm{V}_0\ =\ -\; \frac{\mu_1^4}{2 \lambda_{34}} \; .
\end{equation}
In the following, we will  analyze the topology resulting from the two
neutral  vacuum  solutions  given  in~\eqref{eq:CP3  zeta1  VM  Point}
and~\eqref{eq:CP3 zeta2 VM Point}, respectively.

\subsubsection{CP3 Topology}
\label{CP3 Topology}

It is interesting to discuss  the vacuum topology of the CP3-symmetric
2HDM  for  the  two  solutions  obtained  by  means  of  the  Lagrange
multipliers   $\zeta_1$  and   $\zeta_2$,   given  in~\eqref{CP3zeta1}
and~\eqref{CP3zeta2}, respectively. 

We first note that the  total symmetry group of the CP3-symmetric 2HDM
potential   is   $\mathrm{G}_\mathrm{CP3}   =   \mathrm{CP3}   \otimes
\mathrm{SU}(2)_\mathrm{L} \otimes \mathrm{U}(1)_\mathrm{Y} \simeq {\rm
  Z}_2 \times S^1 \times S^3  \times S^1$, since ${\rm CP3} \cong {\rm
  CP1} \otimes {\rm SO}(2)$.  This means that the ${\rm CP3}$ group is
equivalent  to  the combined,  as  well  as  independent action  of  a
standard  CP1   transformation  and  a   SO(2)  HF  rotation   in  the
$(\phi_1\,,\, \phi_2)$ field space.

Let  us now  consider  the  neutral vacuum  solution  obtained by  the
Lagrange  multiplier~$\zeta_1$.  After electroweak  symmetry breaking,
the  total symmetry  group $\mathrm{G}_\mathrm{CP3}$  breaks  into the
residual group $\mathrm{H}^{(1)}_\mathrm{CP3}  = {\rm CP1}\otimes {\bf
  I} \otimes {\rm U}(1)_{\rm em}$. This can be easily seen, since $\xi
= 0$ in this scenario and  so the CP1 symmetry remains intact, whereas
the  ${\rm  SO}(2)$  HF  symmetry  gets spontaneously  broken  to  the
identity ${\bf  I}$.  As  a consequence, the  vacuum manifold  in the
$\Phi$-space is determined by the coset space
\begin{equation}
\label{eq:CP3 Vacuum Manifold Space} 
\mathcal{M}^\mathrm{CP3}_{\Phi}\ =\ 
\mathrm{G}_\mathrm{CP3}/\mathrm{H}^{(1)}_\mathrm{CP3}\
\simeq\ S^1 \times S^3 \; .  
\end{equation}
Since~$\Pi_1  \left[\mathcal{M}^\mathrm{CP3}_{\Phi}  \right]  =  \Pi_1
\left[S^1 \right]  \neq {\bf I}$,  we conclude that  the CP3-invariant
2HDM  related  to  the  Lagrange  multiplier $\zeta_1$  has  a  vortex
solution  which  is  analyzed  in detail  in  Section~\ref{Topological
  Defects of  the 2HDM}.  Using the result  of \eqref{eq:CP3  zeta1 VM
  Point},  the transitive  action  of the  transformation matrices  of
\eqref{eq:calM+} and \eqref{eq:calM-} result  in the general points on
the vacuum manifold:
\begin{equation}
\phi_{1}\ =\ \frac{1}{\sqrt{2}} \left( \begin {array}{c}
  0\\\noalign{\medskip} v \cos \theta \end {array} \right) \; ,
\qquad \phi_{2} = \frac{1}{\sqrt{2}} \left( \begin
  {array}{c} 0\\\noalign{\medskip} (-1)^n\, v\sin \theta \end {array}
\right) \; , 
\end{equation}
where $v  = \mu_1/\sqrt{\lambda_1}$ and $\theta \in  [0,\pi )$.  There
  is a  relative minus sign for odd  $n$, but this can  be absorbed by
  redefining $\theta$ as $\pi - \theta$.

We may  now determine  the vacuum manifold  of the  CP3-symmetric 2HDM
associated      with      the      second     Lagrange      multiplier
solution~$\zeta_2$~\eqref{CP3zeta2}. In this  case, the total symmetry
group follows  a different  breaking pattern and  the little  group is
$\mathrm{H}^{(2)}_\mathrm{CP3} = {\rm  CP1}\otimes {\rm SO}(2) \otimes
{\rm U}(1)_{\rm  em}$, i.e. neither of  the two symmetries  CP1 and SO(2)
are broken. In  order to see this, we may  consider an SO(2) rotation
of the  vacuum manifold point given in~\eqref{eq:CP3  zeta2 VM Point},
yielding
\begin{equation}
\phi_{1} = \frac{1}{\sqrt{2}} \left( \begin {array}{c}
  0\\\noalign{\medskip} v^\prime e^{\pm i \theta} \end {array} \right)
\; , \qquad \;\; \phi_{2} = \frac{1}{\sqrt{2}} \left( \begin
  {array}{c} 0\\\noalign{\medskip} \pm i v^\prime e^{\pm i
    \theta} \end {array} \right) \; .  
\end{equation}
The phase  $\theta$ can  always be removed  by a ${\rm  U(1)}_{\rm Y}$
hypercharge rotation,  which is a  manifestation of the fact  that the
${\rm  SO(2)}$  symmetry is  not  broken,  after electroweak  symmetry
breaking. Moreover, one could  reparameterize the second Higgs doublet
$\phi_2$  as  $\pm   i\phi_2$,  in  order  to  render   both  VEVs  of
$\phi_{1,2}$ real. Since such a reparameterization does not induce any
additional phase  in the real  quartic couplings of  the CP3-invariant
2HDM potential,  we conclude  that the CP1  symmetry is not  broken as
well.   Thus,  the  vacuum  manifold  determined by  the  coset  space
$\mathrm{G}_\mathrm{CP3}/\mathrm{H}^{(2)}_\mathrm{CP3}$              is
homeomorphic to $S^3$, exactly as  in the SM.  Consequently, there are
no non-trivial topological defects in the 2HDM scenario related to the
second Lagrange multiplier solution~$\zeta_2$.

\section{Topological Defects in the 2HDM}
\label{Topological Defects of the 2HDM}

Using  our analysis  of  the  six accidental  symmetries  of the  2HDM
conducted in Sections \ref{HF  Symmetries} and \ref{CP Symmetries}, we
will now study the topological defects associated with the spontaneous
symmetry breaking  of each of the six  accidental symmetries studied.  
As shown in Table~\ref{table:2HDM Topological Defect Table},  we find that there are
three domain wall, two vortex and one global monopole solutions due to
the  additional symmetries  of the  2HDM, possibly  posing significant
cosmological implications for the model.  A comprehensive introduction
to  the  properties and  formation  of  topological  defects is  given
in~\cite{Vilenkin}.

In our  study of the topological  defects, we assume that  the VEVs of
the two  Higgs doublets $\phi_{1,2}$  are still assigned at  and after
the electroweak symmetry breaking,  such that $(v^0_1)^2 + (v^0_2)^2 =
v^2_{\rm SM}$,  where $v_{\rm  SM}\sim 246$~GeV is  the VEV of  the SM
Higgs doublet.   Due to the  complexity of the  differential equations
that result from  the 2HDM Lagrangian for each  symmetry, our study of
the  scalar  functions  involved  is  carried  out  numerically  using
gradient  flow techniques  which involve  minimizing the  energy  of a
configuration on a  finite grid with initial conditions  that have the
appropriate boundary  conditions. This is  done by defining  an energy
functional  $\mathrm{E} =  \mathrm{E}(f_1, \dots,  f_n)$,  where $f_1,
\dots, f_n$ are the  functions defining the topological solutions, and
then  by solving  the first  order diffusion  equation $\dot{f}_k  = -
\frac{\delta \mathrm{E}}{\delta f_k}$ for $k = 1, \dots, n$.

\begin{table}[t!]
\begin{center}
\begin{tabular}{c||c|c|c|c}
\hline
Symmetry & $\rm G_{HF/CP}$ & $\rm H_{HF/CP}$ & ${\cal M}^{\rm HF/CP}_\Phi$
& Topological Defect \\
\hline
\hline

$\rm Z_2$	& $\rm Z_2$	& 	$\mathbf{I}$	& ${\rm Z_2}$
& Domain Wall	\\ 

\hline

$\rm U(1)_{PQ}$	& ${\rm U(1)_{PQ}} \simeq S^1$	&	$\mathbf{I}$
& $S^1$ &	Vortex	\\ 

\hline

$\rm SO(3)_{HF}$	& ${\rm SO(3)_{HF}}$	&	
${\rm SO(2)_{HF}}$ &  $S^2$	&	Global Monopole	\\

\hline

CP1	& ${\rm CP1} \simeq {\rm Z_2}$	&	$\mathbf{I}$	&
${\rm Z}_2$ & Domain Wall	\\ 

\hline

CP2	& ${\rm Z_2} \otimes \Pi_2$	&
$\Pi_2$	& ${\rm Z}_2$ &	Domain Wall	\\ 

\hline

CP3	& ${\rm CP1} \otimes {\rm SO(2)}$	&
$\rm CP1$	& $S^1$ & Vortex	\\

\hline

\end{tabular}
\caption{\it  Breaking  patterns  of  the total  symmetry  group  $\rm
  G_{HF/CP}$  into  the  little   group  $\rm  H_{HF/CP}$,  after  the
  electroweak  symmetry breaking ${\rm  SU(2)_L} \otimes  {\rm U(1)_Y}
  \to  {\rm U(1)}_{\rm  em}$. The  fourth and  fifth columns  show the
  topology of the vacuum  manifold ${\cal M}^{\rm HF/CP}_\Phi$ and the
  associated topological defect, for  each of the six accidental HF/CP
  symmetries of the 2HDM.\label{table:2HDM Topological Defect Table} }
\end{center} 
\end{table}

\subsection{Domain Walls}

We  begin our  discussion of  topological defects  with  domain walls,
which  have  long been  known  to  have  severe consequences  for  the
evolution  of the  Universe should  they form  at a  symmetry breaking
phase  transition  in the  early  Universe,  since  they can  come  to
dominate               the              Universe's              energy
density~\cite{Zeldovich:1974uw}. Various  mechanisms to reconcile this
undesirable nature of domain walls with current observations have been
discussed, such as the restoration of the broken discrete symmetry and
subsequent  evaporation   of  the  domain  walls  at   a  later  phase
transition~\cite{PhysRevD.9.3357}, the use  of a period of exponential
inflation     to     dilute     the    concentration     of     domain
walls~\cite{PhysRevD.38.1141} and the symmetry of the model being only
an approximate discrete,  exponentially suppressing domain wall energy
density~\cite{Vilenkin:1981zs,Gelmini:1988sf,Larsson:1996sp}.

The  present study of  domain walls  does not  attempt to  analyse the
cosmological  implications,  which  will  be  presented  in  a  future
publication,  rather  it focuses  on  presenting  an  overview of  the
typical domain wall solutions and  analysing whether or not the energy
per unit area  of the domain wall can be made  to be vanishingly small
for specific valid parameter choices.

\subsubsection{$\mathrm{Z}_2$ Domain Walls}
\label{Z2 Domain Walls}

From  our analysis in  Section \ref{Z2  Symmetry}, the  2HDM potential
that is invariant under the  HF $\mathrm{Z}_2$ symmetry can exhibit a disconnected  vacuum manifold, the components  of which are
not linked by  the gauge symmetries of the  theory, provided both VEVs
of  the Higgs  doublets $\phi_{1,2}$  that create  the  neutral vacuum
global  minimum solution  are non-zero,  i.e. $v_{1,2}^0  \neq  0$ and
$v_2^+ = 0$. This scenario  is only apparent within the $\mathrm{Z}_2$
invariant 2HDM  when the determinant  of the matrix $\rm  N_{\mu \nu}$
vanishes, as shown in Section \ref{sec:Z2LMsingular}.

Let us now analyze  an one-dimensional, time-independent kink solution
for the $\mathrm{Z}_2$ symmetry. In  order to find such a solution, we
will use an ansatz for the two Higgs doublets given by:
\begin{equation}
\phi_{1,2}(x)\   =\    \frac{1}{\sqrt{2}}   \left(   \begin   {array}{c}
  0\\\noalign{\medskip}v^0_{1,2}(x)\end {array} \right) \; ,
\end{equation}
where the coordinate $x$ describes the spatial dimension perpendicular
to the  plane of the  domain wall. Using  this ansatz, the  energy per
unit area of the system is
\begin{equation}
\mathrm{E}\ =\ \int_{-\infty}^\infty \!\!\! dx\;\; \mathcal{E}\!
\left( \phi_1, \phi_2 \right) \; , 
\label{eq:Z2 Energy per Unit Area}
\end{equation}
where the energy density for the general 2HDM is given by:
\begin{equation}
\mathcal{E}\! \left( \phi_1, \phi_2 \right)\ =\ (\nabla
\phi_1^\dagger) \cdot (\nabla \phi_1) + (\nabla \phi_2^\dagger) \cdot
(\nabla \phi_2) +\ \mathrm{V}(\phi_1, \phi_2)\ +\ \mathrm{V}_0 \; , 
\end{equation}
where $\nabla$  is the  3-dimensional gradient operator,  expressed in
the relevant coordinate system. Moreover, $\mathrm{V}_0$ is introduced
to normalize the potential contribution  to the energy density to have
a  zero value  at  the global  minimum.   The energy  density for  the
$\mathrm{Z}_2$ invariant 2HDM is given by
\begin{eqnarray}
\mathcal{E}\! (x)  &=& \frac{1}{2} \left(\frac{d v^0_1}{dx} \right)^2
+ \frac{1}{2} \left(\frac{d v^0_2}{dx} \right)^2 - \frac{1}{2} \mu_1^2
v^0_1(x)^2 - \frac{1}{2} \mu_2^2 v^0_2(x)^2 \nonumber \\ 
&& +\ \frac{1}{4} \lambda_1 v^0_1(x)^4 + \frac{1}{4} \lambda_2
v^0_2(x)^4 + \frac{1}{4} \left(\lambda_{34} - |\lambda_5| \right)
v^0_1(x)^2 v^0_2(x)^2\ +\ \mathrm{V}_0 \; . 
\label{eq:Z2 energy density}
\end{eqnarray}
To simplify our study, we introduce the dimensionless quantities
\begin{equation}
\hat{x}\ =\ \mu_2 x \; ,  \qquad \hat{v}_{1,2}^0(x)\ =\
\frac{v^0_{1,2}(x)}{\eta} \; , \qquad \hat{\mathrm{E}}\ =\ \frac{\lambda_2
  \mathrm{E}}{\mu_2^3} \; , 
\label{eq:Z2 DW Rescaling}
\end{equation}
in order  to rescale the energy  per unit area  of \eqref{eq:Z2 Energy
  per  Unit  Area}  to  be  dimensionless.   Here,  we  introduce  the
convention  that  $\,\hat{}\,$  represents a  dimensionless  quantity.
Performing these rescalings leaves  the dimensionless $\rm Z_2$ energy
density, denoted correspondingly  as $\hat{\mathcal{E}}$, dependent on
the following parameters:
\begin{equation}
\mu^2\ =\ \frac{\mu_1^2}{\mu_2^2} \;, \quad \lambda\ =\
\frac{\lambda_1}{\lambda_2} \; , \quad g\ =\ \frac{\lambda_{34} -
  |\lambda_5|}{2 \lambda_2} \; , \quad \eta\ =\ 
\frac{\mu_2}{\sqrt{\lambda_{2}}} \ . 
\end{equation} 
Also, the vacuum manifold  parameters $v_{1,2}^0$ of \eqref{eq:Z2 Zeta
  1 V1} and \eqref{eq:Z2 Zeta 1 V2}, which are the boundary conditions
on the fields $v_{1,2}^0(x)$, are rescaled, such that
\begin{equation}
\lim_{\hat{x} \to \pm \infty} \hat{v}^0_1(\hat{x})\ =\ \sqrt{
  \frac{\mu^2 - g}{\lambda - g^2} } \ , \qquad 
\lim_{\hat{x} \to   \pm \infty} \hat{v}^0_2(\hat{x})\ =\ 
\pm \sqrt{ \frac{\lambda - \mu^2
    g}{\lambda - g^2} } \; . 
\end{equation}
The   equations    of   motion    for   the   two    rescaled   fields
$\hat{v}^0_{1,2}(\hat{x})$ are found to be
\begin{subequations}
\begin{align}
  \label{eq:Z2 V1 EOM} 
\frac{d^2 \hat{v}^0_1}{d\hat{x}^2}\ &=\ \hat{v}^0_1\; \Big[-\mu^2 +
\lambda (\hat{v}^{0}_1)^2 + g (\hat{v}^{0}_2)^2 \Big] \; ,\\ 
  \label{eq:Z2 V2 EOM} 
\frac{d^2 \hat{v}^0_2}{d\hat{x}^2}\ &=\ \hat{v}^0_2\; \Big[ -1 +
(\hat{v}^{0}_2)^2 + g (\hat{v}^{0}_1)^2 \Big]\ . 
\end{align}
\end{subequations}
As no  analytical solutions  are known for  this particular  system of
ordinary   differential  equations,  we   proceed  by   gradient  flow
techniques  to  minimize  the  energy  per unit  area.  To  make  this
possible,  we truncate  the  interval of  integration of  \eqref{eq:Z2
  Energy per Unit Area} from $(-\infty, \infty)$ to $[-R,R]$, ensuring
that $R$ is chosen  to be much larger than the width  of the kink.  By
making  the range  of integration  symmetric about  $\hat{x} =  0$, we
break the translational symmetry usually exhibited by kink solutions.

In order to perform the numerical analysis, a particular parameter set
$\{\mu^2, \lambda, g\}$ must  be chosen that satisfies the constraints
for  a  bounded-from-below  global  minimum,  as these  are  given  in
\eqref{eq:Z2  Zeta1 Global  Minimum Condition}  and  Table \ref{table:HF
  Convexity  Conditions}. For  convenience, we  state these  results in
terms of the rescaled parameter set:
\begin{equation}
  \label{eq:Z2 DW Parameter Constraints 1 - 3}
\lambda\ >\ g^2 \; ,\qquad g\ <\  \mu^2\ <\ \frac{\lambda}{g} \; ,\quad
\quad g\ >\ - \sqrt{\lambda} \; .   
\end{equation}
Additionally, in order  to satisfy that the sum of  the squares of the
two VEVs $v^0_{1,2}$ is $v^2_{\rm  SM}$, we require that the VEV scale
factor $\eta$ have the value given by
\begin{equation}
  \label{eq:Z2 DW Parameter Constraint 4}
\frac{\eta}{\sqrt{2}}\ =\ 
\sqrt{\frac{\lambda - g^2}{\mu^2 - g + \lambda - \mu^2 g }}
\ v_{\rm SM} \; . 
\end{equation}
Here, we make the observation that a value of $\eta$ can always be
found that ensures condition \eqref{eq:Z2 DW Parameter Constraint 4}
is met for any parameter set $\{\mu^2, \lambda, g \}$, provided that
the members of the parameter set remain non-zero, finite and satisfy
conditions \eqref{eq:Z2 DW Parameter Constraints 1 - 3}. 

\begin{figure}[t!]
\begin{center}
\includegraphics[scale=0.195]{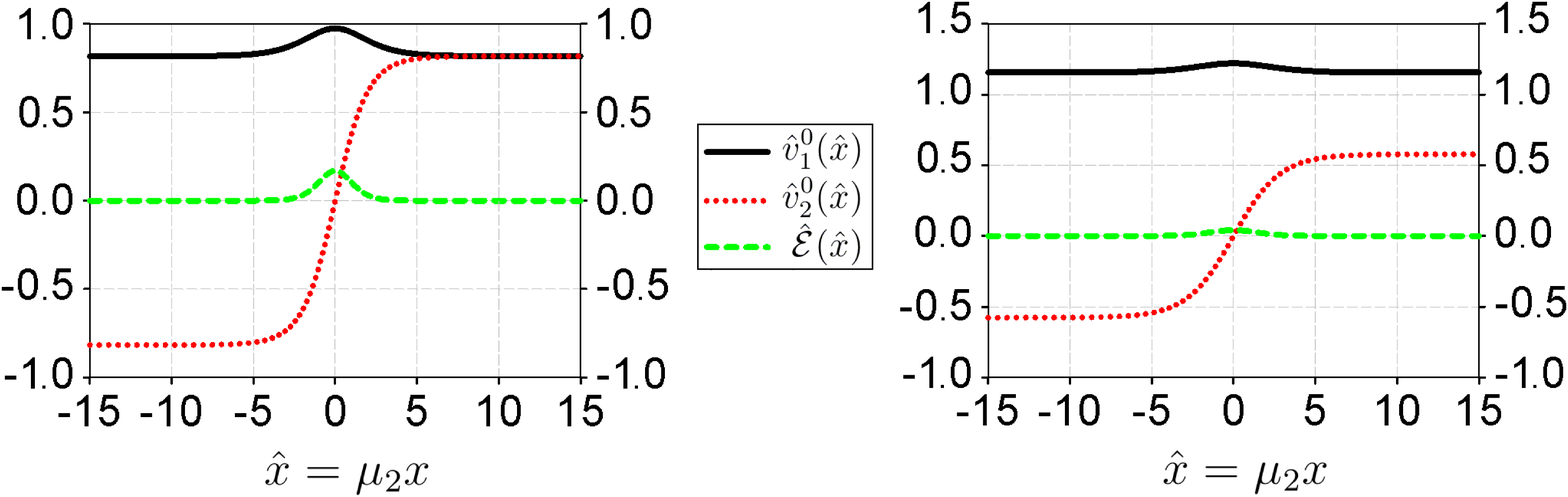} 
\caption{\it  Plots of  $\hat{v}_1^0(\hat{x})$, $\hat{v}_2^0(\hat{x})$
  and  the dimensionless  energy  density $\hat{\mathcal{E}}(\hat{x})$
  for  two  different  valid  parameter  sets  of  the  $\mathrm{Z}_2$
  invariant potential.  The parameter sets  used are $\{1, 1,  0.5 \}$
  (LHS) and $\{1.5, 1, 0.5 \}$ (RHS), and the region of integration
  has $R = 15$.} 
\label{fig:Z2Kink1}
\end{center}
\end{figure}

\begin{figure}[t!]
\begin{center}
\includegraphics[scale=0.225]{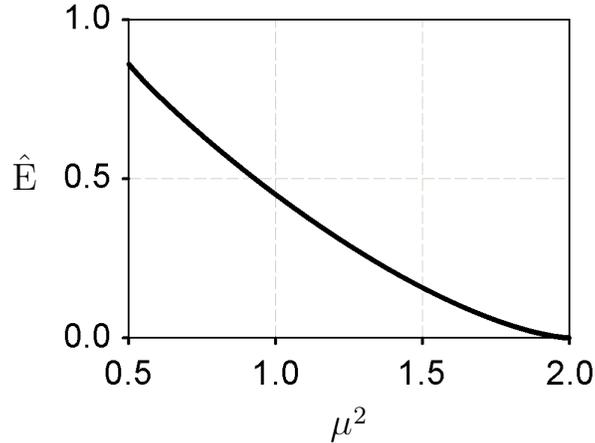}
\caption{\it   Numerical   evaluation  of   the   dependence  of   the
  dimensionless  energy $\hat{\mathrm{E}}$ as  a function  of $\mu^2$,
  for  the  $\mathrm{Z}_2$  invariant  potential.  Here,  we  use  the
  fiducial  values $\lambda  = 1$  and $g  = 0.5$.   Convexity  of the
  potential and global minima conditions for these values require that
  $\mu^2 \in (0.5, 2.0)$.}
\label{fig:Z2KinkEnergy} 
\end{center}
\end{figure}

\begin{figure}[t!]
\begin{center}
\includegraphics[scale=0.195]{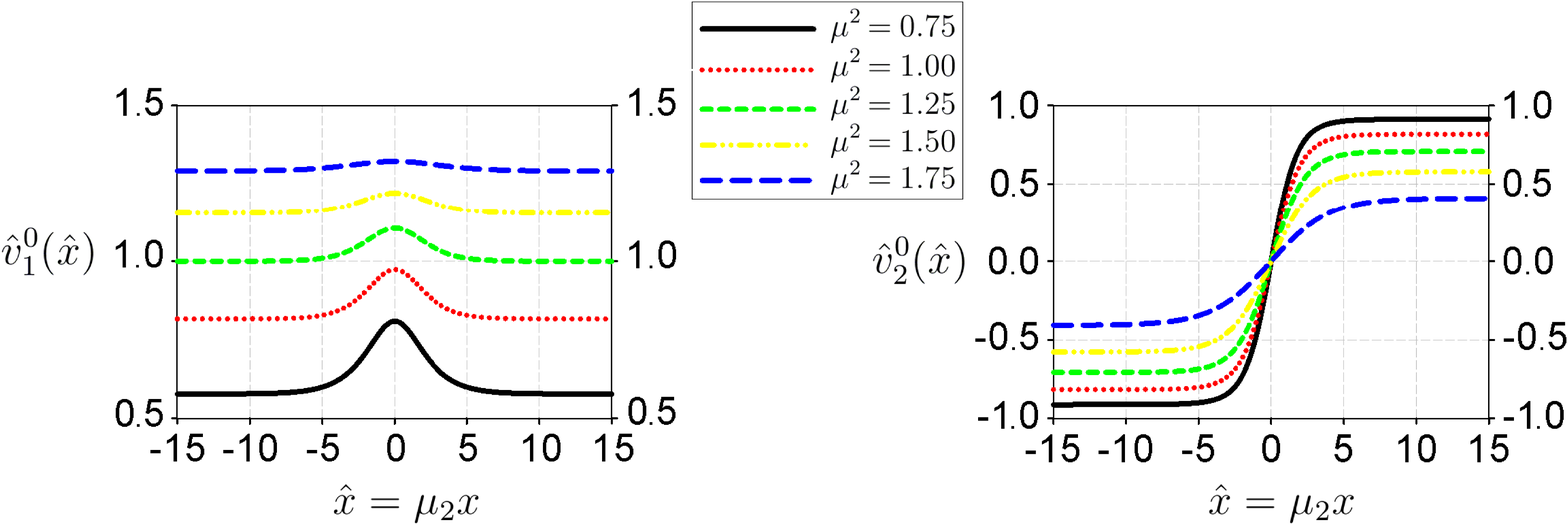} 
\caption{\it Plots comparing various $\hat{v}_1^0(\hat{x})$ (LHS plot)
  and $\hat{v}_2^0(\hat{x})$ (RHS plot) curves for the $\mathrm{Z}_2$
  invariant potential by fixing $\lambda = 1$ and $g = 0.5$, and
  allowing $\mu^2$ to vary, and the region of integration $R = 15$.}
\label{fig:Z2Comparisons}
\end{center}
\end{figure}

We present  two typical solutions in Figure  \ref{fig:Z2Kink1} for the
parameter   sets   $\{\mu^2,   \lambda,   g\}   =   \{1,1,0.5\}$   and
$\{1.5,1,0.5\}$. We  also show the  general form of  the dimensionless
energy  $\rm \hat{E}$  in Figure  \ref{fig:Z2KinkEnergy}  and directly
compare several different solutions in Figure \ref{fig:Z2Comparisons},
as  a function  of $\mu^2$.   From Figures  \ref{fig:Z2KinkEnergy} and
\ref{fig:Z2Comparisons}, we  see that as $\mu^2$  approaches its lower
bound,  $\mu^2 \to  g$, the dimensionless  energy  approaches a
finite value. In  the limit $g \to 0$, we find  that this finite value
is   the  familiar   value  $\frac{2}{3}\sqrt{2}$   [cf.    (1.20)  in
  \cite{Vachaspati}],  and  the   kink  width  decreases  and  becomes
small. Conversely, we see that  as $\mu^2$ approaches its upper bound,
i.e.~as $\mu^2  \to \frac{\lambda}{g}$, the  dimensionless energy $\rm
\hat{E}$ tends towards  zero, the kink width increases  and the energy
density becomes  delocalized. Therefore, the  dimensionless energy can
be  made vanishingly small  for appropriate  choices of  the parameter
set.  This  is a feature  that can be  exploited to avoid  domain wall
domination by making the mass per unit area of the walls ultra-light.

\subsubsection{CP1 Domain Walls}
\label{CP1 Domain Walls}

From our  analysis in Section  \ref{CP1 Symmetry}, the  2HDM potential
which is invariant  under the CP1 symmetry can  exhibit a disconnected
vacuum manifold, the  components of which are not  linked by the gauge
symmetries of  the theory,  provided both VEVs  of the  Higgs doublets
$\phi_{1,2}$ and  the relative phase between the  doublets that create
the   neutral   vacuum   global   minimum   solution   are   non-zero,
i.e.~$v_{1,2}^0  \neq  0$,  $v_2^+  =  0$  and  $\xi  \neq  0$.   This
spontaneous violation of CP  is only apparent within the CP1-invariant
2HDM when the determinant of the matrix $\rm N_{\mu \nu}$ vanishes, as
shown in Section  \ref{Utilizing CP1 zeta}. However, as  we have shown
in  Section  \ref{CP1 Symmetry},  a  neutral  vacuum  solution with  a
CP-conserving global minimum is  also possible, where $\rm det[ N_{\mu
    \nu}]  \neq 0$,  i.e.~a global  minimum with  $v^0_{1,2}  \neq 0$,
$v_2^{+} = 0$ and $\xi  = 0$. Therefore, during our numerical analysis
we  are careful  to choose  parameter  sets that  give global  minimum
neutral vacuum solutions with spontaneous CP violation.

Let  us  now  investigate  an one-dimensional,  time-independent  kink
solution for  the CP1 symmetry. In  order to find such  a solution, we
will use an ansatz for the two Higgs doublets given by
\begin{equation}
\phi_{1}(x)\ =\ \frac{1}{\sqrt{2}} \left( \begin {array}{c}
  0\\\noalign{\medskip}v^0_{1}(x)\end {array} \right) \; , \quad
\phi_{2}(x)\ =\ \frac{1}{\sqrt{2}} \left( \begin {array}{c}
  0\\\noalign{\medskip}v^0_{2}(x)e^{i \xi(x)}\end {array} \right) \; ,
\label{eq:CP1 Domain Wall Ansatz}
\end{equation}
where the coordinate $x$ describes the spatial dimension perpendicular
to the plane of the domain wall. The energy per unit area associated
with the kink solution is again given by \eqref{eq:Z2 Energy per Unit
  Area}, where the energy density for the CP1-invariant 2HDM is given
by
\begin{eqnarray}
\mathcal{E}(x) \! &=& \! \frac{1}{2} \left(\frac{d v_1^0}{dx}
\right)^2 + \frac{1}{2} \left(\frac{d v_2^0}{dx} \right)^2 +
\frac{1}{2} v_2^0(x)^2 \left(\frac{d \xi}{dx} \right)^2 - \frac{1}{2}
\mu_1^2 v^0_1(x)^2 - \frac{1}{2} \mu_2^2 v^0_2(x)^2 \nonumber \\ &&+
\frac{1}{4} \lambda_1 v^0_1(x)^4 + \frac{1}{4} \lambda_2 v^0_2(x)^4 +
\frac{1}{4} \big( \lambda_{34} + \lambda_5 \cos 2\xi(x) \big)
v^0_1(x)^2 v^0_2(x)^2 \nonumber \\ &&+ \Big(-m_{12}^2 +
\frac{1}{2}\lambda_6 v_1^0(x)^2 + \frac{1}{2}\lambda_7 v_2^0(x)^2
\Big) v_1^0(x) v_2^0(x) \cos \xi(x) + \mathrm{V}_0 \; .
\end{eqnarray}
By rescaling \eqref{eq:Z2 Energy per Unit Area} to be dimensionless
for the CP1 energy density, we again make use of the dimensionless
quantities of \eqref{eq:Z2 DW Rescaling}. Performing these rescalings
leaves the dimensionless CP1 energy density dependent on the following
parameters:
\begin{equation}
\mu^2 = \frac{\mu_1^2}{\mu_2^2} \; , \;\; m^2 =
\frac{m_{12}^2}{\mu_2^2} \; , \;\; \lambda =
\frac{\lambda_1}{\lambda_2} \; , \;\; g_{34} =
\frac{\lambda_{34}}{\lambda_2} \; , \;\; g_k =
\frac{\lambda_k}{\lambda_2} \;\; \big({\rm for}\; k = 5,6,7 \big) \; ,
\; \; \eta = \frac{\mu_2}{\sqrt{\lambda_2}} \; . 
\end{equation}
It  is also  useful to  introduce the  parameter $\bar{g}  =  g_{34} -
g_5$. The parameter  set for the CP1-invariant model  then reduces and
becomes $\{\mu^2, m^2, \lambda,  g_{34}, g_5, g_6, g_7 \}$. Similarly,
the vacuum manifold parameters  $v_{1,2}^0$ and $\xi$ of \eqref{eq:CP1
  v1}, \eqref{eq:CP1  v2} and \eqref{eq:CP1  xi value}, which  are the
boundary  conditions on  the fields  $v_{1,2}^0(x)$ and  $\xi(x)$, are
rescaled, so as to give
\begin{subequations}
\begin{align}
\lim_{\hat{x} \to \pm \infty}
\hat{v}^0_1(\hat{x})\ &=\ \sqrt{\frac{2\left( g_6 g_7 - \bar{g}
    g_5\right) + 2\left(2g_5 - g_7^2\right) \mu^2 + 2\left(\bar{g} g_7
    - 2g_6 \right) m^2}{g_5 \left(4\lambda - \bar{g}^2 \right) -
    2\lambda g_6^2 - 2 g_7^2 + 2 \bar{g} g_6 g_7}} \ , \\ 
\lim_{\hat{x} \to \pm \infty}
\hat{v}^0_2(\hat{x})\ &=\ \sqrt{\frac{2\left(2\lambda g_5 - g_6^2
    \right) + 2\left(g_6 g_7 - \bar{g} g_5 \right) \mu^2 +
    2\left(\bar{g} g_6 - 2\lambda g_7 \right) m^2}{g_5 \left(4\lambda
    - \bar{g}^2 \right) - 2\lambda g_6^2 - 2 g_7^2 + 2 \bar{g} g_6
    g_7}} \; , \\ 
\lim_{\hat{x} \to \pm \infty} \xi(\hat{x})\ &=\ \pm \mathrm{arccos}
\left( \frac{2m^2 - g_6 (\hat{v}_1^0)^2 - g_7 (\hat{v}_2^0)^2}{2 g_5
  \hat{v}_1^0 \hat{v}_2^0} \right) \; . \label{eq:CP1 Xi Boundary
  Condition} 
\end{align}
\end{subequations}
The equations of motion for the three rescaled fields
$\hat{v}_{1,2}^0(\hat{x})$ and $\xi(\hat{x})$ are 
\begin{subequations}
\begin{align}
&\frac{d^2 \hat{v}^0_1}{d\hat{x}^2} \ =\  
\left[-\mu^2 + \lambda (\hat{v}^{0}_1)^2 + \frac{1}{2}
\left(g_{34} + g_5 \cos 2 \xi \right) (\hat{v}^{0}_2)^2 +
\frac{3}{2} g_6 \hat{v}^{0}_1 \hat{v}^{0}_2 \cos \xi  \right]\,
\hat{v}^0_1\nonumber\\
& \hspace{2cm} - \left[ m^2 - \frac{1}{2} g_7
(\hat{v}^{0}_2)^2 \right]  \hat{v}^{0}_2 \cos \xi \, , \\ 
&\frac{d^2 \hat{v}^0_2}{d\hat{x}^2} \ =\ \left[ -1  +\
(\hat{v}^{0}_2)^2 + \frac{1}{2} \left(g_{34} + g_5 \cos 2 \xi
\right] (\hat{v}^{0}_1)^2 + \frac{3}{2} g_7 \hat{v}^{0}_1
\hat{v}^{0}_2 \cos \xi \right] \hat{v}^0_2 \nonumber \\ 
&\hspace{2cm}
-\ \left[ m^2 -
\frac{1}{2} g_6 (\hat{v}^{0}_1)^2 \right] \hat{v}^{0}_1 \cos \xi\
+\ \left(\frac{d\xi}{d\hat{x}}\right)^2 \hat{v}^0_2 \; , \\
&(\hat{v}^0_2)^2 \frac{d^2 \xi}{d\hat{x}^2}\ +\ 2 \hat{v}^0_2
\left(\frac{d \xi}{d\hat{x}} \right) \left(\frac{d
  \hat{v}^0_2}{d\hat{x}} \right)\ =\nonumber\\
& \hspace{2cm} -\hat{v}^0_1 \hat{v}^0_2 \sin \xi
\left[ g_5 \hat{v}^0_1 \hat{v}^0_2 \cos \xi - m^2 + \frac{1}{2}
g_6 (\hat{v}^{0}_1)^2 + \frac{1}{2} g_7 (\hat{v}^{0}_2)^2 \right] . 
\end{align}
\end{subequations}

\begin{figure}[t!]
\begin{center}
\includegraphics[scale=0.198]{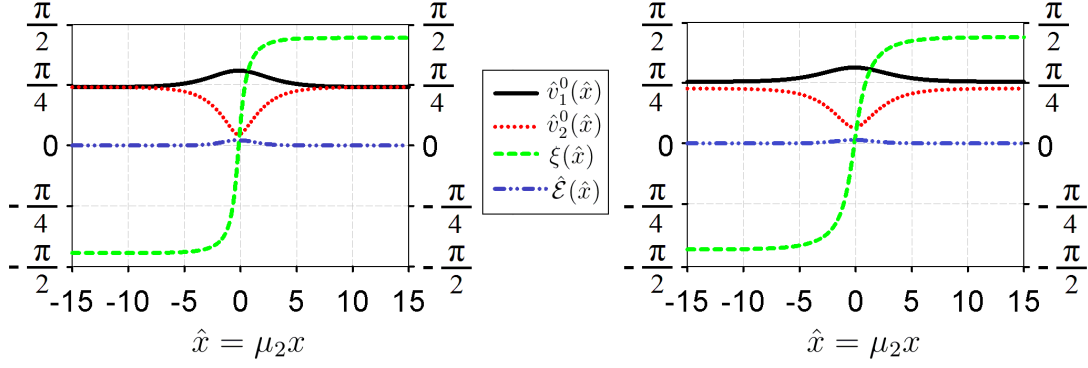} 
\caption{\it Plots of $\hat{v}_1^0(\hat{x})$, $\hat{v}_2^0(\hat{x})$,
  $\xi(\hat{x})$ and the dimensionless energy density
  $\hat{\mathcal{E}}(\hat{x})$ for two different valid parameter sets
  of the CP1-invariant potential. The parameter sets used are $\{1,
  0.1, 1, 2.5, 1, 0, 0\}$ (LHS) and $\{1, 0.1, 1, 2.5, 1, -0.15,
  0.15\}$ (RHS). The region of integration has $R = 15$.} 
\label{fig:CP1Kink1}
\end{center}
\end{figure}

\noindent
As with  the $\rm  Z_2$ domain wall  study, no analytical  solution is
found  to  these equations  of  motion  and  we therefore  proceed  by
gradient flow  techniques. Due to the number  of individual parameters
one  may tune  within the  confines of  the CP1  convexity  and minima
conditions,  relationships  between  the  parameters  are  in  general
complicated and so we end our  CP1 domain wall study by presenting two
typical solutions  in Figure  \ref{fig:CP1Kink1}. However, we  do note
two cases determined by specific choices of the parameter set. For the
case  $\displaystyle  \lim_{\hat{x}  \to  \pm \infty}  \xi(\hat{x})  =
\frac{\pi}{2}$,  which is  guaranteed  if $g_6  (\hat{v}_1^0)^2 +  g_7
(\hat{v}_2^0)^2 = 2m^2$, the CP1  symmetry domain wall reverts back to
a $\rm Z_2$ style domain wall by use of the reparameterization $\phi_2
\to i \phi_2$, as discussed in Section \ref{CP1 Topology}. An explicit
example can  be seen for  the $\rm Z_2$ symmetry  potential parameters
constraints,  $m_{12}^2  =  \lambda_6  =  \lambda_7 =  0$.  Also,  the
dimensionless energy  $\rm \hat{E}$ can be made  vanishingly small for
certain valid  choices of  the parameter set  that leave the  limit of
\eqref{eq:CP1 Xi Boundary  Condition} still finite but $  \ll 1$. This
is  consistent  with the  case  $\displaystyle  \lim_{\hat{x} \to  \pm
  \infty} \xi(\hat{x})  = 0$ in which spontaneous  CP violation ceases
and subsequently  there is  no domain wall  solution, as  discussed in
Section \ref{CP1 Topology}.

\subsubsection{CP2 Domain Walls}

From our  analysis in Section  \ref{CP2 Symmetry}, the  2HDM potential
which is invariant  under the CP2 symmetry can  exhibit a disconnected
vacuum manifold, the  components of which are not  linked by the gauge
symmetries of  the theory,  provided both VEVs  of the  Higgs doublets
$\phi_{1,2}$ that  create the  neutral vacuum global  minimum solution
are non-zero, i.e.~$v_{1,2}^0 \neq 0$, $v_2^+ =  0$. This scenario is
only apparent  within the CP2-invariant  2HDM when the  determinant of
the  matrix   $\rm  N_{\mu  \nu}$   vanishes,  as  shown   in  Section
\ref{Utilizing CP2 zeta}

We now investigate  an one-dimensional, time-independent kink solution
for the CP2 symmetry. In order to find such a solution, we will use an
ansatz for the  two Higgs doublets given by  \eqref{eq:CP1 Domain Wall
  Ansatz}. The energy per unit  area associated with the kink solution
is  again given  by \eqref{eq:Z2  Energy  per Unit  Area}. The  energy
density for the CP2-invariant 2HDM is given by
\begin{eqnarray}
\mathcal{E}(x) \! &=& \! \frac{1}{2} \left(\frac{d v_1^0}{dx}
\right)^2 + \frac{1}{2} \left(\frac{d v_2^0}{dx} \right)^2 +
\frac{1}{2} v_2^0(x)^2 \left(\frac{d \xi}{dx} \right)^2 - \frac{1}{2}
\mu_1^2 \big( v^0_1(x)^2 + v^0_2(x)^2 \big) \nonumber \\ 
&&+\ \frac{1}{4} \lambda_1 \big( v^0_1(x)^4 + v^0_2(x)^4 \big) +
\frac{1}{4} \big( \lambda_{34} + R_5 \cos 2\xi(x) - I_5 \sin 2\xi(x)
\big) v^0_1(x)^2 v^0_2(x)^2 \nonumber \\ 
&&+\ \frac{1}{2} v_1^0(x) v_2^0(x) \big( v_1^0(x)^2 - v_2^0(x)^2 \big)
\big( R_6 \cos \xi(x) - I_6 \sin \xi(x) \big)\ +\ \mathrm{V}_0 \; . 
\end{eqnarray}
Again, it is useful to rescale \eqref{eq:Z2 Energy per Unit Area} to be
dimensionless for the CP2 energy density, and so we introduce 
\begin{equation}
\hat{x}\ =\ \mu_1 x \; ,  \qquad \hat{v}_{1,2}^0(x)\ =\
\frac{v^0_{1,2}(x)}{\eta} \; ,\qquad \hat{\mathrm{E}}\ =\ \frac{\lambda_1
  \mathrm{E}}{\mu_1^3} \; . 
\label{eq:CP2 DW Rescaling}
\end{equation}
Performing  these  rescalings  leaves  the  dimensionless  CP2  energy
density $\hat{\cal E}$ dependent on the following parameters:
\begin{equation}
g_k\ =\ \frac{R_k}{\lambda_1} \ \ \big({\rm for}\; k = 3,4,5,6 \big) \; , 
\quad h_k\ =\ \frac{I_k}{\lambda_1} \ \ \big({\rm for}\; k = 5,6 \big) \;, 
\quad \eta\ =\ \frac{\mu_1}{\sqrt{\lambda_1}} \; .
\end{equation}
Therefore,  the  parameter set  for  the  CP2-invariant model  becomes
$\{g_3,  g_4,  g_5,  g_6,  h_5,  h_6\}$.  Also,  the  vacuum  manifold
parameters $v_{1,2}^0$  and $\xi$ of  \eqref{eq:CP2 v1}, \eqref{eq:CP2
  v2} and \eqref{eq:CP2 xi}, which  are the boundary conditions on the
fields $v_{1,2}^0(x)$ and $\xi(x)$, are rescaled as follows:
\begin{subequations}
\begin{align}
\hat{v}^0_1(\hat{x})\ &\to\ \left\{ \begin {array}{cl} 
\displaystyle \sqrt{\left(\frac{2}{2 + g_3 - \hat{\zeta}} \right)
  \left(1 - \frac{1}{\hat{\mathrm{B}}} \, \frac{2 h_5 g_6 - 2 h_6 (g_4
    + g_5 + \hat{\zeta})}{4 h_6 g_6 - h_5(2 - g_3 + \hat{\zeta})}
  \right)},\ 
\mbox{as $\hat{x} \to -\infty$} \bigskip \\ 
\displaystyle \sqrt{\left(\frac{2}{2 + g_3 - \hat{\zeta}} \right)
  \left(1 + \frac{1}{\hat{\mathrm{B}}} \, \frac{2 h_5 g_6 - 2 h_6 (g_4
    + g_5 + \hat{\zeta})}{4 h_6 g_6 - h_5(2 - g_3 + \hat{\zeta})}
  \right)},\  \mbox{as $\hat{x} \to +\infty$} \end {array} \right. \; ,\\ 
& \nonumber \\
\hat{v}^0_2(\hat{x})\ &\to\ \left\{ \begin {array}{cl} 
\displaystyle -\sqrt{\left(\frac{2}{2 + g_3 - \hat{\zeta}} \right)
  \left(1 + \frac{1}{\hat{\mathrm{B}}} \, \frac{2 h_5 g_6 - 2 h_6 (g_4
    + g_5 + \hat{\zeta})}{4 h_6 g_6 - h_5(2 - g_3 + \hat{\zeta})}
  \right)},\ \mbox{as $\hat{x} \to -\infty$} \bigskip \\  
\displaystyle \sqrt{\left(\frac{2}{2 + g_3 - \hat{\zeta}} \right)
  \left(1 - \frac{1}{\hat{\mathrm{B}}} \, \frac{2 h_5 g_6 - 2 h_6 (g_4
    + g_5 + \hat{\zeta})}{4 h_6 g_6 - h_5(2 - g_3 + \hat{\zeta})}
  \right)},\ \mbox{as $\hat{x} \to +\infty$} \end {array} \right. \; ,\\ 
& \nonumber \\
\lim_{\hat{x} \to \pm \infty} \xi(\hat{x})\ &=\ \mathrm{arctan} \left(
\frac{(g_4 + g_5 + \hat{\zeta}) h_6 - h_5 g_6}{h_5 h_6 - (g_4 - g_5 +
  \hat{\zeta}) g_6} \right) \; , 
\end{align}
\end{subequations}
where the parameter $\hat{\mathrm{B}}$ is defined as
\begin{equation}
\hat{\mathrm{B}} = \ \sqrt{\left(\frac{h_5 g_6 - h_6(g_4 + g_5 +
    \hat{\zeta})}{g_6 (g_4 - g_5 + \hat{\zeta}) - h_5 h_6} \right)^2
  +\ \left(\frac{2 h_5 g_6 - 2 h_6 (g_4 + g_5 + \hat{\zeta})}{4 h_6
    g_6 - h_5(2 - g_3 + \hat{\zeta})} \right)^2 +\ 1}\  .
\end{equation}
These   boundary  conditions  depend   on  the   non-trivial  Lagrange
multiplier implemented  to produce  the neutral vacuum  solution. This
Lagrange multiplier satisfies the cubic equation:
\begin{eqnarray}
&&\hat{\zeta}^3 + \big( 2 - g_3 + 2 g_4 \big) \hat{\zeta}^2 + \big(
  2g_4 (2 - g_3) + g_4^2 - g_5^2 - h_5^2 - 4g_6^2 - 4h_6^2 \big)
  \hat{\zeta} \nonumber \\ 
&& + (2 - g_3)(g_4^2 - g_5^2 - h_5^2) - 4 (g_6^2 + h_6^2) (g_4 - g_5) 
  + 8 h_6 (h_5 g_6 - g_5 h_6)\ =\ 0 \; .  
\label{eq:CP2 Domain Wall Singular Condition}
\end{eqnarray}
In order to find a valid parameter set, we start by choosing parameter
values  that satisfy  the  CP2 convexity  conditions  (shown in  Table
\ref{table:CP Convexity Conditions}) and then solve \eqref{eq:CP2 Domain
  Wall  Singular  Condition} to  find  the  three  possible values  of
$\hat{\zeta}$. We  then find  the rescaled vacuum  manifold parameters
which  correspond to  each  $\hat{\zeta}$ solution,  and determine  if
these solutions  correspond to local minima, i.e.~we  require that the
CP2    Hessian   matrix   ${\rm    H}$   in    \eqref{eq:CP2   Hessian
  H11}--\eqref{eq:CP2 Hessian  H33} be positive  definite.  If~they do
indeed relate  to minima, we calculate  the value of  the potential at
these  extremal  points  to  determine  which  $\hat{\zeta}$  solution
generates the global minimum.

\begin{figure}[t!]
\begin{center}
\includegraphics[scale=0.225]{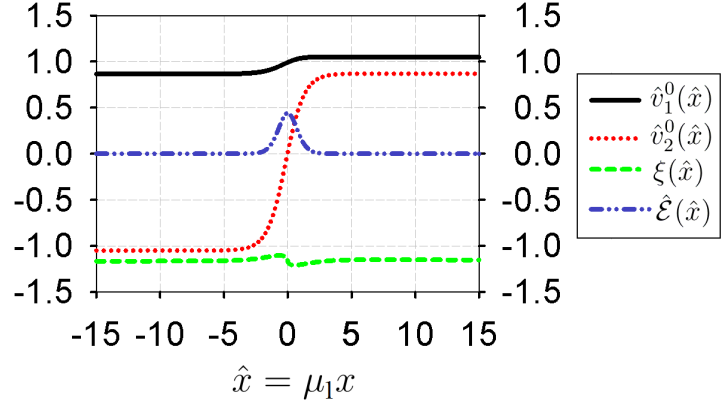}
\caption{\it    Numerical    estimates   of    $\hat{v}_1^0(\hat{x})$,
  $\hat{v}_2^0(\hat{x})$, $\xi(\hat{x})$  and the dimensionless energy
  density $\hat{\mathcal{E}}(\hat{x})$,  for a valid  parameter set of
  the CP2-invariant  potential. The  input parameter set  is $\{0.125,
  0.375, 0.125,  0.125, -0.25, -0.25\}$. The region  of integration
  has $R = 15$.} 
\label{fig:CP2Kink1}
\end{center}
\end{figure}

The   equations   of   motion    for   the   three   rescaled   fields
$\hat{v}_{1,2}^0(\hat{x})$ and $\xi(\hat{x})$ are:
\begin{subequations}
\begin{align}
&\frac{d^2 \hat{v}^0_1}{d\hat{x}^2}\ =\ 
\hat{v}^0_1 \left[-1 + (\hat{v}^{0}_1)^2 + \frac{1}{2} \left(g_{34} + 
g_5 \cos 2 \xi - h_5 \sin 2 \xi \right) (\hat{v}^{0}_2)^2 \right] \nonumber \\
& \hspace{2cm} +\ \frac{1}{2} \hat{v}^{0}_2 \left(3 (\hat{v}^{0}_1)^2
\ -\ (\hat{v}^{0}_2)^2 \right) \left(g_6 \cos \xi - h_6 \sin \xi
\right) \; , \\ 
&\frac{d^2 \hat{v}^0_2}{d\hat{x}^2}\ =\ \hat{v}^0_2 \left[-1 + 
(\hat{v}^{0}_2)^2 + \frac{1}{2} \left(g_{34} + 
g_5 \cos 2 \xi - h_5 \sin 2 \xi \right) (\hat{v}^{0}_1)^2 + 
\left( \frac{d\xi}{d\hat{x}} \right)^2 \right] \nonumber \\
& \hspace{2cm} +\ \frac{1}{2} \hat{v}^{0}_1 \left((\hat{v}^{0}_1)^2 - 
3(\hat{v}^{0}_2)^2 \right) \left(g_6 \cos \xi - h_6 \sin \xi \right) \; , \\
& (\hat{v}^0_2)^2 \frac{d^2 \xi}{d\hat{x}^2}\ +\ 
2 \hat{v}_2^0 \left(\frac{d \hat{v}^0_2}{d\hat{x}} \right) 
\left(\frac{d \xi}{d\hat{x}}\right) \ =\ - \frac{1}{2} \hat{v}^0_1
\hat{v}_2^0 \left[\hat{v}^0_1 \hat{v}^0_2 \left(g_5 \sin 2 \xi + h_5
\cos 2\xi \right) \right. \nonumber \\ 
&\hspace{2cm}
\left.
+\! \left((\hat{v}^{0}_1)^2 \!-\! (\hat{v}^{0}_2)^2 \right) \left(g_6
\sin \xi \!+\! h_6 \cos \xi \right) \right] \; .  
\end{align}
\end{subequations}
We can  obtain numerical  solutions to these  equations of  motion, by
making use of gradient flow techniques.  Since there is a large number
of individual  parameters that could  vary within the confines  of the
CP2  convexity  and   minima  conditions,  relationships  between  the
parameters are, in general, complicated. Instead, we present a typical
example of  CP2 domain walls,  as shown in  Figure \ref{fig:CP2Kink1}.
As  with the  previous two  case,  we observe  that the  dimensionless
energy $\rm \hat{E}$  can be made vanishingly small  for certain valid
choices of the parameter set, such as allowing $g_{5,6}$ and $h_{5,6}$
to tend to zero.

\subsection{Vortices}

We now turn our attention  to other topological defects which may form
in  the 2HDM,  such as  vortices.  Whilst  vortex solutions  have been
discussed  in  the  2HDM  \cite{Bimonte:1994qh}, these  vortices  were
generated by the SM gauge  group, whereas the vortices we discuss here
are  generated  solely  by  the  spontaneous  breaking  of  the  ${\rm
  U(1)}_{\rm  PQ}$ and CP3  accidental symmetries  which the  2HDM can
exhibit for specific constraints on the parameters of the potential.

Vortices are often regarded as the most favourable topological defect,
since their  energy density does  not grow relative to  the background
and  so  for sufficiently  small  initial  energy densities,  vortices
behave   benignly   and   can   comply   with   current   cosmological
observations. Due to the  axially symmetric, one-dimensional and typically high mass density characteristics of a cosmic string, strings can act as a gravitational lens~\cite{1984ApJ...282L..51V,1985ApJ...288..422G}  and  searches are
already under way to detect  possible vortices which may be within the
current horizon, e.g. using precision cosmic microwave background data
from experiments such as WMAP~\cite{jeong-2005-624,Battye:2010xz}.

As  with our  domain  wall study,  we  do not  study the  cosmological
implications of  the 2HDM's  vortices, which we  reserve for  a future
study, but  focus on presenting  an overview of typical  solutions and
determining whether  or not the energy  per unit length  of the vortex
can be made to be vanishingly  small for specific and valid choices of
the model parameters.

\subsubsection{$\mathrm{U(1)}_{\mathrm{PQ}}$ Vortices}
\label{U1 Strings}

From our  analysis in Section \ref{U(1) Symmetry},  the 2HDM potential
which  is invariant under  the global  Peccei-Quinn U(1)  symmetry can
exhibit a  non-simply connected vacuum manifold provided  both VEVs of
the Higgs doublets $\phi_{1,2}$  that create the neutral vacuum global
minimum solution  are non-zero, i.e.  $v_{1,2}^0 \neq 0$ and  $v_2^+ =
0$.     This    scenario     is    only     apparent     within    the
$\mathrm{U(1)}_{\mathrm{PQ}}$ invariant  2HDM when the  determinant of
the  matrix   $\rm  N_{\mu  \nu}$   vanishes,  as  shown   in  Section
\ref{sec:U1LMsingular}.

In order  to find  a two-dimensional time-independent  vortex solution
for the $\rm U(1)_{PQ}$ symmetry, we  use the ansatz for the two Higgs
doublets:
\begin{equation}
\phi_1 (r)\ =\ \frac{1}{\sqrt{2}} \left( \begin {array}{c}
  0\\\noalign{\medskip}v^0_1(r)\end {array} \right) \; , \qquad 
\phi_2 (r,\chi)\ =\ \frac{1}{\sqrt{2}} \left( \begin {array}{c}
  0\\\noalign{\medskip}v^0_2(r) e^{i n \chi}\end {array} \right) \; , 
\end{equation}
where the coordinate $r$ describes the space radially outward from the
core of  the vortex, and $\chi$  is an azimuthal  angle which accounts
for the  winding of  the vortex, with  winding number $n$.  Using this
ansatz, the energy per unit length of the system is then:
\begin{equation}
\mathrm{E}\ =\ 2 \pi \int_0^\infty \!\!\!\!\! rdr\;\; \mathcal{E}\!
\left( \phi_1, \phi_2 \right) \; , 
\label{eq:U1 Energy Integral}
\end{equation}
where  the   energy  density  for   the  $\mathrm{U(1)}_{\mathrm{PQ}}$
invariant 2HDM is given by
\begin{eqnarray}
\mathcal{E}\! \left(r \right) &=& \frac{1}{2}
\left(\frac{d v^0_1}{dr} \right)^2 + \frac{1}{2} \left(\frac{d
  v^0_2}{dr} \right)^2 + \frac{n^2}{2r^2} v^0_2(r)^2 - \frac{1}{2}
\mu_1^2 v^0_1(r)^2 - \frac{1}{2} \mu_2^2 v^0_2(r)^2 \nonumber \\ 
&&+ \frac{1}{4} \lambda_1 v^0_1(r)^4 + \frac{1}{4} \lambda_2
v^0_2(r)^4 + \frac{1}{4} \lambda_{34} v^0_1(r)^2 v^0_2(r)^2 +
\mathrm{V}_0 \; . 
\end{eqnarray}
For this type  of energy density, the integral  of \eqref{eq:U1 Energy
  Integral} is  logarithmically divergent in  $r$, so we  truncate the
region of  integration from $[0,\infty)$ to  $[0, R]$, where  $R$ is a
  cut off radius~\cite{Vilenkin}. To once again simplify our study, we
  rescale the  energy per  unit length of  the vortex  in \eqref{eq:U1
    Energy   Integral}  to   be  dimensionless   by   introducing  the
  dimensionless quantities
\begin{equation}
\hat{r}\ =\ \mu_2 r \; , \quad  
\hat{v}_{1,2}^0(r)\ =\ \frac{v_{1,2}^0(r)}{\eta} \; ,\quad
\hat{\mathrm{E}}\ =\ \frac{\mathrm{E}}{2 \pi \eta^2} \; .
\end{equation}
With the  help of these rescalings, the  dimensionless $\rm U(1)_{PQ}$
energy density derived from  \eqref{eq:U1 Energy Integral} depends now
on the following parameters:
\begin{equation}
\mu^2\ =\ \frac{\mu_1^2}{\mu_2^2} \; , \quad  \lambda\ =\
\frac{\lambda_1}{\lambda_2} \; , \quad g\ =\
\frac{\lambda_{34}}{2\lambda_2} \; , \quad \eta\ =\
\frac{\mu_2}{\sqrt{\lambda_{2}}} \; . 
\end{equation}
Having      rescaled      the      vacuum     manifold      parameters
$\hat{v}_{1,2}^0(\hat{r})$,  we  require  that  these  approach  their
corresponding VEVs given in~\eqref{eq:VEV1 U1} and~\eqref{eq:VEV2 U1},
as $r \to \infty$.  To be precise, we impose the boundary conditions:
\begin{subequations}
\begin{align}
\left. \frac{d \hat{v}^0_1}{d\hat{r}}\right|_{\hat{r} = 0}\ =\ 0 \;,
&\qquad \lim_{\hat{r} \to \infty} \hat{v}^0_1(\hat{r})\ =\ 
\sqrt{ \frac{\mu^2 - g}{\lambda - g^2} } \ , \\
\lim_{\hat{r} \to 0} \hat{v}^0_2(\hat{r})\ =\ 0 \; ,&\qquad 
\lim_{\hat{r} \to \infty} \hat{v}^0_2(\hat{r})\ =\ 
\sqrt{ \frac{\lambda - \mu^2 g}{\lambda - g^2} } \ .
\end{align}
\end{subequations}
These conditions  force $\hat{v}^0_2(\hat{r})$  to be regular  for all
values   of  $\hat{r}$  and   require  $\hat{v}^0_1(\hat{r})$   to  be
continuous and radially symmetric. The equations of motion for the two
rescaled fields $\hat{v}^0_{1,2}(\hat{r})$ are found to be
\begin{subequations}
\begin{align}
\frac{d^2 \hat{v}^0_1}{d\hat{r}^2}\ &=\ 
\hat{v}^0_1\; \big(-\mu^2 + \lambda (\hat{v}^{0}_1)^2 + g
(\hat{v}^{0}_2)^2 \big)\ , \\
\frac{d^2 \hat{v}^0_2}{d\hat{r}^2} \ +\ \frac{1}{\hat{r}} \frac{d
  \hat{v}^0_2}{d\hat{r}}\ &=\ \hat{v}^0_2\; \Big(-1 +
\frac{n^2}{\hat{r}^2} + (\hat{v}^{0}_2)^2 + g (\hat{v}^{0}_1)^2 \Big)\; . 
\end{align}
\end{subequations}
As  is typical  of  vortex  studies, no  analytical  solutions to  the
equations of  motion are  found and  so we make  use of  gradient flow
numerical techniques.

For  our numerical  analysis,  we choose  a  particular parameter  set
$\{\mu^2,  \lambda,  g, n\}$  that  satisfies  the  constraints for  a
bounded-from-below global minimum, which  are of exactly the same form
as for the $\rm Z_2$ symmetry in \eqref{eq:Z2 DW Parameter Constraints
  1 - 3}.  In order to  satisfy that $(v^0_1)^2 + (v^0_2)^2 = v^2_{\rm
  SM}$, we  require that  the VEV scale  factor $\eta$ have  the value
given by \eqref{eq:Z2 DW Parameter Constraint 4}. We note that a value
of $\eta$ can  always be found that ensures  condition \eqref{eq:Z2 DW
  Parameter  Constraint 4}  is  met for  any  parameter set  $\{\mu^2,
\lambda,  g, n \}$,  provided that  the members  of the  parameter set
remain non-zero and finite, and satisfy the conditions in \eqref{eq:Z2
  DW Parameter Constraints 1 - 3}.

\begin{figure}[t!]
\begin{center}
\includegraphics[scale=0.22]{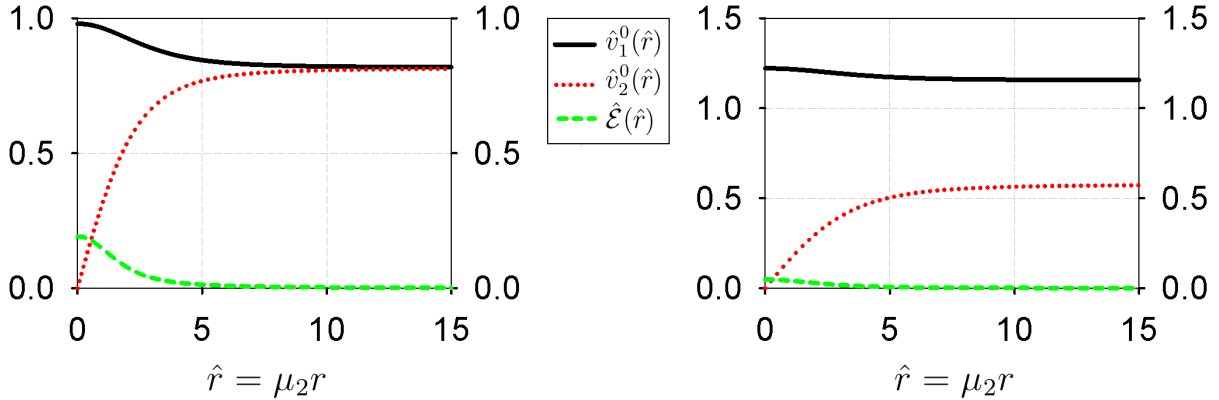}
\caption{\it Plots of $\hat{v}_1^0(\hat{r})$, $\hat{v}_2^0(\hat{r})$
  and the dimensionless energy density $\hat{\mathcal{E}}(\hat{r})$
  for two different valid parameter sets of the
  $\mathrm{U(1)}_{\mathrm{PQ}}$ invariant potential. The parameter
  sets used are $\{1, 1, 0.5, 1 \}$ (LHS) and $\{1.5, 1, 0.5, 1 \}$
  (RHS). The cut off radius used for both plots is $R = 15$.}
\label{fig:U1Vortices}
\end{center}
\end{figure}

\begin{figure}[t!]
\begin{center}
\includegraphics[scale=0.225]{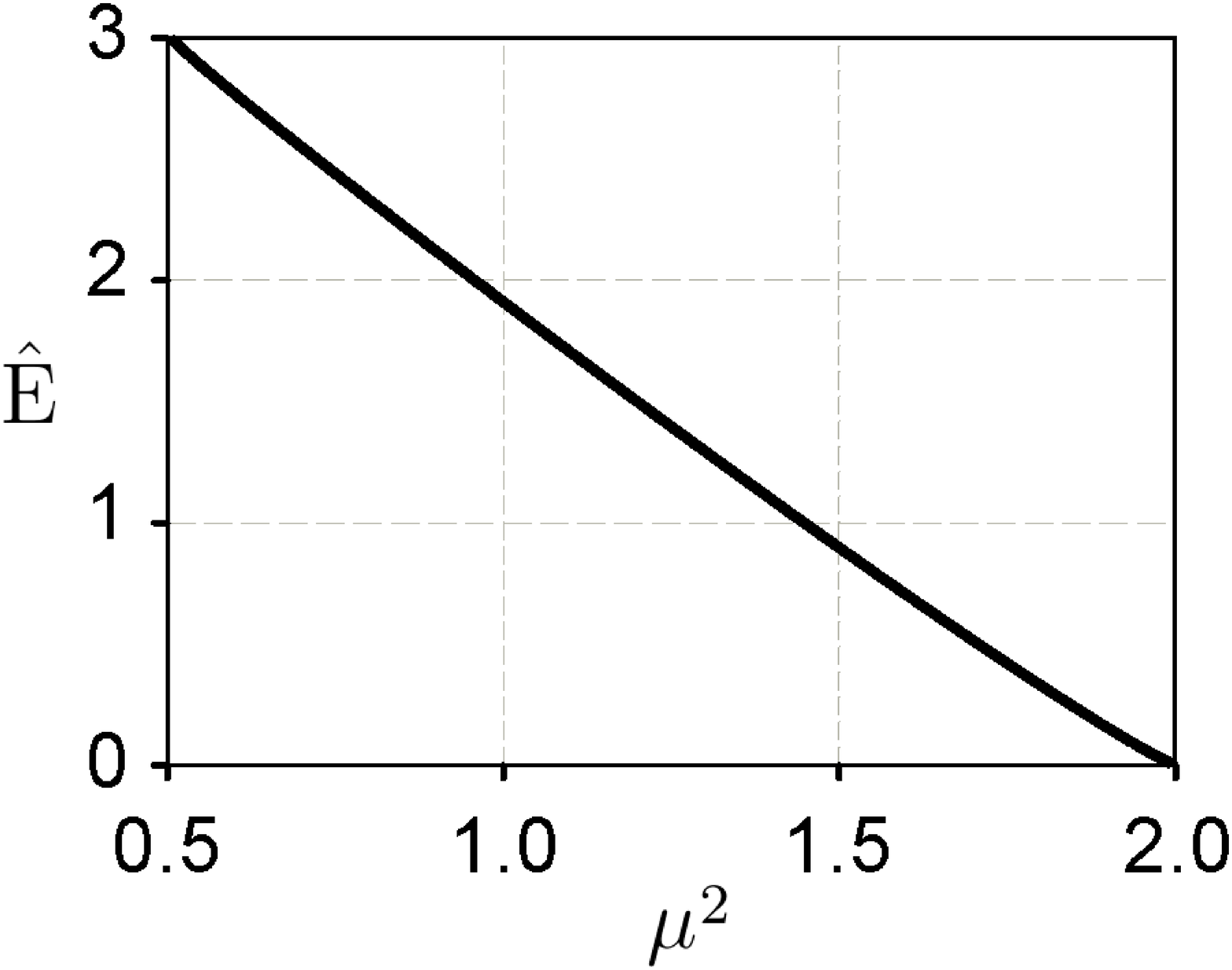}
\caption{\it Numerical evaluation of the dimensionless energy
  $\hat{\mathrm{E}}$, as a function of $\mu^2$ for the
  $\mathrm{U(1)}_{\mathrm{PQ}}$ invariant potential. Here, we use the
  fiducial values $\lambda = 1$, $g = 0.5$ and $n = 1$. Convexity of
  the potential and global minima conditions for these values require
  that $\mu^2 \in (0.5, 2.0)$. }
\label{fig:U1VortexEnergy}
\end{center}
\end{figure}

\begin{figure}[t!]
\begin{center}
\includegraphics[scale=0.188]{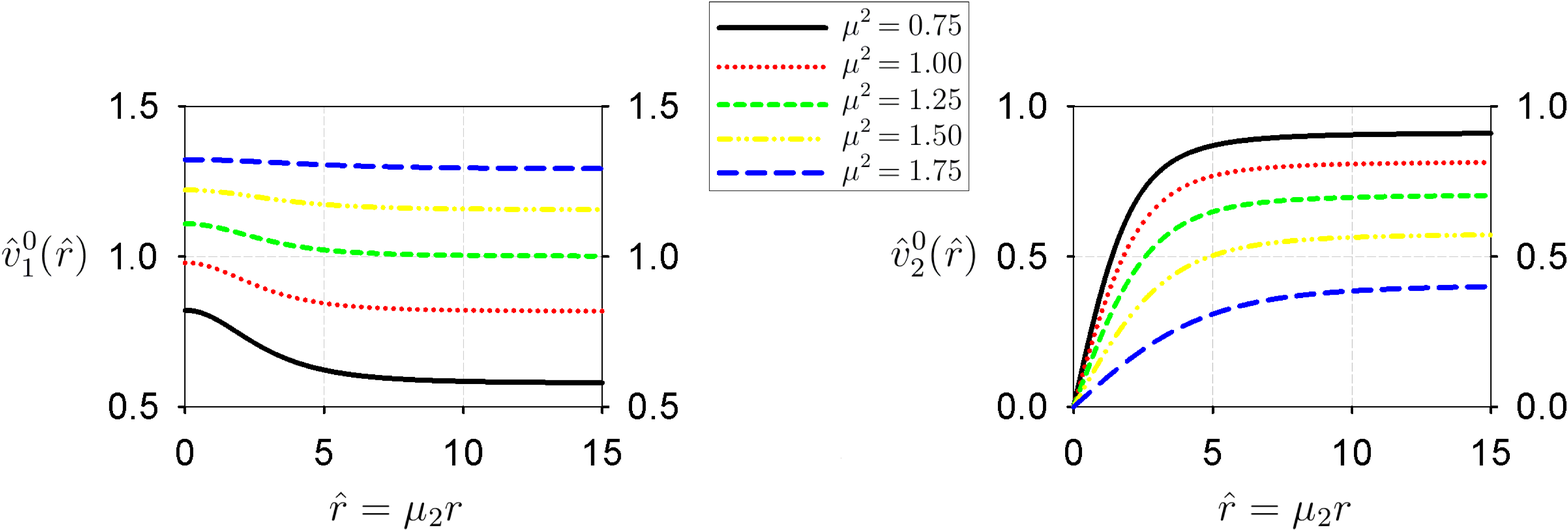}
\caption{\it Plots comparing various $\hat{v}_1^0(\hat{r})$ (LHS plot)
  and $\hat{v}_2^0(\hat{r})$ (RHS plot) curves for the
  $\mathrm{U(1)}_{\mathrm{PQ}}$ invariant potential. Here, we fix
  $\lambda = 1$, $g = 0.5$ and $n = 1$, and allow $\mu^2$ to vary. The
  cut off radius used for both plots is $R = 15$.}
\label{fig:U1Comparison1}
\end{center}
\end{figure}

\begin{figure}[t!]
\begin{center}
\includegraphics[scale=0.18]{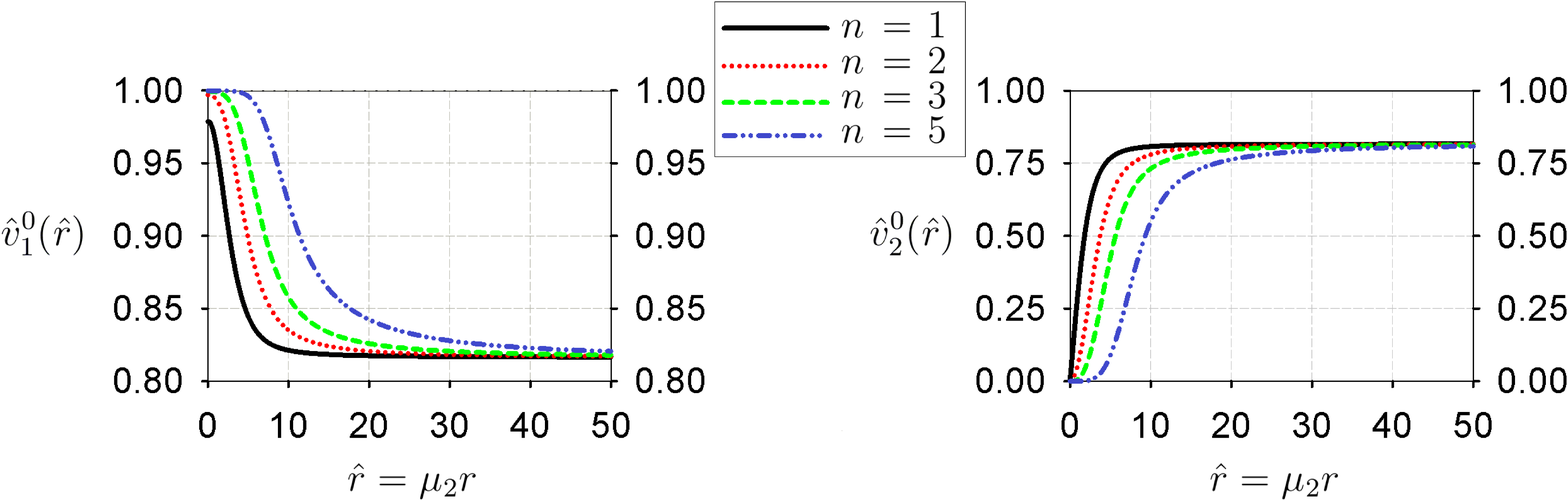}
\caption{\it Plots comparing various $\hat{v}_1^0(\hat{r})$ (LHS plot)
  and $\hat{v}_2^0(\hat{r})$ (RHS plot) curves for the
  $\mathrm{U(1)}_{\mathrm{PQ}}$ invariant potential for various
  winding numbers $n$. Here, we fix $\lambda = 1$, $g = 0.5$ and
  $\mu^2 = 1$, and allow $n$ to vary. The cut off radius used for both
  plots is $R = 50$}
\label{fig:U1Comparison2}
\end{center}
\end{figure}

We   conclude  our   $\mathrm{U(1)}_{\mathrm{PQ}}$  vortex   study  by
presenting two  typical solutions in  Figure~\ref{fig:U1Vortices}.  We
show the  general form  of the dimensionless  energy $\rm  \hat{E}$ in
Figure~\ref{fig:U1VortexEnergy}, as a function of $\mu^2$, noting that
the dimensionless energy tends to zero as $\mu^2$ approaches its upper
limit, i.e.  $\mu^2 \to  \frac{\lambda}{g}$.  We also directly compare
several  different solutions  in  Figures \ref{fig:U1Comparison1}  and
\ref{fig:U1Comparison2} by varying $\mu^2$  and the winding number $n$
respectively.      From     Figures    \ref{fig:U1Comparison1}     and
\ref{fig:U1Comparison2}  in   particular,  we  see   that  as  $\mu^2$
increases, the  width of the  vortex core increases and  similarly, as
the winding number increases, so does the width of the vortex core.

\subsubsection{CP3 Vortices}

As  discussed in  Section \ref{CP3  Symmetry}, the  CP3-symmetric 2HDM
potential can exhibit a non-simply connected vacuum manifold, provided
a neutral vacuum  global minimum solution exists where  the sum of the
squares of the two VEVs of the doublets $\phi_{1,2}$ is non-zero. Such
a  scenario can  be realized  within  the CP3-invariant  2HDM, if  the
matrix $\rm N_{\mu \nu}$ happens to be singular.  However, as shown in
Section  \ref{CP3 Singular  Section}, there  are two  possible neutral
vacuum  solutions  that  could   form  the  global  minimum  solution,
depending on  the relative magnitudes of the  quantities $2 \lambda_1$
and $\lambda_{34}$.  If $2 \lambda_1 > \lambda_{34}$, we find that any
possible  vortex solution  can  be removed  by gauge  transformations,
whereas for  cases with  $\lambda_{34} > 2  \lambda_1$, no  such gauge
transformations are  possible, allowing a vortex  solution.  Hence, we
study  cases of  the latter  type to  ensure vortex  formation  in the
CP3-invariant potential.

In  order to  obtain a  time-independent vortex  solution for  the CP3
symmetry, the following ansatz for the two Higgs doublets is used: 
\begin{equation}
\phi_1 (r, \chi)\ =\ \frac{1}{\sqrt{2}} \left( \begin {array}{c}
  0\\\noalign{\medskip}v(r) \cos (n \chi) \end {array} \right) \;,\qquad
\phi_2 (r,\chi)\ =\ \frac{1}{\sqrt{2}} \left( \begin {array}{c}
  0\\\noalign{\medskip} -v(r) \sin (n \chi) \end {array} \right) \; , 
\end{equation}
where the coordinate $r$ describes the space radially outward from the
core of  the vortex, and $\chi$  is an azimuthal  angle which accounts
for the  winding of  the vortex, with  winding number $n$.  Using this
ansatz,  the  energy  per  unit  length  of the  system  is  given  by
\eqref{eq:U1  Energy  Integral}  where  the  energy  density  for  the
CP3-invariant 2HDM is given by
\begin{equation}
\mathcal{E}\! \left( r \right)\ =\ \frac{1}{2} \left(\frac{d v}{dr}
\right)^2 + \frac{n^2}{2r^2} v(r)^2 - \frac{1}{2} \mu_1^2 v(r)^2 +
\frac{1}{4} \lambda_1 v(r)^4\ +\ \mathrm{V}_0 \; . 
\end{equation}
As before, \eqref{eq:U1  Energy Integral} is logarithmically divergent
for  the CP3  energy density  and so  we truncate  \eqref{eq:U1 Energy
  Integral} to a cut off radius  $r = R$. Our study can be simplified,
if we rescale the energy per unit length of the vortex in \eqref{eq:U1
  Energy Integral} to be dimensionless and introduce the dimensionless
quantities
\begin{equation}
\hat{r}\ =\ \mu_1 r \; , \qquad  \hat{v}(r)\ =\ \frac{v(r)}{\eta} \; ,\qquad
\hat{\mathrm{E}}\ =\ \frac{\mathrm{E}}{2 \pi \eta^2} \; ,
\end{equation}

\begin{figure}[t!]
\begin{center}
\includegraphics[scale=0.195]{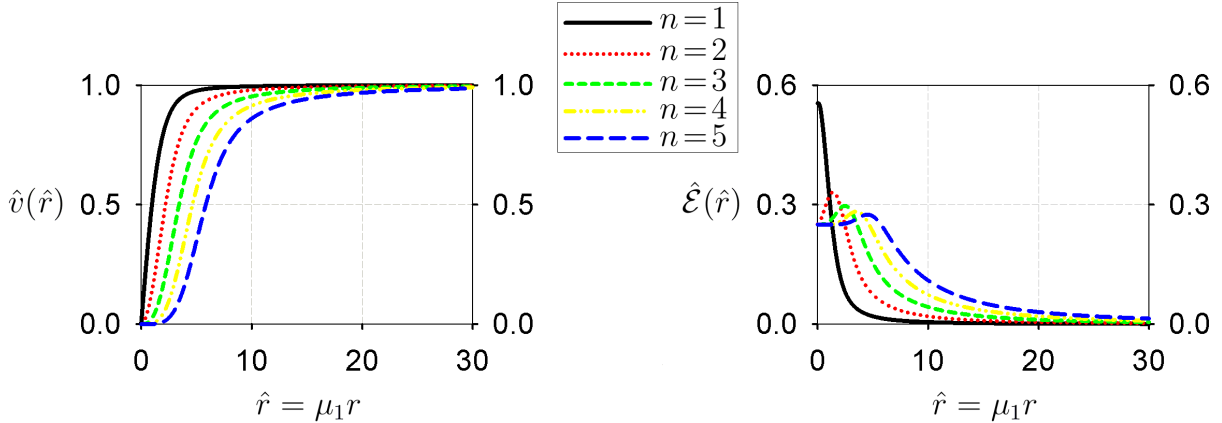}
\caption{\it Plots of $\hat{v} (\hat{r})$ (LHS plot) and the
  dimensionless energy density $\hat{\mathcal{E}}(\hat{r})$ (RHS plot)
  for the CP3-invariant potential. The winding number $n$ is varied
  from 1 to 5 and the cut off radius for both plots is $R = 30$.}
\label{fig:CP3Comparison}
\end{center}
\end{figure}

\noindent
with $\eta  = \frac{\mu_1}{\sqrt{\lambda_{1}}}$. We  then require that
$(v^0_1)^2 + (v^0_2)^2  = v^2_{\rm SM}$ and also  $\eta = v_{\rm SM}$.
Under  the above  parameter  re-definitions and  provided the  winding
number $n$  is non-zero, the  boundary conditions on the  vacuum field
$\hat{v}(\hat{r})$,  which  follows  from  \eqref{eq:CP3  zeta1  R^0},
become
\begin{equation}
\lim_{\hat{r} \to 0} \hat{v}(\hat{r})\ =\ 0 \; ,\qquad 
\lim_{\hat{r} \to  \infty} \hat{v}(\hat{r})\ =\ 1 \; . 
\end{equation}
These conditions force $\hat{v}(\hat{r})$ to be regular for all values
of $\hat{r}$ and ensure  that the dimensionful field $v(r)$ approaches
its VEV  in the limit $r \to  \infty$. The equation of  motion for the
rescaled field $\hat{v}(\hat{r})$ is
\begin{equation}
\frac{d^2 \hat{v}}{d\hat{r}^2}\ +\ 
\frac{1}{\hat{r}} \frac{d \hat{v}}{d\hat{r}}\ =\ 
\hat{v}\; \Big(-1 + \frac{n^2}{\hat{r}^2} + \hat{v}^2 \Big) \; .
\end{equation}
The  above differential equation  is solved  numerically, by  means of
gradient  flow methods.  Our  numerical analysis  only depends  on the
choice  of   the  single  parameter  $n$,   i.e.~the  winding  number.
Figure~\ref{fig:CP3Comparison}  presents the  dependence  of $\hat{v}$
and the corresponding energy density  $\hat{\cal E}$, as a function of
$\hat{r}$, for various values of $n$.  We observe that as the value of
$n$ increases, the  width of the vortex core  increases and the energy
density radially spreads out, giving the characteristic volcano shape.

\subsection{Global Monopoles}

We complete our study of the  topological defects that may form in the
2HDM due  to the spontaneous  breaking of the 6  accidental symmetries
with the  global monopole.  This topological solution  arises from the
symmetry  breaking of the  $\rm SO(3)_{HF}$  symmetry to  its subgroup
$\rm SO(2)_{HF}$.   In spite  of being intrinsically  unstable, global
monopoles may  have important  cosmological implications, as  they can
provide  a  mechanism  for  structure formation  within  the  Universe
\cite{PhysRevLett.65.1709}.

As with our previous topological defect studies, we do not analyse the
cosmological implications of the  2HDM's global monopole, but focus on
presenting an overview of possible solutions.

\subsubsection{SO$(3)_{\mathrm{HF}}$ Global Monopoles}

Our  analysis in  Section \ref{SU2  Symmetry}  has shown  that a  $\rm
SO(3)_{HF}$-invariant  2HDM potential  can exhibit  a  vacuum manifold
containing  non-contractible  2-spheres,  provided  a  neutral  vacuum
solution for a global minimum exists, such that $(v^0_1)^2 + (v^0_2)^2
=  v^2_{\rm  SM}$.   This  scenario  is  only  possible  in  the  $\rm
SO(3)_{HF}$-invariant 2HDM,  for a  singular matrix $\rm  N_{\mu \nu}$
[cf.~Section \ref{sec:SO3HFLMsingular}].

We may  seek a time-independent spherically  symmetric global monopole
solution  for the  $\rm SO(3)_{HF}$  symmetry,  by making  use of  the
following ansatz for the two Higgs doublets:
\begin{equation}
\phi_1 (r, \chi)\ =\ \frac{1}{\sqrt{2}} \left( \begin {array}{c}
  0\\\noalign{\medskip}v(r) \sin \chi \end {array} \right) \;, \qquad
\phi_2 (r,\chi, \psi)\ =\ \frac{1}{\sqrt{2}} \left( \begin {array}{c}
  0\\\noalign{\medskip} v(r) e^{i \psi} \cos \chi \end {array} \right)
\; , 
\end{equation}
where  the coordinate $r$  describes the  space radially  outward from
axis of  symmetry of  the monopole, $\chi$  is an azimuthal  angle and
$\psi$ is  a polar angle. Using  this ansatz, the  energy per monopole
is
\begin{equation}
\mathrm{E}\ =\ 4 \pi \int_0^\infty \!\!\! r^2 dr\;\; \mathcal{E}\! 
\left(\phi_1, \phi_2 \right) \; ,
\label{eq:SO(3) Energy Integral}
\end{equation}
where the  energy density for  the $\rm SO(3)_{HF}$ invariant  2HDM is
given by
\begin{equation}
\mathcal{E}\! \left( r \right)\ =\ 
\frac{1}{2} \left(\frac{d v}{dr} \right)^2 + \frac{1}{r^2} v(r)^2 -
\frac{1}{2} \mu_1^2 v(r)^2 + \frac{1}{4} \lambda_1
v(r)^4\ +\ \mathrm{V}_0 \; , 
\end{equation}
For  this type  of  energy density,  the  integral of  \eqref{eq:SO(3)
  Energy Integral}  is linearly divergent  in $r$, so we  truncate the
region of  integration from $[0,\infty)$ to  $[0, R]$, where  $R$ is a
  cut  off   radius~\cite{Vilenkin}.   Our  study   gets  considerably
  simplified, if we rescale \eqref{eq:SO(3) Energy Integral} to become
  dimensionless and introduce the dimensionless quantities
\begin{equation}
\hat{r}\ =\ \mu_1 r \; ,  \qquad \hat{v}(r)\ =\ \frac{v(r)}{\eta} \;
, \qquad \hat{\mathrm{E}}\ =\ \frac{\lambda_1
  \mathrm{E}}{4 \pi \mu_1} \; .
\end{equation}

\begin{figure}[t!]
\begin{center}
\includegraphics[scale=0.225]{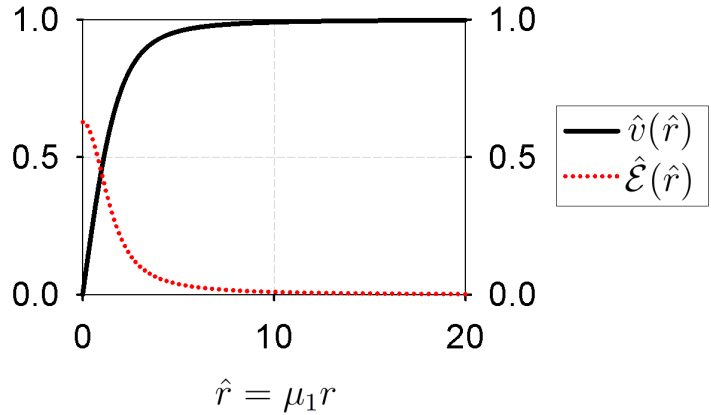}
\caption{\it Plots of $\hat{v} (\hat{r})$ and the dimensionless energy
  density $\hat{\mathcal{E}}(\hat{r})$ for the $\rm SO(3)_{HF}$
  invariant potential. The cut off radius for this plot is $R = 20$.} 
\label{fig:SO(3) Monopole}
\end{center}
\end{figure}

\noindent
We then  define $\eta =  \frac{\mu_1}{\sqrt{\lambda_{1}}}$ and require
that  $\eta  = v_{\rm  SM}$.   Under  these  rescalings, the  boundary
conditions on  the vacuum field $\hat{v}(\hat{r})$,  which follow from
\eqref{SO3:R0}, become
\begin{equation}
\lim_{\hat{r} \to 0} \hat{v}(\hat{r})\ =\ 0 \;,\qquad \lim_{\hat{r} \to
  \infty} \hat{v}(\hat{r})\ =\ 1 \; . 
\end{equation}
These conditions force $\hat{v}(\hat{r})$ to be regular for all values
of $\hat{r}$ and ensures  the dimensionful field $v(r)$ approaches its
VEV  in the  limit $r  \to  \infty$. The  equation of  motion for  the
rescaled field $\hat{v}(\hat{r})$ is
\begin{equation}
\frac{d^2 \hat{v}}{d\hat{r}^2}\ +\ \frac{2}{\hat{r}} \frac{d
  \hat{v}}{d\hat{r}}\ =\ \hat{v}\; \Big(-1 + \frac{2}{\hat{r}^2} +
\hat{v}^2 \Big) \; . 
\end{equation}
As with  the majority  of monopole studies,  we rely on  gradient flow
techniques to  numerically solve the above  differential equation.  In
Figure~\ref{fig:SO(3) Monopole}, we present the single solution for an
$\rm    SO(3)_{HF}$    global     monopole,    by    displaying    the
$\hat{r}$-dependence  of the vacuum  field $\hat{v}(\hat{r})$  and its
respective dimensionless energy density $\hat{\cal E}$.

\section{The U(1)$_\mathrm{Y}$-Violating 2HDM}
\label{Beyond the 2HDM}

In  this   section  we  discuss   the  application  of   our  Majorana
scalar-field formalism to 2HDM potentials which are not restricted by
the ${\rm U(1)_Y}$ hypercharge  group. Even though such potentials may
not be viable within the context of the SM, they may still be realized
in              models             describing             cosmological
inflation~\cite{Garbrecht:2006az,Battye:2008wu}.     Furthermore,   we
classify  all possible  15  symmetries  that may  occur  in a  general
$\rm U(1)_Y$-violating 2HDM potential, within the 6-dimensional bilinear field
space.

If conservation  under some ${\rm  U(1)_Y}$ hypercharge group  is lifted
from the theory, then  additional SU(2) gauge-invariant bilinears can,
in   principle,  be   present   in  the   2HDM   potential,  such   as
$\phi_1^{\mathrm{T}} i  \sigma^2 \phi_2$ and  its Hermitian conjugate,
$-\phi_2^\dagger i  \sigma^2 \phi_1^{*}$. Counting the  number of real
independent parameters, the resulting  potential would have 6 bilinear
mass terms and  20 quartic terms. Its explicit  analytic form is given
by
\begin{eqnarray}
 \label{eq:U1 less potential}
\mathrm{V} \!\!\!\!\!\!\!\!\! &&= -\mu_1^2 (\phi_1^{\dagger} \phi_1) -
\mu_2^2 (\phi_2^{\dagger} \phi_2) - m_{12}^2 (\phi_1^{\dagger} \phi_2)
- m_{12}^{*2}(\phi_2^{\dagger} \phi_1) - m_{34}^2 (\phi_1^{\mathrm{T}}
i \sigma^2 \phi_2) + m_{34}^{*2} (\phi_2^\dagger i \sigma^2
\phi_1^{*}) \nonumber \\ 
&&+ \lambda_1 (\phi_1^{\dagger} \phi_1)^2 + \lambda_2
(\phi_2^{\dagger} \phi_2)^2 + \lambda_3 (\phi_1^{\dagger}
\phi_1)(\phi_2^{\dagger} \phi_2) + \lambda_4 (\phi_1^{\dagger}
\phi_2)(\phi_2^{\dagger} \phi_1) + \frac{\lambda_5}{2}
(\phi_1^{\dagger} \phi_2)^2 + \frac{\lambda_5^{*}}{2}
(\phi_2^{\dagger} \phi_1)^2 \nonumber \\ 
&&+ \lambda_6 (\phi_1^{\dagger} \phi_1) (\phi_1^{\dagger} \phi_2) +
\lambda_6^{*} (\phi_1^{\dagger} \phi_1)(\phi_2^{\dagger} \phi_1) +
\lambda_7 (\phi_2^{\dagger} \phi_2) (\phi_1^{\dagger} \phi_2) +
\lambda_7^{*} (\phi_2^{\dagger} \phi_2) (\phi_2^{\dagger} \phi_1)
\nonumber \\ 
&&+ \lambda_8 (\phi^\dagger_1 \phi_1) (\phi_1^{\mathrm{T}} i \sigma^2
\phi_2) - \lambda_8^{*} (\phi^\dagger_1 \phi_1) (\phi_2^\dagger i
\sigma^2 \phi_1^{*}) + \lambda_9 (\phi^\dagger_2 \phi_2)
(\phi_1^{\mathrm{T}} i \sigma^2 \phi_2) - \lambda_9^{*}
(\phi^\dagger_2 \phi_2) (\phi_2^\dagger i \sigma^2 \phi_1^{*})
\nonumber \\ 
&&+ \lambda_{10} (\phi^\dagger_1 \phi_2) (\phi_1^{\mathrm{T}} i
\sigma^2 \phi_2) \!-\! \lambda_{10}^{*} (\phi^\dagger_2 \phi_1)
(\phi_2^\dagger i \sigma^2 \phi_1^{*}) \!+\! \lambda_{11}
(\phi^\dagger_2 \phi_1) (\phi_1^{\mathrm{T}} i \sigma^2 \phi_2) \!-\!
\lambda_{11}^{*} (\phi^\dagger_1 \phi_2) (\phi_2^\dagger i \sigma^2
\phi_1^{*}) \nonumber \\ 
&&+ \frac{\lambda_{12}}{2} (\phi_1^{\mathrm{T}} i \sigma^2 \phi_2)^2
+ \frac{\lambda_{12}^{*}}{2} (\phi_2^\dagger i \sigma^2
\phi_1^{*})^2 \; .
\end{eqnarray}
We note that the quartic couplings $\lambda_{1,2,3,4}$ are real and
$\lambda_{5,6, \dots, 12}$ are complex.

In order to account for the additional bilinear and quartic terms that
occur in the ${\rm U(1)_Y}$-violating 2HDM potential, we need to promote
the  4-vector  $\widetilde{\mathrm{R}}^\mu$  in \eqref{Rmu}  into  the
6-vector $\mathrm{R}^\mathrm{A}$, with $\mathrm{A} = 0,1,2,3,4,5$. The
individual components of $\mathrm{R}^\mathrm{A}$ read:
\begin{equation}
\mathrm{R}^\mathrm{A}\ =\ \left( \begin {array}{c} \phi_1^{\dagger}
 \phi_1+\phi_2^{\dagger} \phi_2\\\noalign{\medskip}\phi_1^{\dagger}
 \phi_2+\phi_2^{\dagger}
 \phi_1\\\noalign{\medskip}-i\left[\phi_1^{\dagger}
 \phi_2-\phi_2^{\dagger}
 \phi_1\right]\\\noalign{\medskip}\phi_1^{\dagger}
 \phi_1-\phi_2^{\dagger} \phi_2\\\noalign{\medskip}
 \phi_1^{\mathrm{T}} i \sigma^2 \phi_2 - \phi_2^{\dagger} i \sigma^2
 \phi_1^{*} \\\noalign{\medskip} -i \left[ \phi_1^{\mathrm{T}} i
 \sigma^2 \phi_2 + \phi_2^{\dagger} i \sigma^2 \phi_1^{*}
 \right] \end {array} \right) \; .
\end{equation}
As with the 4-vector $\mathrm{R}^\mu$, we can construct the 6-vector
$\mathrm{R}^\mathrm{A}$ using the 8-dimensional complex multiplet $\Phi$ as
\mbox{$\mathrm{R}^\mathrm{A} = \Phi^\dagger \Sigma^\mathrm{A} \Phi$}.
To determine the structure of $\Sigma^\mathrm{A}$, we start again with
the general ${\rm GL(8,\mathbb{C})}$ covariant (and ${\rm
 SU(2)_L}$-invariant) ansatz
\begin{equation}
\Sigma^\mathrm{A}\ =\ \Sigma^\mathrm{A}_{\;\; \alpha \beta}\:
\sigma^\alpha \otimes \sigma^\beta \otimes \sigma^0\; .
\end{equation}
The particular form of $\Sigma^\mathrm{A}_{\;\; \alpha \beta}$ is now
only constrained by the Majorana condition on $\Sigma^\mathrm{A}$,
namely $(\Sigma^\mathrm{A})^\mathrm{T} = \mathrm{C}^{-1}
\Sigma^\mathrm{A} \mathrm{C}$, in close analogy with~(\ref{eq:A
 transpose}). In terms of the tensor $\Sigma^\mathrm{A}_{\;\; \alpha
 \beta}$, the Majorana condition requires that
\begin{equation}
\Sigma^\mathrm{A}_{\;\; \alpha \beta}\ =\ \Sigma^\mathrm{A}_{\;\; \mu
 \nu} \eta^{\mu}_{\;\alpha} (\delta_-)^\nu_{\;\beta} \; .
\end{equation}
Only  6 elements  of $\Sigma^\mathrm{A}_{\;\;  \alpha  \beta}$ survive
this constraint: $\Sigma^\mathrm{A}_{\;\; 00}, \Sigma^\mathrm{A}_{\;\;
  01},   \Sigma^\mathrm{A}_{\;\;  03},   \Sigma^\mathrm{A}_{\;\;  12},
\Sigma^\mathrm{A}_{\;\;    22}$   and~$\Sigma^\mathrm{A}_{\;\;   32}$.
Hence,  the   six  components  of   the  6-vector  $\Sigma^\mathrm{A}$
compatible with the Majorana condition have the tensorial structure
\begin{eqnarray}
 \label{eq:sigma M form} 
\Sigma^{\mu} \!& = &\! \frac{1}{2}\left( \begin
 {array}{cc}\sigma^{\mu}& {\bf 0}_2\\
\noalign{\medskip} {\bf 0}_2 &(\sigma^\mu)^{\mathrm{T}} 
\end{array} \right) \otimes \sigma^0\; ,\nonumber\\[3mm]
\Sigma^4 \!& = &\! \frac{1}{2} \left( \begin {array}{cc}
 {\bf 0}_2 & i \sigma^2 \\
\noalign{\medskip}-i \sigma^2 & {\bf 0}_2 \end{array} \right) 
\otimes \sigma^0\; ,\qquad
\Sigma^5\ =\ \frac{1}{2} \left( \begin {array}{cc}
 {\bf 0}_2 & - \sigma^2 \\
\noalign{\medskip}- \sigma^2 & {\bf 0}_2 \end{array} \right) 
\otimes \sigma^0\; .
\end{eqnarray}
Comparing \eqref{eq:sigma  M form}  with \eqref{eq:sigma mu  form}, we
notice that the imposition  of the ${\rm U(1)_Y}$ hypercharge symmetry
on the SU(2)-invariant potential   restricts $\Sigma^{4,5} = {\bf 0}_8$
and     so    effectively    reduces     $\mathrm{R}^\mathrm{A}$    to
$\mathrm{R}^\mu$, as it should.

In  the  absence  of   the  ${\rm  U(1)_Y}$  hypercharge  symmetry,  the
transformation  matrix ${\rm M}$  no longer  splits into  two distinct
parts, but  takes on the general form  as determined in~(\ref{Mmunu}).
Under    a    ${\rm    SU(2)_L}$-invariant    reparameterization-group
transformation   ${\rm   M}  \in   {\rm   GL}(4,\mathbb{C})$  of   the
scalar-field multiplet $\Phi$,  with ${\rm M}^* = {\rm  C}\, {\rm M}\,
{\rm  C}$ [cf.~\eqref{eq:MajM}], the  6-vector $\mathrm{R}^\mathrm{A}$
transforms as
\begin{equation}
\mathrm{R}^\mathrm{A}\ \ \mapsto\ \ \mathrm{R}^{\prime\,\mathrm{A}}\ =\
e^{\sigma/8}\, \Lambda^\mathrm{A}_{\;\;\mathrm{B}}\;\mathrm{R}^\mathrm{B}\; ,
\end{equation}
where $e^\sigma = {\rm det}[\mathrm{M}^\dagger \mathrm{M}] > 0$ is a
real scale factor and $\Lambda^\mathrm{A}_{\;\;\mathrm{B}}$ is related
to the transformation matrix ${\rm M}$ by
\begin{equation}
 \label{eq:Lambda M equation for U1 less potential}
e^{\sigma/8}\, \Lambda^\mathrm{A}_{\;\;\mathrm{B}}\;
\Sigma^\mathrm{B}\ =\ \mathrm{M}^{\dagger} \Sigma^\mathrm{A}
\mathrm{M}\; .
\end{equation}
Note  that  the  matrix  $\Lambda^\mathrm{A}_{\;\;\mathrm{B}}$  is  an
element  of  SO(1,5). This  last  fact  may  be verified  by  defining
$\overline{\Sigma}^\mathrm{A}        \equiv        (\Sigma^0,        -
\Sigma^{1,2,3,4,5})$,  in direct  analogy  with $\overline{\sigma}^\mu
\equiv  (\sigma^0,   -\sigma^{1,2,3})$,  and  checking   the  Clifford
algebra:
\begin{equation}
  \label{eq:Clifford}
\Sigma^\mathrm{A}\, \overline{\Sigma}^\mathrm{B}\ +\
\Sigma^\mathrm{B}\, \overline{\Sigma}^\mathrm{A}\ =\ \frac{1}{2}\,\eta^{\rm
  AB}\, {\bf I}_8\; ,
\end{equation}
where ${\bf I}_8$ is  the 8-dimensional identity matrix and $\eta^{\rm
  AB} = {\rm diag}\,  (1,-1,-1,-1,-1,-1)$ is the respective metric for
the  $(1+5)$-dimensional   Minkowski  flat  space.    As  a  byproduct
of~\eqref{eq:Clifford}, we obtain that
\begin{equation}
\mathrm{tr}\,\big[\,\Sigma^\mathrm{A} \, \overline{\Sigma}^\mathrm{B}\,
 \big]\ =\ 2\,\eta^{\mathrm{AB}}\; .
\end{equation}
The       latter       can        be       used       to       compute
$\Lambda^\mathrm{A}_{\;\;\mathrm{B}}$ as
\begin{equation}
 \label{eq:LambdaAB}
\Lambda^\mathrm{A}_{\;\;\mathrm{B}}\ =\ \frac{1}{2}\; e^{-\sigma/8}\, \eta_{\rm BC}\; 
{\rm tr}\big[\, \mathrm{M}^{\dagger} \Sigma^\mathrm{A}
\mathrm{M} \overline{\Sigma}^\mathrm{C}\,\big]\; .
\end{equation}

With the aid of the newly introduced 6-vector $\mathrm{R}^\mathrm{A}$,
the potential of \eqref{eq:U1 less potential} can be written down in a
quadratic form similar to \eqref{eq:Pot before LM}:
\begin{equation}
\mathrm{V}\ =\ -\frac{1}{2}\; \mathrm{M}_\mathrm{A} \mathrm{R}^\mathrm{A}\:
+\: 
\frac{1}{4}\; \mathrm{L}_{\mathrm{AB}} \mathrm{R}^\mathrm{A} \mathrm{R}^\mathrm{B} 
\label{eq:Pot without U1 before LM}\; ,
\end{equation}
where the 6-vector $\mathrm{M}_\mathrm{A}$ containing the mass terms
and the $6\times 6$ quartic coupling matrix $\mathrm{L}_{\mathrm{AB}}$
read:

\vspace{-2mm}
{\scriptsize \begin{subequations}
\begin{align}
\mathrm{M}_\mathrm{A} \! &= \! \left( \begin {array}{cccccc} \mu_1^2 +
 \mu_2^2\,,&2\mathrm{Re}(m_{12}^2)\,,&-2\mathrm{Im}(m_{12}^2)\,,&\mu_1^2 -
 \mu_2^2\,,&2\mathrm{Re}(m_{34}^2)\,,&-2\mathrm{Im}(m_{34}^2) \end {array}
\right)\; , \\[2mm]
\mathrm{L}_{\mathrm{AB}} \!&=\! \left(\! \begin {array}{cccccc} \lambda_1 +
 \lambda_2 + \lambda_3&\mathrm{Re}(\lambda_6 +
 \lambda_7)&-\mathrm{Im}(\lambda_6 + \lambda_7)&\lambda_1 - \lambda_2
 &\mathrm{Re}(\lambda_8 + \lambda_9)&-\mathrm{Im}(\lambda_8 +
 \lambda_9)\\\noalign{\medskip}\mathrm{Re}(\lambda_6 +
 \lambda_7)&\lambda_4+\mathrm{Re}(\lambda_5)&
 -\mathrm{Im}(\lambda_5)&\mathrm{Re}(\lambda_6 
 - \lambda_7)&\mathrm{Re}(\lambda_{10} +
 \lambda_{11})&-\mathrm{Im}(\lambda_{10} + \lambda_{11}) 
\\\noalign{\medskip}-\mathrm{Im}(\lambda_6 +
\lambda_7)&-\mathrm{Im}(\lambda_5)&\lambda_4 -
\mathrm{Re}(\lambda_5)&-\mathrm{Im}(\lambda_6 -
\lambda_7)&-\mathrm{Im}(\lambda_{10} -
\lambda_{11})&-\mathrm{Re}(\lambda_{10} -
\lambda_{11})\\\noalign{\medskip}\lambda_1 -
\lambda_2&\mathrm{Re}(\lambda_6 - \lambda_7)&-\mathrm{Im}(\lambda_6 -
\lambda_7)&\lambda_1+\lambda_2-\lambda_3 &\mathrm{Re}(\lambda_8 -
\lambda_9)&-\mathrm{Im}(\lambda_8 - \lambda_9)\\\noalign{\medskip}
\mathrm{Re}(\lambda_8 + \lambda_9)&\mathrm{Re}(\lambda_{10} +
\lambda_{11})&-\mathrm{Im}(\lambda_{10} -
\lambda_{11})&\mathrm{Re}(\lambda_8 -
\lambda_9)&\mathrm{Re}(\lambda_{12})&-\mathrm{Im}(\lambda_{12})
\\\noalign{\medskip} -\mathrm{Im}(\lambda_8 +
\lambda_9)&-\mathrm{Im}(\lambda_{10} +
\lambda_{11})&-\mathrm{Re}(\lambda_{10} -
\lambda_{11})&-\mathrm{Im}(\lambda_8 -
\lambda_9)&-\mathrm{Im}(\lambda_{12})&-\mathrm{Re}(\lambda_{12}) \end
 {array} \!\right)\; .
\end{align}
\end{subequations} }
Note that in  the ${\rm U(1)_Y}$-symmetric limit, ${\rm  M_A} \to {\rm
  M}_\mu$  and  ${\rm  L_{AB}}  \to  {\rm  L}_{\mu\nu}$,  whereas  the
elements of ${\rm  M_A}$ and ${\rm L_{AB}}$ vanish  for the components
${\rm A,B} = 4, 5$.

We may now use an approach analogous to~\cite{Ivanov:2007de}, in order
to identify  all possible accidental symmetries that  could take place
within  a general  $\rm U(1)_Y$-violating  2HDM.  Requiring  that the  kinetic
terms  remain invariant  under  ${\rm GL(8,\mathbb{R})}$  scalar-field
transformations, we are restricted to consider unitary rotations ${\rm
  U} \in  {\rm U}(4)$ in  the $\Phi$-space, subject into  the Majorana
constraint:  ${\rm U}^*\  =\  {\rm  C}\, {\rm  U}\,  {\rm C}$.   These
Majorana-constrained  ${\rm U(4)}$  transformations  induce orthogonal
rotations  ${\rm  SO(5)} \subset  {\rm  SO(1,5)}$,  which  act on  the
`spatial'  components  ${\rm  A}   =  1,2,\dots  5$  of  the  6-vector
$\mathrm{R}^\mathrm{A}$.   In  detail, we  may  classify all  possible
symmetries derived from ${\rm SO}(5)$, which include ${\rm SO}(5)$ and
its proper, improper and  semi-simple subgroups. If~${\rm Z}_2$ is the
reflection group  for one of  the spatial components of  ${\rm R}^{\rm
  A}$, we may now list all the symmetries starting from the larger and
going  to the  smaller  group.  In  this  way, the  symmetries may  be
grouped into the following five categories:
\begin{eqnarray}
  \label{eq:O5}
\mbox{~~I.} \!\!&&\!\! {\rm SO}(5);\nonumber\\[3mm]  
\mbox{~II.} \!\!&&\!\! {\rm  O}(4)\otimes {\rm Z}_2;\ {\rm  SO}(4);\ 
                                                        \nonumber\\[3mm]  
\mbox{III.} \!\!&&\!\! {\rm O}(3)\otimes {\rm O}(2);\ {\rm SO}(3)\otimes ({\rm
      Z}_2)^2;\ {\rm  O}(3)\otimes  {\rm  Z}_2;\  {\rm  SO}(3); \\[3mm]
\mbox{~IV.} \!\!&&\!\! 
{\rm O}(2)\otimes {\rm O}(2)\otimes {\rm Z}_2;\  
{\rm O}(2)\otimes {\rm O}(2);\ {\rm O}(2)\otimes ({\rm Z}_2)^3\,;\ 
{\rm SO}(2)\otimes ({\rm Z}_2)^2;\nonumber\\ 
\!\!&&\!\! {\rm O}(2)\otimes {\rm Z}_2;\ {\rm SO}(2);\nonumber\\[3mm]  
\mbox{~~V.}\!\!&&\!\!  ({\rm Z}_2)^4;\ ({\rm Z}_2)^2\ .\nonumber
\end{eqnarray}
Note that  all the symmetry transformations have  determinant equal to
$+1$. With this restriction, we  get 15 distinct symmetries that could
act on a general  tree-level $\rm U(1)_Y$-violating 2HDM potential.  Moreover,
the above classification  in~\eqref{eq:O5} contains the ${\rm U(1)_Y}$
group.  More explicitly, the six accidental symmetries reported in the
literature are: the first symmetry  under Category III and the first 5
symmetries    under   Category~IV,   i.e.~${\rm    O}(3)\otimes   {\rm
  O}(2);\  {\rm  O}(2)\otimes   {\rm  O}(2)\otimes  {\rm  Z}_2;\  {\rm
  O}(2)\otimes {\rm  O}(2);\ {\rm O}(2)\otimes  ({\rm Z}_2)^3\,;\ {\rm
  SO}(2)\otimes  ({\rm  Z}_2)^2;\ {\rm  O}(2)\otimes  {\rm Z}_2$.   In
Table~\ref{table:Parameter  Conditions  for  first  6 symmetries  of  U1
  violating potential},  we show  the parameter restrictions  of these
six HF/CP  symmetries for the  full $\rm U(1)_Y$-violating 2HDM  potential, as
these are realized in a specific basis where the spatial part of ${\rm
  L}_{\rm  AB}$ (with  ${\rm  A},\, {\rm  B}  = 1,2\dots  5$) is  made
diagonal by an SO(5) rotation.  In such a diagonally reduced basis, we
have 
\begin{equation}
  \label{eq:reduced}
{\rm Im}\, \lambda_5\ =\ 0\;,\quad \lambda_6\ =\ \lambda_7\;,\quad
\lambda_8\ =\ \lambda_9\;,\quad
\lambda_{10}\ =\ \lambda_{11}\ =\ 0\;, 
\quad {\rm  Im}\,\lambda_{12}\ =\ 0\; .
\end{equation}

\begin{table}[!t]
{\small 
\begin{center}
\begin{tabular}{c||cccccccccccc}
\hline
Symmetry & 	$\mu_1^2$	&	$\mu_2^2$	&
$m_{12}^2$	&	$m_{34}^2$	&	$\lambda_1$	&
$\lambda_2$	&	$\lambda_3$	&	$\lambda_4$	&
${\rm Re}\,\lambda_5$	&	$\lambda_6 = \lambda_7$	&	$\lambda_8 =
\lambda_9$	\\
\hline
\hline

$\mathrm{Z}_2$	&	--	&	--	&	0	&
0	&	--	&	--	&	--	&	--
&	--	&	0	&	0	\\ 

\hline

$\mathrm{U}(1)_{\mathrm{PQ}}$	&	--	&	--	&
0	&	--	&	--	&	--	&	--
&	--	&	0	&	0	&	--	\\  

\hline

$\mathrm{SO(3)}_{\mathrm{HF}}$	&	--	&	$\mu_1^2$
&	0	&	--	&	--	&	$\lambda_1$
&	--	&	$2\lambda_1 - \lambda_3$	&	0
&	0	&	0	\\ 

\hline

CP1	&	--	&	--	&	Real	&	Real
&	--	&	--	&	--	&	--	&
--	&	Real	&	Real	\\ 

\hline

CP2	&	--	&	$\mu_1^2$	&	0	&
Real	&	--	&	$\lambda_1$	&	--	&
--	&	--	& 0  &	Real \\ 

\hline

CP3	&	--	&	$\mu_1^2$	&	0	&
Real	&	--	&	$\lambda_1$	&	--	&
--	&	$2\lambda_1 - \lambda_{34}$	 & 0	&
0 \\  

\hline
\end{tabular} 
\end{center} }
\caption{\it Parameter relations in the general $\rm U(1)_Y$-violating
  2HDM potential that result from the imposition of the six accidental
  symmetries, in the diagonally  reduced basis ${\rm Im}\, \lambda_5 =
  0$, $\lambda_{10} = \lambda_{11}  = 0$ and ${\rm Im}\,\lambda_{12} =
  0$  [cf.\  \eqref{eq:reduced}].  The  quartic  coupling ${\rm  Re}\,
  \lambda_{12}$  remains  unconstrained by  the  six considered  HF/CP
  symmetries.    Finally,  a   dash   indicates  the   absence  of   a
  constraint. \label{table:Parameter  Conditions for first  6 symmetries
    of U1 violating potential} }
\end{table}

Given the classification  in~\eqref{eq:O5}, we observe that symmetries
higher than ${\rm O}(3)$, which  contain the ${\rm U(1)_Y}$ group, can
still occur.  For instance, one  such symmetry is ${\rm SO}(5)$, which
is  obtained when $2\lambda_1  = 2\lambda_2  = \lambda_3$,  $\mu^2_1 =
\mu^2_2$, and all other  parameters vanish. The symmetry ${\rm SO}(5)$
is  equivalent  to ${\rm  O}(8)$~\cite{Deshpande:1977rw}  in the  real
field space and includes the gauge-group rotation ${\rm SU}(2)_{\rm L}
\otimes  {\rm   U(1)}_{\rm  Y}$.   In  the   extended  bilinear  ${\rm
  R^A}$-space,  ${\rm SO}(5)$ breaks  down to  ${\rm SO}(4)$  or ${\rm
  O(4)}\times {\rm Z}_2$, giving rise to four pseudo-Goldstone bosons,
as it  should be.  Notice that within the $\rm SU(2)_L$ and $\rm U(1)_Y$ constrained bilinear
formalism, it is not possible  to clearly make the distinction between
the  ${\rm SO(3)}_{\rm  HF}$ symmetry  and the  possible  higher HF/CP
symmetry~${\rm SO(5)}$.

\begin{table}[!t]
{\small 
\begin{center}
\begin{tabular}{c||cccccccccccc}
\hline
Symmetry & 	$\mu_1^2$	&	$\mu_2^2$	&
$m_{12}^2$	&	$\lambda_1$	&
$\lambda_2$	&	$\lambda_3$	&	$\lambda_4$	&
${\rm Re}\,\lambda_5$	&	$\lambda_6 = \lambda_7$	&	\\
\hline
\hline

SO(5)	&	--	&	$\mu_1^2$	&	0	&	--	&	$\lambda_1$	&	$2\lambda_1$	&	0	&	0	&	0		\\ 

\hline

$\rm O(4) \times Z_2$	&	--	&	$\mu_1^2$	& 0
&	--	&	$\lambda_1$	& --	& 0	& 0 & 0	\\  

\hline

$\mathrm{SO(4)}$	&	--	&	--	&	0
& --	&	--	& --	&	0	&	0
&	0	\\ 

\hline

$\rm O(3) \times O(2)$	&	--	&	$\mu_1^2$	&	0	&	-- &	$\lambda_1$	&	$2\lambda_1$	&	--	&	0	& 	0	\\ 

\hline

$\rm SO(3) \times (Z_2)^2$	&	--	&	$\mu_1^2$
&	0	&	--	&	$\lambda_1$	& --	&
--	&	$\lambda_4$	&	0	 \\  

\hline

$\rm O(3) \times Z_2$	&	--	&	$\mu_1^2$	&
Real	& --	&	$\lambda_1$	& -- &	--	& $\lambda_4$ 	&	Real	\\  

\hline

$\rm SO(3)$	& --	& --	& Real & --	& -- & 	--& --&
$\lambda_4$ &	Real	\\  

\hline
\end{tabular} 
\end{center} }
\caption{\it Parameter relations in the general $\rm U(1)_Y$-invariant
  2HDM  potential that result  from the  imposition of  the additional
  accidental  symmetries  shown  in   Categories  I,  II  and  III  of
  \eqref{eq:O5}, in the  reduced basis ${\rm Im}\, \lambda_5  = 0$ and
  $\lambda_6 =  \lambda_7$~[cf.~\eqref{eq:reduced}].  A dash indicates
  the  absence  of a  constraint.  \label{table:Parameter Conditions  of
    higher symmetries for U(1) invariant potential} }
\end{table} 

Another interesting  example is the  symmetry ${\rm SO}(4)$,  which is
obtained  from  a  ${\rm  U(1)_Y}$-  and  ${\rm  Z}_2$-invariant  2HDM
potential, with the additional  constraint that $\lambda_4 = \lambda_5
= 0$.  This  model is equivalent to the  model ${\rm O(4)}\otimes {\rm
  O}(4)$~\cite{Deshpande:1977rw} in the  scalar-field space, where the
second ${\rm O}(4)$ describes  the gauge group rotations. The symmetry
${\rm  SO(4)}$  breaks  into  ${\rm  SO(3)}$,  giving  rise  to  three
pseudo-Goldstone  bosons.   Again, this  breaking  scenario cannot  be
distinguished within a $\rm SU(2)_L$ and $\rm U(1)_Y$ constrained bilinear formalism, and can
be easily confused with  the CP3 symmetry. In Table \ref{table:Parameter
  Conditions of  higher symmetries  for U(1) invariant  potential}, we
display the  7 additional  accidental symmetries that  may occur  in a
$\rm   U(1)_Y$-invariant   2HDM   potential,  along   with   parameter
restrictions      obtained     in      the      diagonally     reduced
basis~[cf.~\eqref{eq:reduced}].    Note  that   all   symmetries  lead to
CP-invariant scalar  potentials.  Further details  of these additional
HF/CP symmetries will be given elsewhere.

\section{Conclusions}

Unlike the  SM, the 2HDM has  a rich landscape of  discrete and global
symmetries,  whose  spontaneous   breaking  may  lead  to  non-trivial
topological solutions.  In this  paper, we have  taken the  first step
towards analyzing  a number of generic symmetries  for their resulting
vacuum topology within the  2HDM. For definiteness, we have considered
the three HF symmetries: $\mathrm{Z}_2$, $\mathrm{U(1)}_{\mathrm{PQ}}$
and $\mathrm{SO(3)}_{\mathrm{HF}}$, and  the three CP symmetries: CP1,
CP2 and  CP3 (cf.~Table~2). In order  to study the  vacuum topology of
these  six  symmetries, we  have  introduced  a Majorana  scalar-field
formalism based on two subgroups of $\mathrm{GL(8,\mathbb{C})}$, where
the  HF  and CP  transformations  may  act  on a  single  scalar-field
multiplet representation.

Using Sylvester's criterion, we have derived the general conditions in
order   to  have   a  convex,   stable  and   bounded-from-below  2HDM
potential. Given  these convexity  and stability constraints,  we have
solved  analytically   the  minimization  conditions   of  the  scalar
potential, by  making use of  the Lagrange multiplier method.  We have
thus  obtained  all two  non-zero  solutions  for  the neutral  vacuum
expectation values of the Higgs doublets for the aforementioned six HF
and CP symmetries,  in terms of the gauge-invariant  parameters of the
theory.

In order to identify the  nature of the topological defects associated
with  the spontaneous  symmetry breaking  for  each of  the above  six
symmetries,  we have  studied  the homotopy  groups  of the  resulting
vacuum manifold after spontaneous symmetry breaking. In particular, we
have  found  the  existence  of  domain walls  from  the  breaking  of
$\mathrm{Z}_2$, CP1  and CP2  discrete symmetries, vortices  in models
with  broken $\mathrm{U(1)}_{\mathrm{PQ}}$  and CP3  symmetries  and a
global monopole in  a model with $\mathrm{SO(3)}_{\mathrm{HF}}$-broken
symmetry.  We  have  then  studied the  topological  defect  solutions
numerically, as functions of the  potential parameters of the 2HDM. We
have given numerical examples for each topological defect, showing the
energy of the defect for typical situations.

As we  have explicitly demonstrated in Section  \ref{Beyond the 2HDM},
our  Majorana scalar-field  formalism  can be  applied  to identify  7
further accidental symmetries in the 2HDM potential, which include the
maximal symmetries O(8) and ${\rm O(4)\otimes O(4)}$ in the real field
space~\cite{Deshpande:1977rw}.  These  symmetries remain undetected by
the constrained  ${\rm SU(2)}$  bilinear field approach  considered so
far in the literature.

Our Majorana  scalar-field formalism  can also be  used to  study more
general   scalar  potentials   which  are   not  constrained   by  the
$\mathrm{U(1)}_{\rm Y}$ hypercharge symmetry and can realize a maximal
number of  15 distinct  symmetries.  Such 2HDM  potentials may  not be
directly related  to the  observable SM gauge  group, but may  form an
independent  hidden  sector,  as  it  is, for  example,  the  case  in
supersymmetric              theories             of             hybrid
inflation~\cite{Garbrecht:2006az,Battye:2008wu}.    The  formation  of
topological defects, such as  domain walls, cosmic strings, monopoles,
or  textures, through  the  spontaneous symmetry  breaking of  global,
local or  discrete symmetries may have important  implications for the
analysis of  the cosmological data. It would  be therefore interesting
to analyze the cosmological  constraints on the fundamental parameters
of  the 2HDM,  using  the formalism  and  the computational  framework
developed in this~paper.

\newpage
\begin{appendix}
\appendixpage	
\addappheadtotoc	

\section{$\sigma^\mu$ Matrix Identities} \label{sigma Matrices Identities}

Here we list a number of useful identities for the matrices
$\sigma^\mu = (\sigma^0, \sigma^{1,2,3})$, where $\sigma^0 \equiv {\bf
 1}_2$ and $\sigma^{1,2,3}$ are the standard Pauli matrices. These
identities are used in Appendix~\ref{The Form of S} to derive the
explicit form of $\Sigma^\mu$. Under transposition and complex
conjugation, the individual components of $\sigma^\mu$ transform as
\begin{center}
\begin{tabular}{lll}
$(\sigma^{0})^\mathrm{T}\ =\ \sigma^0\; ,$ &\hspace*{10mm}&
 $(\sigma^{0})^*\ =\ \sigma^0\; ,$ \\ 
$(\sigma^{1})^\mathrm{T}\ =\ \sigma^1\; ,$ &\hspace*{10mm}&
 $(\sigma^{1})^*\ =\ \sigma^1\; ,$ \\ 
$(\sigma^{2})^\mathrm{T}\ =\ -\sigma^2\; ,$ &\hspace*{10mm}&
 $(\sigma^{2})^*\ =\ -\sigma^2\; ,$ \\ 
$(\sigma^{3})^\mathrm{T}\ =\ \sigma^3\; ,$ &\hspace*{10mm}&
 $(\sigma^{3})^*\ =\ \sigma^3\; .$ \\ 
\end{tabular} 
\end{center}
Hence, the above identities may be cast into the more compact form:
\begin{subequations}
\begin{align}
(\sigma^{\mu})^\mathrm{T}\ &=\ (\delta_{-})^{\mu}_{\;\nu} \sigma^{\nu}\; ,\\ 
 \label{eq:sigma star}
(\sigma^{\mu})^*\ &=\ (\delta_{-})^{\mu}_{\;\nu} \sigma^{\nu} \; ,
\end{align}
\end{subequations}
with
\begin{equation}
\mathrm{(\delta_{\pm})}^{\mu}_{\;\nu}\ \equiv\ \mathrm{diag} (1,1,\pm 1,1)\; .
\end{equation}
We will also frequently use the sandwich products
\begin{subequations}
\begin{align}
\sigma^1 \sigma^{\mu} \sigma^1\ &=\ \mathrm{(J_1)}^{\mu}_{\;\nu} \sigma^{\nu}\;, \\
\sigma^2 \sigma^{\mu} \sigma^2\ &=\ \mathrm{(J_2)}^{\mu}_{\;\nu} \sigma^{\nu}\;, \\
\sigma^3 \sigma^{\mu} \sigma^3\ &=\ \mathrm{(J_3)}^{\mu}_{\;\nu} \sigma^{\nu}\;,
\end{align}
\end{subequations}
where the tensors ${\rm J}_{1,2,3}$ are defined as
\begin{subequations}
\begin{align}
\mathrm{(J_1)}^{\mu}_{\;\nu}\ &\equiv\ \mathrm{diag} (1,1, -1, -1) \; , \\
\mathrm{(J_2)}^{\mu}_{\;\nu}\ &\equiv\ \mathrm{diag} (1, -1,1, -1) \; ,\\
\mathrm{(J_3)}^{\mu}_{\;\nu}\ &\equiv\ \mathrm{diag} (1, -1, -1,1) \; .
\end{align}
\end{subequations}
Finally, it is interesting to note the identity
\begin{equation}
\mathrm{(J_2)}^{\mu}_{\;\lambda}
(\delta_{-})^{\lambda}_{\;\nu}\ =\ \eta^\mu_{\;\nu}\; .
\label{eq:J2 Delta Nu Relation}
\end{equation}

\section{The Form of $\Sigma^\mu$ and the Transformation
 Matrices}\label{The Form of S} 

In order to derive the explicit form of $\Sigma^\mu$ in ${\rm
 GL}(8,\mathbb{C})$, we start with the following general ansatz:
\begin{equation}
 \label{eq:Ansatz}
\Sigma^\mu\ =\ \Sigma^\mu_{\;\alpha \beta}\; \sigma^\alpha \otimes
\sigma^\beta\; ,
\end{equation}
where we have suppressed the ${\rm SU(2)_L}$ gauge-group space for
convenience. Then, we need to apply two constraints to determine
the tensor coefficients $\Sigma^\mu_{\;\alpha \beta}$: the
$\mathrm{U(1)}_\mathrm{Y}$ constraint and the Majorana constraint.

\subsection{The $\mathrm{U(1)}_\mathrm{Y}$ Constraint on
 $\Sigma^\mu$} \label{U(1) Constraint}

Under a $\mathrm{U(1)}_\mathrm{Y}$ transformation, the 4-component
multiplet $\Phi$ defined in~(\ref{eq:Phi Definition}) transforms as follows:
\begin{equation}
\Phi^{'}\ =\ \mathrm{U}_\mathrm{Y} \Phi\; ,
\end{equation}
where
\begin{equation}
 \label{eq:U Y Definition}
\mathrm{U}_\mathrm{Y}\ =\ e^{i \mathrm{Y} \theta (\sigma^3 \otimes
 \sigma^0)} = \mathrm{diag} \left(e^{i \mathrm{Y} \theta}, e^{i
 \mathrm{Y} \theta}, e^{-i \mathrm{Y} \theta}, e^{-i \mathrm{Y}
 \theta} \right) = \mathrm{B}_{\nu} \sigma^{\nu} \otimes \sigma^{0}\; ,
\end{equation}
with
\begin{equation}
\mathrm{B}_{\nu}\ =\ \left[\cos\ (\mathrm{Y} \theta),\,0,\,0,\,i\sin
 (\mathrm{Y} \theta)\right]\; . 
\end{equation}
Invariance of the 4-vector $\mathrm{R}^\mu = \Phi^\dagger \Sigma^\mu \Phi$
[cf.~(\ref{Rmu})] under a ${\rm U(1)_Y}$ transformation implies the
following double equality constraint on $\Sigma ^\mu$:
\begin{equation}
 \label{eq:U1 A Constraint}
\Sigma^{\mu}\ =\ \mathrm{U}_\mathrm{Y}^* \Sigma^{\mu}
\mathrm{U}_\mathrm{Y}\ =\ \mathrm{U}_\mathrm{Y} \Sigma^{\mu}
\mathrm{U}_\mathrm{Y}^*\; . 
\end{equation}
Given the ansatz of $\Sigma^\mu$ in~\eqref{eq:Ansatz}, the above
double constraint gets translated into:
\begin{subequations}
 \begin{align}
\Sigma^\mu\ =\ \mathrm{U}_\mathrm{Y}^* \Sigma^{\mu}
\mathrm{U}_\mathrm{Y}\ &=\ \Sigma^{\mu}_{\;\alpha \beta}
\mathrm{B}^*_\nu \mathrm{B}_\lambda \left[(\sigma^\nu)^* \sigma^\alpha
 \sigma^\lambda \right] \otimes \sigma^\beta \; , \\ 
\Sigma^\mu\ =\ \mathrm{U}_\mathrm{Y} \Sigma^{\mu} \mathrm{U}_\mathrm{Y}^*\ &=\
\Sigma^{\mu}_{\;\alpha \beta} \mathrm{B}_\nu \mathrm{B}^*_\lambda
\left[\sigma^\nu \sigma^\alpha (\sigma^\lambda)^* \right] \otimes
\sigma^\beta \; .
\end{align}
\end{subequations}
Using the identity (\ref{eq:sigma star}), the above two relations can
be rewritten as
\begin{subequations}
\begin{align}
 \label{eq:U star A U}
\mathrm{U}_\mathrm{Y}^* \Sigma^{\mu} \mathrm{U}_\mathrm{Y}\ &=\
\Sigma^{\mu}_{\;\alpha \beta} \mathrm{B}^*_\nu \mathrm{B}_\lambda
\big[(\delta_{-})^{\nu}_{\;\gamma} \,\sigma^\gamma \sigma^\alpha
\sigma^\lambda \big] \otimes \sigma^\beta \; , \\
 \label{eq:U A U star} 
\mathrm{U}_\mathrm{Y} \Sigma^{\mu} \mathrm{U}_\mathrm{Y}^*\ &=\
\Sigma^{\mu}_{\;\alpha \beta} \mathrm{B}_\nu \mathrm{B}^*_\lambda
\big[(\delta_{-})^{\lambda}_{\;\gamma} \, \sigma^\nu \sigma^\alpha
\sigma^\gamma \big] \otimes \sigma^\beta \; .
\end{align} 
\end{subequations}
Substituting the explicit forms of $\mathrm{B}_{\mu}$ and
$(\delta_{-})^{\mu}_{\;\nu}$, \eqref{eq:U star A U} and \eqref{eq:U A
 U star} become respectively:
\begin{subequations}
\begin{align}
\Sigma^{\mu}\ &=\ \Sigma^{\mu}_{\;\alpha \beta} \left( \cos^2
(\mathrm{Y} \theta) \sigma^\alpha + \sin^2 (\mathrm{Y} \theta)
\mathrm{(J_3)}^{\alpha}_{\;\rho} \sigma^\rho + i \sin (\mathrm{Y}
\theta) \cos (\mathrm{Y} \theta) \left[\sigma^\alpha, \sigma^3 \right]
\right) \otimes \sigma^\beta\; ,\quad \\
\Sigma^{\mu}\ &=\ \Sigma^{\mu}_{\;\alpha \beta} \left(
\cos^2 (\mathrm{Y} \theta) \sigma^\alpha + \sin^2 (\mathrm{Y} \theta)
\mathrm{(J_3)}^{\alpha}_{\;\rho} \sigma^\rho - i \sin (\mathrm{Y}
\theta) \cos (\mathrm{Y} \theta) \left[\sigma^\alpha, \sigma^3 \right]
\right) \otimes \sigma^\beta\; .\quad
\end{align}
\end{subequations}
Evidently, in order that the above two constraints are satisfied, the
commutator term must vanish, i.e.
\begin{equation}
\left[\sigma^\alpha, \sigma^3 \right]\ =\ 0\; .
\end{equation}
This can only happen for the choices $\alpha = 0,3$, implying that
\begin{equation}
\Sigma^{\mu}_{\;1\beta}\ =\ \Sigma^{\mu}_{\;2 \beta}\ =\ 0\; ,
\end{equation}
independently of the Lorentz indices $\mu$ and $\beta$. As a
consequence, the ${\rm U(1)_Y}$ constraint leads to the block diagonal
form for the matrix $\Sigma^\mu$:
\begin{equation}
 \label{SmuU1Y}
\Sigma^\mu\ =\ 
\left( \begin {array}{cc}\Sigma^\mu_{\;0 \beta} \sigma^\beta
 &0\\\noalign{\medskip}0&\Sigma^\mu_{\;3 \beta} \sigma^\beta 
\end {array} \right)\; . 
\end{equation}

\subsection{The Majorana Constraint on $\Sigma^\mu$}

The Majorana condition~\eqref{MajoranaPhi} on the scalar multiplet
$\Phi$ gives rise to another important constraint on the form of
$\Sigma^\mu$. Specifically, the condition~\eqref{MajoranaPhi} implies
the invariance of vector $\mathrm{R}^\mu$ defined in~(\ref{Rmu}) under
charge conjugation. Thus, when $\Phi \to {\rm C}\, \Phi^*$,
$\mathrm{R}^\mu$ transforms as
\begin{equation}
\mathrm{R}^\mu\ =\ \Phi^{\dagger} \Sigma^{\mu} \Phi \ \ \to \ \
\mathrm{R}^\mu_{\rm C}\ =\ \Phi^\mathrm{T} \mathrm{C}^{\dagger}
\Sigma^{\mu} \mathrm{C}\Phi^* 
\ =\ 
\Phi^{\dagger}\mathrm{C}^\mathrm{T} (\Sigma^{\mu})^\mathrm{T}\mathrm{C}^*\Phi\; .
\end{equation}
Requiring that $\mathrm{R}^\mu = \mathrm{R}^\mu_{\rm C}$ yields the
Majorana constraint: 
\begin{equation}
 \label{eq:A transpose}
(\Sigma^{\mu})^\mathrm{T}\ =\ \mathrm{C}^{-1} \Sigma^{\mu}
 \mathrm{C}\; .
\end{equation}
For the general ansatz \eqref{eq:Ansatz}, the last constraint is
equivalent to
\begin{equation}
 \label{eq:CAC}
\Sigma^{\mu}_{\;\alpha \beta} (\sigma^\alpha)^\mathrm{T} \otimes 
(\sigma^\beta)^\mathrm{T} \ =\ \Sigma^{\mu}_{\;\alpha \beta} 
\left(\sigma^2 \sigma^{\alpha} \sigma^2 \right) \otimes
\sigma^{\beta}\; .
\end{equation}
Employing the identities of Appendix \ref{sigma Matrices Identities},
we obtain the constraining equation on $\Sigma^{\mu}_{\;\alpha
 \beta}$:
\begin{equation}
\Sigma^{\mu}_{\;\alpha \beta}\ =\ \Sigma^{\mu}_{\;\lambda \rho}
\eta^{\lambda}_{\;\alpha} (\delta_{-})^{\rho}_{\;\beta}\; . 
\end{equation}
Assuming that $\Sigma^\mu$ has the ${\rm U(1)_Y}$-invariant
form~(\ref{SmuU1Y}) and using the identity~\eqref{eq:J2 Delta Nu
 Relation} allows us to express $\Sigma^{\mu}_{\;\alpha \beta}$ as
follows:
\begin{equation}
 \label{B 0 Condition}
\Sigma^{\mu}_{\;\alpha \beta}\ =\ \left\{ \begin {array}{cl} 
\displaystyle \Sigma^{\mu}_{\;0 \rho} (\delta_{-})^\rho_{\;\beta}\;, & 
 \mbox{for $\alpha = 0$} \\ 
\displaystyle -\Sigma^{\mu}_{\;3 \rho} (\delta_{-})^\rho_{\;\beta}\;, & 
 \mbox{for $\alpha = 3$} 
\end {array} \right. 
\end{equation}
From this last expression, we find that the two non-zero parts of the
$\Sigma^\mu_{\;\alpha\beta}$ tensor are then, in general, proportional
to the following matrices:
\begin{subequations}
\begin{align}
\Sigma^{\mu}_{\;0 \rho}\ &\propto\ (\delta_{+})^\mu_{\;\rho} +
(\delta_{-})^{\mu}_{\;\rho}\; , \\ 
\Sigma^{\mu}_{\;3 \rho}\ &\propto\ (\delta_{+})^\mu_{\;\rho} -
(\delta_{-})^{\mu}_{\;\rho}\; .
\end{align}
\end{subequations}
This can be written down in the covariant form:
\begin{equation}
\Sigma^{\mu}_{\;\alpha \beta}\ =\ \mathrm{a}_{\alpha}
(\delta_{+})^{\mu}_{\;\beta} + \mathrm{b}_\alpha
(\delta_{-})^{\mu}_{\;\beta}\; , 
\end{equation}
where the vectors $\mathrm{a}_{\alpha}$ and $\mathrm{b}_\alpha$ 
are defined as
\begin{subequations}
\begin{align}
\mathrm{a}_{\alpha}\ &\equiv\ \frac{1}{4} \left(1,0,0,-1\right) \; , \label{eq:a}\\
\mathrm{b}_{\alpha}\ &\equiv\ \frac{1}{4} \left(1,0,0,1\right) \; . \label{eq:b}
\end{align}
\end{subequations}
Implementing all the above results, the ${\rm U(1)_Y}$-invariant
vector $\mathrm{R}^\mu$ compatible with the Majorana constraint takes on the
simple form:
\begin{equation}
 \label{eq:sigma mu form}
\Sigma^{\mu}\ =\ \frac{1}{2}\left( 
\begin {array}{cc}\sigma^{\mu}&0\\
\noalign{\medskip}0&(\sigma^\mu)^{\mathrm{T}} 
\end {array} \right)\; .
\end{equation}

\subsection{The Majorana Constraint on
 $\mathrm{GL(8,\mathbb{C})}$}\label{The Majorana Constraint on M}

It is interesting to discuss the reduction of the
$\mathrm{GL(8,\mathbb{C})}$ group under the Majorana constraint
$\mathrm{M}^{*} = \mathrm{C\,M\,C}$ for HF symmetries, where
$\mathrm{M} = M_{\mu \nu} \sigma^\mu \otimes \sigma^\nu$ (with
$M_{\mu\nu} \in \mathbb{C}$) becomes a general member of
$\mathrm{GL(4,\mathbb{C})}$ after suppressing the ${\rm SU(2)_L}$
gauge group space. The Majorana reduction, $\mathrm{M}$, pertinent to
CP transformations is analogous and will not be repeated here.
Applying the Majorana constraint on ${\rm M}$, we obtain the
expression
\begin{equation}
 \label{eq:M3} 
\mathrm{M}^{*}\ =\ M_{\mu \nu}^{*} \left(\sigma^{\mu} \right)^{*}
\otimes \left(\sigma^{\nu} \right)^{*}\ =\ M_{\mu \nu} \left(\sigma^2
\otimes \sigma^0 \right) \left(\sigma^{\mu} \otimes \sigma^{\nu}
\right) \left(\sigma^2 \otimes \sigma^0 \right)\; .
\end{equation}
We may now use the so-called mixed-product identity: $\left(\mathrm{A}
\otimes \mathrm{B}\right) \left(\mathrm{C} \otimes \mathrm{D} \right)
= \left(\mathrm{AC}\right) \otimes \left(\mathrm{BD}\right)$ and the
identity \eqref{eq:sigma star}, in order to rewrite \eqref{eq:M3} as
follows:
\begin{equation}
M_{\mu \nu}^{*} (\delta_{-})^{\mu}_{\;\alpha}
(\delta_{-})^{\nu}_{\;\beta} \sigma^{\alpha} \otimes
\sigma^{\beta}\ =\ M_{\mu \nu} \left(\sigma^2 \sigma^{\mu} \sigma^2
\right) \otimes \sigma^{\nu}\; . 
\end{equation}
Further use of the sandwich products given in Appendix \ref{sigma
 Matrices Identities} implies
\begin{equation}
 \label{eq:M4}
M_{\mu \nu}^{*} (\delta_{-})^{\mu}_{\;\alpha}
(\delta_{-})^{\nu}_{\;\beta} \sigma^{\alpha} \otimes
\sigma^{\beta}\ =\ M_{\mu \beta} \mathrm{(J_2)}^{\mu}_{\;\alpha}
\sigma^{\alpha} \otimes \sigma^{\beta}\; , 
\end{equation}
which translates into the constraining equation:
\begin{equation}
M_{\lambda \rho}^{*}\ =\ M_{\mu \nu} \eta^{\mu}_{\;\lambda}
(\delta_{-})^{\nu}_{\;\rho}\; . 
\end{equation}
Solving this last equation term by term results in 
the following constraints:
\begin{center}
\begin{tabular}{lllllll}
$M_{00} = M_{00}^{*}$ &\hspace*{10mm}& $M_{01} = M_{01}^{*}$
 &\hspace*{10mm}& $M_{02} = -M_{02}^{*}$ &\hspace*{10mm}& $M_{03} =
 M_{03}^{*}$ \\ 
$M_{10} = -M_{10}^{*}$ && $M_{11} = -M_{11}^{*}$ && $M_{12} =
 M_{12}^{*}$ && $M_{13} = -M_{13}^{*}$ \\ 
$M_{20} = -M_{20}^{*}$ && $M_{21} = -M_{21}^{*}$ && $M_{22} =
 M_{22}^{*}$ && $M_{23} = -M_{23}^{*}$ \\ 
$M_{30} = -M_{30}^{*}$ && $M_{31} = -M_{31}^{*}$ && $M_{32} =
 M_{32}^{*}$ && $M_{33} = -M_{33}^{*}$ \\ 
\end{tabular} 
\end{center}
Hence, from the 32 independent parameters of ${\rm M}$, half are
eliminated by the Majorana condition. The resulting 16 free
parameters generate a group which is isomorphic to
$\mathrm{GL(4,\mathbb{R})}$ acting on a complex four-dimensional
vector space.

\section{Trace and Determinant Relations for $\mathrm{N}_{\mu \nu}$
 and $\mathrm{L}_{\mu \nu}$}\label{The Determinant of N}

Relations involving the traces and determinants of $\mathrm{N}_{\mu
 \nu}$ and $\mathrm{L}_{\mu \nu}$ play an important role in the
calculation of the VEVs of the Higgs doublets and in the derivation of
stability and convexity conditions for the 2HDM potential.

To facilitate our presentation, we use the shorthand notation
$\mathrm{N} \equiv \mathrm{N}_{\mu \nu}$, $\mathrm{L} \equiv
\mathrm{L}_{\mu\nu}$ and $\eta \equiv \eta_{\mu\nu}$ to represent the
2-rank tensors as $4\times 4$ matrices. We also assume the standard
multiplication law between matrices, e.g.~$(\mathrm{N}^2)_{\mu\nu} =
\mathrm{N}_{\mu \alpha} \mathrm{N}_{\alpha \nu}$, $(\mathrm{L}
\eta)_{\mu\nu} = \mathrm{L}_{\mu \alpha} \eta_{\alpha \nu}$ etc. In
the above notation, the determinant of $\mathrm{N}$ may be written as
\begin{equation}
\mathrm{det} \left[\mathrm{N} \right]\ =\ \mathrm{det}
\left[\mathrm{L} - \zeta \eta \right]\; ,
\end{equation}
which can be calculated by the following determinant-trace identity:
\begin{equation}
\mathrm{det} \left[\mathrm{N} \right]\ =\ \frac{1}{24} \left\{
\mathrm{tr}^4 \left[\mathrm{N} \right] - 6 \mathrm{tr}^2
\left[\mathrm{N} \right] \mathrm{tr} \left[\mathrm{N}^2 \right] + 3
\mathrm{tr}^2 \left[\mathrm{N}^2 \right] + 8 \mathrm{tr}
\left[\mathrm{N} \right] \mathrm{tr} \left[\mathrm{N}^3 \right] - 6
\mathrm{tr} \left[\mathrm{N}^4 \right] \right\}\; . 
\end{equation}
The trace relations between $\mathrm{N}$ and $\mathrm{L}$ are found to
be
\begin{subequations}
\begin{align}
\mathrm{tr} \left[\mathrm{N} \right] &= \mathrm{tr} \left[\mathrm{L}
 \right] + 2 \zeta\; , \\ 
\mathrm{tr} \left[\mathrm{N}^2 \right] &= \mathrm{tr}
\left[\mathrm{L}^2 \right] - 2 \zeta \mathrm{tr} \left[\mathrm{L} \eta
 \right] + 4 \zeta^2\; , \\
\mathrm{tr} \left[\mathrm{N}^3 \right] &= \mathrm{tr}
\left[\mathrm{L}^3 \right] - 3\zeta \mathrm{tr} \left[\mathrm{L}^2
 \eta \right] + 3 \zeta^2 \mathrm{tr} \left[\mathrm{L} \right] + 2
\zeta^3\; , \\
\mathrm{tr} \left[\mathrm{N}^4 \right] &= \mathrm{tr}
\left[\mathrm{L}^4 \right] - 4 \zeta \mathrm{tr} \left[\mathrm{L}^3
 \eta \right] + 3 \zeta^2 \mathrm{tr} \left[\mathrm{L}^2 \right] + 2
\zeta^2 \mathrm{tr} \left[\mathrm{L} \eta \mathrm{L} \eta \right] - 4
\zeta^3 \mathrm{tr} \left[\mathrm{L} \eta \right] + 4 \zeta^4\; . 
\end{align}
\end{subequations}
Thus, the determinant of $\mathrm{N}$ is given by
\begin{equation}
\mathrm{det}[\mathrm{N}]\ =\ -\zeta^4 - A \zeta^3 - B \zeta^2 - C
\zeta - D\; ,
\label{eq:quartic function}
\end{equation}
where 
\begin{subequations}
\begin{align}
&\!\!\!\!A =\! -\mathrm{tr}[\mathrm{L} \eta]\;, \\
&\!\!\!\!B =\! \mathrm{tr} [\mathrm{L}^2] \!-\! \frac{1}{2} \mathrm{tr}^2
[\mathrm{L} \eta] \!+\! 2 \mathrm{tr} [\mathrm{L}^2 \eta] \!+\! \frac{1}{2}
\mathrm{tr} [\mathrm{L} \eta \mathrm{L} \eta] \!-\! \mathrm{tr}
 [\mathrm{L}] \left( 2\mathrm{tr}[\mathrm{L} \eta] \!+\!
 \mathrm{tr} [\mathrm{L}] \right) \;, \\ 
&\!\!\!\!C =\! -\mathrm{tr} [\mathrm{L}^3 \eta] \!+\! \mathrm{tr} [\mathrm{L}]
\left( \mathrm{tr} [\mathrm{L}^2 \eta] \!+\! \mathrm{tr} [\mathrm{L}^2]
\right) \!+\! \frac{1}{2} \mathrm{tr} [\mathrm{L} \eta] \left(
\mathrm{tr} [\mathrm{L}^2] \!-\! \mathrm{tr}^2 [\mathrm{L}] \right) \!-\!
\frac{1}{3} \mathrm{tr}^3 [\mathrm{L}] \!-\! \frac{2}{3} \mathrm{tr}
 [\mathrm{L}^3] \; ,  \\ 
&\!\!\!\!D =\! -\mathrm{det}[\mathrm{L}]\; . 
\end{align}
\end{subequations}
Notice that the coefficients $A$, $B$, $C$ and $D$ are entirely
expressed in terms of traces of powers of ${\rm L}$ and the
determinant of ${\rm L}$. These latter expressions depend explicitly
on the quartic couplings of the 2HDM potential as follows:
\begin{subequations}
\begin{align}
\mathrm{tr} \left[\mathrm{L} \right] &= 2\lambda_1 + 2\lambda_2 +
2\lambda_4\; ,\\
\mathrm{tr} \left[\mathrm{L}^2 \right] &= 4\lambda_1^2 + 4\lambda_2^2
+ 2\lambda_3^2 + 2\lambda_4^2 + 8|\lambda_5|^2 + 4|\lambda_6|^2 +
4|\lambda_7|^2\; , \\
\mathrm{tr} \left[\mathrm{L}^3 \right] &= 8 \lambda_1^3 + 8
\lambda_2^3 + 6(\lambda_1 + \lambda_2) \lambda_3^2 + 2 \lambda_4^3 +
24 \lambda_4 |\lambda_5|^2 + 12 \lambda_1 |\lambda_6|^2 + 12 \lambda_2
|\lambda_7|^2 \nonumber \\ 
&+ 12 \lambda_3 (R_6 R_7 + I_6 I_7) + 6(\lambda_4 + 2 R_5) (R_6^2 +
R_7^2) + 6(\lambda_4 - 2R_5)(I_6^2 + I_7^2) \nonumber \\ 
&+ 24I_5 (R_6 I_6 + R_7 I_7)\; , \\ 
\mathrm{tr} \left[\mathrm{L}^4 \right] &= 16 \lambda_1^4 + 16
\lambda_2^4 + 2 \lambda_3^4 + 16(\lambda_1^2 + \lambda_2^2 + \lambda_1
\lambda_2) \lambda_3^2 + 2 \lambda_4^4 + 16|\lambda_5|^2 (3
\lambda_4^2 + 2 |\lambda_5|^2) \nonumber \\ 
&+ 8 |\lambda_6|^2 (4 \lambda_1^2 + 2 \lambda_1 \lambda_4 +
\lambda_3^2 + \lambda_4^2 + 4 |\lambda_5|^2 + |\lambda_6|^2) \nonumber
\\ 
&+ 8 |\lambda_7|^2 (4 \lambda_2^2 + 2 \lambda_2 \lambda_4 +
\lambda_3^2 + \lambda_4^2 + 4 |\lambda_5|^2 + |\lambda_7|^2) \nonumber
\\ 
&+ 32 (\lambda_1 + \lambda_4)\left[R_5(R_6^2 - I_6^2) + 2I_5 R_6 I_6
 \right] + 32 (\lambda_2 + \lambda_4)\left[R_5(R_7^2 - I_7^2) + 2I_5
 R_7 I_7 \right] \nonumber \\ 
&+ 16 (R_6 R_7 + I_6 I_7)^2 + 16 \lambda_3 (2\lambda_1 + 2\lambda_2 +
\lambda_4)(R_6 R_7 + I_6 I_7) \nonumber \\ 
&+ 32 \lambda_3 \left[R_5(R_6 R_7 - I_6 I_7) + I_5 (R_6 I_7 + I_6 R_7)
 \right]\; , \\
\mathrm{tr} \left[\mathrm{L} \eta \right] &= 2\lambda_3 - 2\lambda_4\;,\\
\mathrm{tr} \left[\mathrm{L}^2 \eta \right] &= 4(\lambda_1 +
\lambda_2)\lambda_3 -2\lambda_4^2 -8|\lambda_5|^2 - 2 (R_6 - R_7)^2 -
2(I_6 - I_7)^2\; ,\\ 
\mathrm{tr} \left[\mathrm{L}^3 \eta \right] &= 4(2\lambda_1^2 +
2\lambda_2^2 + 2\lambda_1 \lambda_2 + |\lambda_6|^2 +
|\lambda_7|^2)\lambda_3 -4\lambda_1 |\lambda_6|^2 - 4 \lambda_2
|\lambda_7|^2 \nonumber \\ 
&+ 4(2\lambda_1 + 2\lambda_2 - \lambda_3)(R_6 R_7 + I_6 I_7) + 2
\lambda_3^3 - 2 \lambda_4^3 - 24 \lambda_4 |\lambda_5|^2 \nonumber \\ 
&- 4(\lambda_4 + 2R_5)(R_6^2 + R_7^2 - R_6 R_7) - 4(\lambda_4 -
2R_5)(I_6^2 + I_7^2 - I_6 I_7) \nonumber \\ 
&+ 8I_5 (I_6 R_7 + R_6 I_7 - 2R_7 I_7 - 2R_6 I_6)\; ,\\
\mathrm{tr} \left[\mathrm{L} \eta \mathrm{L} \eta \right] &= 8
\lambda_1 \lambda_2 + 2 \lambda_3^2 + 2 \lambda_4^2 + 8|\lambda_5|^2 -
8(R_6 R_7 + I_6 I_7)\; , \\ 
 \label{detL}
\mathrm{det}[\mathrm{L}] &= (4 \lambda_1 \lambda_2 -
\lambda_3^2)(\lambda_4^2 - 4 |\lambda_5|^2) - 4 \lambda_4 (\lambda_1
|\lambda_7|^2 + \lambda_2 |\lambda_6|^2) + 4 |\lambda_6|^2
|\lambda_7|^2 \nonumber \\ 
&- 4 (R_6 R_7 + I_6 I_7)^2 + 8 \lambda_1 \left[R_5 (R_7^2 - I_7^2) + 2
 I_5 R_7 I_7 \right] \nonumber \\ 
&+ 8 \lambda_2 \left[R_5 (R_6^2 - I_6^2) + 2 I_5 R_6 I_6 \right] + 4
\lambda_3 \lambda_4 (R_6 R_7 + I_6 I_7) \nonumber \\ 
&- 8 \lambda_3 \left[R_5 (R_6 R_7 - I_6 I_7) + I_5 (I_6 R_7 + R_6 I_7)
 \right]\; .
\end{align}
\end{subequations}

To find the values for the Lagrange multiplier $\zeta$ that lead to a
singular ${\rm N}$ matrix with $\mathrm{det}[\mathrm{N}] = 0$, we need
to solve a quartic equation. To do so, we first apply the standard
linear transformation to $\zeta$,
\begin{equation}
\rho\ =\ \zeta + \frac{A}{4}\ ,
\end{equation}
which enables one to reduce the quartic order polynomial of
\eqref{eq:quartic function} to the incomplete quartic equation
\begin{equation}
\rho^4 + \alpha \rho^2 + \beta \rho + \gamma\ =\ 0\;,
\label{eq:Incomplete Quartic}
\end{equation}
where 
\begin{subequations}
\begin{align}
\alpha\ &=\ B - \frac{3A^2}{8}\ , \\
\beta\ &=\ \frac{A^3}{8} - \frac{A B}{2} + C\; ,\\
\gamma\ &=\ D - \frac{A C }{4} + \frac{A^2 B}{16} -
\frac{3A^4}{256}\ . 
\end{align}
\end{subequations}
In terms of the quartic couplings $\lambda_{1,2,\dots 7}$, the
coefficients $\alpha$, $\beta$ and $\gamma$ are given by
\begin{subequations}
\begin{align}
\alpha\ &=\ -4 \lambda_1 \lambda_2 - \frac{1}{2} (\lambda_3 +
\lambda_4)^2 - 4 |\lambda_5|^2 + 4(R_6 R_7 + I_6 I_7)\; , \\
\beta\ &=\ (4|\lambda_5|^2 - 4 \lambda_1 \lambda_2) (\lambda_3 +
\lambda_4) + 4 \lambda_1 |\lambda_7|^2 + 4 \lambda_2 |\lambda_6|^2
\nonumber \\ 
& \ - 8 R_5 (R_6 R_7 - I_6 I_7) - 8 I_5 (I_6 R_7 + R_6 I_7)\; , \\
\gamma\ &=\ 16 \lambda_1 \lambda_2 |\lambda_5|^2 + \frac{1}{16}
(\lambda_3 + \lambda_4)^4 - (\lambda_3 + \lambda_4)^2 (\lambda_1
\lambda_2 + |\lambda_5|^2 + R_6 R_7 + I_6 I_7) \nonumber \\ 
& \ + 2(\lambda_3 + \lambda_4)(\lambda_1 |\lambda_7|^2 + \lambda_2
|\lambda_6|^2) - 4 |\lambda_6|^2 |\lambda_7|^2 + 4(R_6 R_7 + I_6
I_7)^2\nonumber \\ 
& \ - 8 \lambda_1 \left[R_5(R_7^2 - I_7^2) + 2I_5 R_7 I_7 \right] - 8
\lambda_2 \left[R_5(R_6^2 - I_6^2) + 2I_5 R_6 I_6 \right] \nonumber\\ 
& \ + 4(\lambda_3 + \lambda_4) \left[R_5 (R_6 R_7 - I_6 I_7) + I_5 (R_6
 I_7 + I_6 R_7)\right]\; . 
\end{align}
\end{subequations}
The analytical solutions to the incomplete quartic equation can now be
found by making using of the Descartes--Euler method. To this end, we
first construct the cubic resolvent equation of~\eqref{eq:Incomplete
 Quartic}, which is
\begin{equation}
x^3 + 2 \alpha x^2 + (\alpha^2 - 4 \gamma)x - \beta^2\ =\ 0\; , 
\end{equation}
whose roots are determined by the standard formulae:
\begin{subequations}
\begin{align}
x_1\ &=\ \frac{\delta}{6} + \frac{2 \alpha^2 + 24 \gamma}{3\delta} -
\frac{2\alpha}{3}\ , \\
x_2\ &=\ -\frac{(1 + i\sqrt{3})\delta}{12} - \frac{(1 - i
 \sqrt{3})(\alpha^2 + 12 \gamma)}{3\delta} - \frac{2\alpha}{3}\ , \\
x_3\ &=\ -\frac{(1 - i\sqrt{3})\delta}{12} - \frac{(1 + i
 \sqrt{3})(\alpha^2 + 12 \gamma)}{3\delta} - \frac{2\alpha}{3}\ ,
\end{align}
\end{subequations}
where
\begin{eqnarray}
\delta^3 \!&=&\! 8 \alpha^3 - 288\alpha \gamma + 108\beta^2\nonumber\\
 \!&&\!+ 12\sqrt{-48\alpha^4 \gamma + 384 \alpha^2 \gamma^2 - 768\gamma^3 +
 12 \alpha^3 \beta^2 - 432 \alpha \beta^2 \gamma + 81 \beta^4}\ . 
\end{eqnarray}
Having thus obtained the cubic roots $x_{1,2,3}$, the four roots
$\zeta_{1,2,3,4}$ of the original quartic equation
$\mathrm{det}[\mathrm{N}] = 0$ are then given by
\begin{subequations}
\begin{align}
\zeta_1\ &=\ -\frac{1}{2} \left(\sqrt{x_1} + \sqrt{x_2} + \sqrt{x_3}
\right) - \frac{A}{4}\ ,\\
\zeta_2\ &=\ -\frac{1}{2} \left(\sqrt{x_1} - \sqrt{x_2} - \sqrt{x_3}
\right) - \frac{A}{4}\ ,\\
\zeta_3\ &=\ -\frac{1}{2} \left(-\sqrt{x_1} + \sqrt{x_2} - \sqrt{x_3}
\right) - \frac{A}{4}\ ,\\
\zeta_4\ &=\ -\frac{1}{2} \left(-\sqrt{x_1} - \sqrt{x_2} + \sqrt{x_3}
\right) - \frac{A}{4} \ .
\end{align}
\end{subequations}

\section{Inverting the Transformation Matrix Relations}
\label{Inverting the Transformation Matrix Relations}

It would be useful to give the relations between the transformation
matrices $\mathrm{M}$ and the SO(1,3) matrices
$\Lambda^{\mu}_{\;\nu}$, by assuming that the scale factor is
$e^\sigma = 1$.

As was discussed in Section~\ref{subsec:Majorana}, the general matrix
$\mathrm{M}$ may describe both the HF and CP transformations by the
matrices ${\rm M}_+$ and ${\rm M}_-$, respectively, which contain the
reduced two-by-two matrices $\mathrm{T}_\pm \in {\rm
 SL}(2,\mathbb{C})$. Following \cite{nla.cat-vn651907}, we first
note that
\begin{equation}
 \label{eq:Trace Identity 1}
\sigma_\mu\, \mathrm{T}_{\pm}\, \bar{\sigma}^\mu\ =\
2\mathrm{tr}[\mathrm{T}_{\pm}]\; \sigma^0 \; ,
\end{equation}
where $\bar{\sigma}^\mu = (\sigma^0, -\sigma^{1,2,3} )$. On the other
hand, contracting \eqref{eq:M + Lambda Main Relation} and \eqref{eq:M
 - Lambda Main Relation} from the RHS by
$\overline{\Sigma}^\mu\ =\ {\rm diag}\, [ \bar{\sigma}^\mu\, , \,
(\bar{\sigma}^\mu)^{\rm T}]$ yields the relations
\begin{equation}
 \label{eq:Succinct Form}
(\Lambda_{\pm})^{\mu}_{\;\nu}\,\, (\delta_\pm)^\nu_{\;\lambda}
 \sigma^\lambda \bar{\sigma}_\mu\ =\ \mathrm{T}^\dagger_{\pm}
 \sigma_\mu \mathrm{T}_\pm \bar{\sigma}^\mu \; .
\end{equation}
Making use now of the identity~\eqref{eq:Trace Identity 1}, we can
solve for ${\rm T}^\dagger_\pm$:
\begin{equation}
\mathrm{T}^\dagger_{\pm}\ =\ \frac{1}{2 \mathrm{tr}[\mathrm{T}_\pm]}
\,\, (\delta_\pm)^\nu_{\;\lambda} (\Lambda_{\pm})^{\mu}_{\;\nu}\,\,
\sigma^\lambda \bar{\sigma}_\mu\ . 
\end{equation}
To  remove  the $\sigma^\mu$-dependence  from  the  RHS  of the  above
equation, we use  a relationship derived by taking  the determinant on
both sides of \eqref{eq:Succinct Form}:
\begin{equation}
\mathrm{det} [(\Lambda_{\pm})^{\mu}_{\;\nu}\,\,
 (\delta_\pm)^\nu_{\;\lambda} \sigma^\lambda \bar{\sigma}_\mu ]\ =\ 
\mathrm{det} [\mathrm{T}^\dagger_{\pm}
 (2\mathrm{tr}[\mathrm{T}_\pm]) ]\ =\ 
(2\mathrm{tr}[\mathrm{T}_\pm])^2\; \mathrm{det}
[\mathrm{T}^\dagger_{\pm} ]\; .
\end{equation}
Since ${\rm det}\,[\mathrm{T}_\pm ] = 1$ for $\mathrm{T}_\pm \in {\rm
 SL}(2,\mathbb{C})$, one arrives at
\begin{equation}
2\mathrm{tr}[\mathrm{T}_\pm]\ =\ \left\{ \mathrm{det}
\left[(\Lambda_{\pm})^{\mu}_{\;\nu}\,\, (\delta_\pm)^\nu_{\;\lambda}
 \sigma^\lambda \bar{\sigma}_\mu \right] \right\}^{\frac{1}{2}}\; .
\end{equation}
Here, we have omitted the negative solution from the square root as
this is accounted for by the U$(1)_\mathrm{Y}$ invariance of the
theory. Thus, one ends up with the expression
\begin{equation}
 \label{eq:T in terms of Lambda}
\mathrm{T}_{\pm}^\dagger\ =\ \frac{1}{\left\{ \mathrm{det}
 \left[(\Lambda_{\pm})^{\mu}_{\;\nu}\,\, (\delta_\pm)^\nu_{\;\lambda}
 \sigma^\lambda \bar{\sigma}_\mu \right] \right\}^{\frac{1}{2}}}
\,\, (\delta_\pm)^\nu_{\;\lambda} (\Lambda_{\pm})^{\mu}_{\;\nu}\,\,
\sigma^\lambda \bar{\sigma}_\mu \; .
\end{equation}
The determinant in the denominator of the above equation can be
calculated using the relation: $2\mathrm{det} [\mathrm{G}] =
\mathrm{tr[G]}^2 - \mathrm{tr}[\mathrm{G}^2]$, which results in
\begin{eqnarray}
\mathrm{D}_{\pm}^2 \!& =& \! \mathrm{det}\!
\left[(\Lambda_{\pm})^{\mu}_{\;\nu} (\delta_\pm)^\nu_{\;\lambda}
 \sigma^\lambda \bar{\sigma}_\mu \right] \nonumber\\
\!& =& \! 4 +
\mathrm{tr}[\Lambda_\pm \delta_\pm]^2 - \mathrm{tr}[(\Lambda_\pm
 \delta_\pm)^2] -i \epsilon^{\lambda \mu \rho \alpha}
(\Lambda_{\pm})_{\mu \nu} (\Lambda_{\pm})_{\alpha \beta}
(\delta_\pm)^{\nu}_{\;\lambda} (\delta_\pm)^{\beta}_{\;\rho}\; . 
\end{eqnarray}
Here we use the convention $\epsilon^{0123} = +1$ for the Levi--Civita
tensor. We can now use the identity
\begin{equation}
\sigma^\lambda \bar{\sigma}_\mu\ =\ \eta^\lambda_{\;\mu} \sigma^0 +
\eta_{\mu}^{\;\,0} \sigma^\lambda - \eta^{\lambda 0} \sigma_\mu + i
\epsilon^{0\lambda}_{\;\;\;\;\mu \alpha} \sigma^\alpha  \; ,
\end{equation}
to write down the numerator of \eqref{eq:T in terms of Lambda} in the form
\begin{equation}
(\delta_\pm)^\nu_{\;\lambda} (\Lambda_{\pm})^{\mu}_{\;\nu}\,\,
 \sigma^\lambda \bar{\sigma}_\mu = \mathrm{tr}[\Lambda_\pm
 \delta_\pm] \sigma^0 + \left\{(\delta_\pm)^\mu_{\;i}
 (\Lambda_{\pm})^{0}_{\;\mu} - (\delta_\pm)_{\mu}^{\;\,0}
 (\Lambda_{\pm})_{i}^{\;\,\mu} + i \epsilon^{0 \nu}_{\;\;\;\; \mu i}
 (\delta_\pm)^\alpha_{\;\nu} (\Lambda_{\pm})^{\mu}_{\;\alpha}
 \right\} \sigma^i\; . 
\end{equation}
Using the representation $\mathrm{T}_\pm = (\mathrm{T}_\pm)_\mu
\sigma^\mu$, the individual components of $({\rm T}_\pm)_\mu$ derived
from~(\ref{eq:T in terms of Lambda}) are given by
\begin{subequations}
\begin{align}
(\mathrm{T}_{\pm})_0 &= \frac{1}{\mathrm{D}_\pm}
 \mathrm{tr}[\Lambda_\pm \delta_\pm]\; , \\
(\mathrm{T}_{\pm})_i &= \frac{1}{\mathrm{D}_\pm}
 \left[\, (\delta_\pm)^\mu_{\;\;i} (\Lambda_{\pm})^{0}_{\;\mu} -
 (\delta_\pm)_{\mu}^{\;\,0} (\Lambda_{\pm})_{i}^{\;\,\mu} - i
 \epsilon^{0 \nu}_{\;\;\;\; \mu i} (\delta_\pm)^\alpha_{\;\nu}
 (\Lambda_{\pm})^{\mu}_{\;\alpha} \right] \; .
\end{align}
\end{subequations}

\end{appendix}

\newpage
\bibliographystyle{JHEP-2}
\bibliography{biblio}
\end{document}